\documentclass[a4paper,12pt]{article} 
\usepackage{amsmath, amssymb, amsfonts,amstext, amsbsy, array, graphicx} 
\usepackage{a4,cite,float, epic, epsfig,latexsym, curves, wrapfig, subfigure}
\usepackage{graphics}
\usepackage{colortbl}
\usepackage{dsfont}
\usepackage{rotating}
\usepackage{varioref}
\usepackage{citesort}

\usepackage{xcolor}

\usepackage{pstricks}

\usepackage{pst-node}

\usepackage{pst-plot}

\usepackage{pst-coil} 

\usepackage[english]{babel}

\newcmykcolor{lightorange}{0 0.1 0.2 0}
\newcmykcolor{orange}{0 0.5 1 0}
\newcmykcolor{redorange}{0 0.75 1 0}
\newcmykcolor{lightred}{0 0.125 0.125 0}
\newcmykcolor{lightgreen}{0.125 0 0.125 0}
\newcmykcolor{lightcyan}{0.25 0 0 0}
\newcmykcolor{lightredorange}{0 0.125 0.133 0}
\newlength{\unit}

\psset{xunit=\unit,yunit=\unit,runit=\unit}
\newlength{\dlinewidth}
\newlength{\linew}
\newlength{\doublesep}
\psset{doublesep=\doublesep}
\psset{linewidth=\linew}
\psset{dotscale=0.8}

\numberwithin{equation}{section}

\def\beq{\begin{equation}} \def\eeq{\end{equation}}
 
\newcommand{\bea}{\begin{eqnarray}}
\newcommand{\eea}[1]{\label{#1}\end{eqnarray}}
\renewcommand{\Re}{\mbox{Re}\,} \renewcommand{\Im}{\mbox{Im}\,}
 \def\tm{\theta_-}

 \def\N{{\cal N}}

\newcommand{\gr}[1]{\underline{#1}}
\newcommand{\col}{~,}
\newcommand{\pnt}{~.}
\newcommand{\AdS}{\text{AdS}}

\newcommand{\twob}{{\text{II}\,\text{B}}}
\newcommand{\DBI}{\mathrm{DBI}}
\newcommand{\CS}{\mathrm{CS}}
\newcommand{\E}{\mathrm{E}}
\renewcommand{\S}{\mathrm{S}}

\newcommand{\preparderiv}[1]{\frac{\partial}{\partial #1}}

\newcommand{\comm}[2]{\left[#1\smash[b]{\mathbin{,}}#2\right]}

\newcommand{\de}{\operatorname{d}\!}

\newcommand{\e}{\operatorname{e}}

\newlength{\neglength}

\DeclareMathOperator{\tr}{tr}

\DeclareMathOperator{\re}{Re}
\DeclareMathOperator{\im}{Im}
\DeclareMathOperator{\vol}{vol}
\begin{document}


\thispagestyle{empty}

\begin{flushright}
\vspace{-2cm}
{\small MPP-2006-116 \\
LMU-ASC 61/06 \\
IFUM 875-FT \\
\texttt{hep-th/0610276}}
\end{flushright}
\vspace{1.0cm}
\begin{center} \noindent \Large \bf
Adding flavour to the Polchinski-Strassler background
\end{center}
\renewcommand{\thefootnote}{\fnsymbol{footnote}}

\vspace{1cm}
\centerline{Riccardo Apreda ${}^a$, Johanna
Erdmenger ${}^{a,b}$, Dieter L\"ust ${}^{a,b}$, Christoph Sieg ${}^c$
\footnote[1]{\noindent \tt apreda@mppmu.mpg.de, \\
\hspace*{6.3mm}jke@mppmu.mpg.de, \\
\hspace*{6.3mm}luest@mppmu.mpg.de, \\ 
\hspace*{6.3mm}csieg@mi.infn.it .\\}}

 \vspace{1cm}
\begin{center}
\emph{ $^a$ Max Planck-Institut f\"ur Physik (Werner Heisenberg-Institut),\\ 
 F\"ohringer Ring 6, 80805 M\"unchen, Germany}
\\
\vspace{0.2cm}
\emph{$^b$ Arnold-Sommerfeld-Center for Theoretical Physics, Department f\"ur Physik,
Ludwig-Maximilians-Universit\"at, Theresienstra\ss e 37, 80333 M\"unchen, Germany}\\
\vspace{0.2cm}
\emph{$^c$  Dipartimento di Fisica, Universit\`a degli Studi di Milano, \\
Via Celoria 16, 20133 Milano, Italy}
\end{center}

\medskip
 
\begin{abstract}
\medskip
\noindent 
As an extension of holography with flavour,
we analyze in detail the embedding of a  
$\text{D}7$-brane probe into the Polchinski-Strassler gravity background,
in which the breaking of conformal symmetry is induced by 
a $3$-form flux $G_3$.  This corresponds to giving masses to the
adjoint chiral multiplets.
We consider the $\N=2$ supersymmetric case in which one of the adjoint chiral 
multiplets is kept massless while the masses of the other two are equal. 
This setup requires a generalization of the known expressions for the 
backreaction of $G_3$ in the case of three equal masses to generic mass 
values. We work to second order in the masses to obtain the embedding of 
$\text{D}7$-brane probes in the background. At this order, 
the $2$-form potentials 
corresponding to the background flux induce an $8$-form potential
which couples to the worldvolume of the $\text{D}7$-branes. We show
that the embeddings preserve an $SU(2) \times SU(2)$ symmetry. 
We study possible embeddings both analytically in a particular
approximation, as well as numerically.
The embeddings preserve supersymmetry, as we investigate using the
approach of holographic
renormalization. The meson spectrum associated to one of the
embeddings found reflects
the presence of the adjoint masses by displaying a mass gap.
\end{abstract}

\renewcommand{\thefootnote}{\arabic{footnote}}
\newpage
\section{Introduction and summary}

\setcounter{equation}{0}

Over the last years, substantial progress has been made in the context 
of the AdS/CFT correspondence \cite{Maldacenaoriginal} towards a
gravity dual description of QCD-like theories, in particular also for
theories which involve fields in the fundamental representation of
the gauge group, i.e.\ quarks \cite{Bertolini} - \cite{Karch:2002sh}. 
Quark fields in the fundamental representation 
can be introduced for instance by adding
$\text{D}7$-brane probes in addition to the  $\text{D}3$-branes 
responsible for the adjoint degrees of freedom. Moreover, supersymmetry can
be broken further by turning on additional background fields, i.e. by 
embedding the branes into less supersymmetric backgrounds.
For theories of this kind, holographic descriptions
in particular  of meson spectra and decay constants \cite{krumy,meson}
chiral symmetry breaking by quark
condensates \cite{BEEGK} - \cite{chiral} and 
thermal phase transitions \cite{AEEG,thermal} 
have been found, using a
variety of brane constructions in different supergravity backgrounds.
In a number of examples, there is astonishing agreement with
experimental results.
There have also been phenomenological `bottom-up' approaches inspired
by the string-theoretical results \cite{AdSQCD,Hong}. Moreover from a more
theoretical point of view there have been embeddings of brane probes
in the Klebanov-Strassler \cite{KS} and 
Maldacena-Nu\~nez
\cite{MN} backgrounds \cite{SakaiSonnenschein,KSMN}, and progress towards 
holographic models of flavour beyond the probe approximation has been
made \cite{NunezParedes,beyond}.

All of these holographic models have been perfectioned in a number of
respects. However, even if remaining in the supergravity
approximation and in the probe limit, 
there are still aspects which are desirable to improve. For instance,
it is desirable to embed a $\text{D}7$-brane probe into a gravity background 
with a well-controlled infrared behaviour in the interior, 
which in addition returns to a well-controlled four-dimensional field theory 
in the ultraviolet near the boundary. As we discuss in this paper, 
this is achieved by embedding a $\text{D}7$-brane
probe into the $\N=2 $ version of the Polchinski-Strassler background
\cite{Polchinski:2000uf}. The field theory dual to this background is
known as $\N=2^*$ theory and 
corresponds to giving mass to the adjoint $\N=2$
hypermultiplet in the $\N=4$ theory.

Moreover, intrinsically the models considered for holography with flavour
have only one scale parameter,
usually associated to the supergravity background, which sets the
scale of both supersymmetry breaking and conformal symmetry 
breaking\footnote{Throughout this paper we remain in the
  supergravity approximation, such that the string tension $\alpha'{}^{-1}$
  remains large. This paragraph merely refers to the fact that the holographic
  models yield e.g.\ meson masses of the order of the SUSY breaking scale.}. 
Generically the observables calculated in these models are also of the
order of magnitude of this same scale. This is unsatisfactory from the
phenomenological point of view, since the meson masses, for instance, 
are known to be much smaller than the SUSY breaking scale. 
A possible approach to separating the two scales (i.e.\ meson masses and SUSY
breaking scale) may be an
appropriately adapted version of the Giddings-Kachru-Polchinski
mechanism \cite{GKP,Kachru} in which scales are separated by fluxes. 
As a precursor to such a mechanism, in this paper we study 
$\AdS/\text{CFT}$
with flavour in a supergravity background where the symmetry breaking
is generated by the $3$-form flux $G_3$. 
These are our two main motivations for embedding a $\text{D}7$-brane probe 
into a suitable form of the Polchinski-Strassler background.

For embedding a $\text{D}7$-brane probe, it turns out that a sufficiently
symmetric and thus tractable background is the $\N=2$ version where
two of the adjoint  chiral multiplet masses are equal, while the third
vanishes. 
We consider this background in an expansion in the adjoint masses 
$m_p$, $p=1,2,3$
to second order. For the $\N=1^*$ case, such extensions have been computed
in  \cite{Freedman:2000xb} and to third order in 
\cite{LopesCardoso:2004ni}, {which lead to a dynamical formation of
a gaugino condensate.} Here, we modify
these results in order to obtain the $\N=2^*$ case 
with $m_1=0$, $m_2=m_3=m$. 

To all orders in the adjoint masses,
the corresponding supergravity solution has been constructed in
\cite{Warner2000,Pilch:2003jg,Brandhuber}, 
however without explicit reference to
the fluxes. Thus it is a slightly different approach from the one
considered here. On the field theory side, the theory corresponding to
this background flows to the Donagi-Witten integrable 
field theory \cite{Donagi} in the infrared.

We discuss the structure of our order $\mathcal{O}(m^2)$
metric in the deep interior 
of the space. We find that two overlapping 2-spheres form whose radius is 
of order $\mathcal{O}(m)$. Denoting the ten dimensions by 
$x^\mu$, $\mu=0,\dots,3$ and $y^i$, $i=4,\dots,9$, the two spheres form in the
$y^5, y^6, y^7$ and $y^7, y^8, y^9$ directions, respectively. They give rise
to an $SU(2)\times SU(2)$ symmetry of the background, which is isomorphic to
$SO(4)$. 

For the $\text{D}7$-brane embedding in the $\N=2$ version of
Polchinski-Strassler, we find that these symmetries are sufficient to
ensure that the differential equation determining the embedding is
ordinary. This is achieved by embedding the $\text{D}7$-brane probe in the 
$y^5, y^6, y^8, y^9$ directions which correspond to the adjoint matter 
with mass $m$ in the $\mathcal{N}=2^*$ theory. 
The variable $r$ given by $r^2 = \rho^2+(y^4)^2+(y^7)^2$ 
is then the direction perpendicular to the
boundary of the deformed AdS space, which may be interpreted as the
energy scale. 

The background generically breaks the $U(1)$ symmetry in the $y^4, y^7$ plane
perpendicular to the $\text{D}7$-brane. 
We find that there are solutions
for the embedding for which the angular coordinate in this subspace is
constant, such that $y^7=0$ and $y^4= y^4(\rho)$. Another type of
solutions has  $y^4=0$ and $y^7= y^7(\rho)$.
Since there is no background two-sphere in the
$y_4$ directions, the embeddings of the form $y_4(\rho)$ are
repelled by the singularity at $r=0$. 
By applying the methods of holographic
renormalization, we confirm that these particular embeddings
preserve supersymmetry.

The other type of embedding solutions of the form $y^7=y^7(\rho)$ with
$y_4=0$ feel the effect of the shell of polarized $\text{D}3$-branes forming
the background. At small values of the quark
mass, they merge with the shell of polarized $\text{D}3$-branes in the deep
interior of the space. These embeddings are supersymmetric too. 

Although both of the above embeddings have similar behaviour, the fields 
living on their worldvolumes are different. The pullback of 
$B$ induces source terms in the equations of motion for $F$. Since $F$
lives on the four-dimensional brane volume transverse to $x^\mu$, the 
equations of motion derived from the combined 
Dirac-Born-Infeld and Chern-Simons action only contain the primitive
$(1,1)$ components 
of $F$. Thus, there must not appear source terms for the $(2,0)$ and $(0,2)$
components. In other words, the field strength components
$\de(P[B]_{2,0}+P[B]_{0,2})$ along the $\text{D}7$-brane directions
derived from the $(2,0)$ and $(0,2)$ components 
of $B$ must vanish. This is a constraint on the embedding. In the 
$\mathcal{N}=2$ case, for our choices of the embedding,
$P[B]_{(2,0)}$ and $P[B]_{(0,2)}$ vanish themselves and do not give 
non-trivial constraints.
In \cite{Minasian} the absence of the $(2,0)$ and $(0,2)$ components 
was found as a condition for supersymmetry to be preserved.

Finally, for the embedding of type $y^7=y^7(\rho)$, $y^4 = 0$ we
calculate the lowest-lying radial meson mode by considering small
fluctuations about the embedding.
In the range of parameters for which our order $m^2$ approximation to the
Polchinski-Strassler background is valid, 
we find that the meson mass satisfies 
$M= \sqrt{b \, m_{\text{q}}^2+c}$, with
$m_\text{q}$ the quark mass and $b$, $c$ some constants. 
This behaviour coincides with expectations from field theory:
The offset $c$ results from
the presence of the adjoint hypermultiplet masses and corresponds to a mass
gap for the mesons.

In this paper we are mainly concerned with the technical aspects of embedding
a $\text{D}7$-brane in the Polchinski-Strassler background, and leave physical
applications for the future. Still, let us mention the interesting
possibility of D-term supersymmetry breaking in the dual field theory
by switching on a non-commutative instanton on the $\text{D}7$-brane, 
along the lines
of \cite{Nekrasov,SeibergNC} (see also \cite{Lerda}). 
This may provide a gravity dual realization of metastable SUSY vacua 
\cite{Intriligator}, complementary to \cite{Bertolini2}.
The effect of commutative
instantons on the $\text{D}7$-brane in the AdS/CFT context was studied in
\cite{Johannes,AEEG}. 

The outline of the paper is as follows. In section \ref{PSbackground}, 
we obtain the $\N=2$ 
background to order $\mathcal{O}(m^2)$ in the flux perturbation, 
adapting the $\N=1^*$
results of  \cite{Freedman:2000xb,LopesCardoso:2004ni}. 
Moreover we discuss the structure of the
metric in the deep interior of the space, which is helpful for
understanding the
symmetries and the infrared behaviour of the embedding.

In section \ref{forms} we present the necessary
Ramond-Ramond forms for the DBI analysis, and in particular calculate the form
$C_8$.  

In section \ref{analytics} we perform the embedding by
establishing the Dirac-Born-Infeld and Chern-Simons actions, deriving the
equations of motion for the embedding, and discussing the solutions.
Moreover we discuss the role of the gauge and $B$ fields.
By expanding the embedding functions to second order in the adjoint
masses, we find analytic solutions for the embedding.
We show that
they are consistent with supersymmetry by applying holographic 
renormalization.

In section \ref{numerics} 
we present a numerical analysis of the embeddings. Moreover, as an
example for an associated meson mass, we
calculate the meson mass obtained from small radial fluctuations about
the embedding $y^7=y^7(\rho)$, $y^4=0$.

We conclude in section \ref{conclusions}. 
A number of lengthy and involved
calculations are relegated to a series of appendices.

\newpage

\section{Polchinski-Strassler background to order $m^2$}
\label{PSbackground}
\subsection{Metric}
\renewcommand\thefootnote{\arabic{footnote}}
\setlength{\skip\footins}{8mm}
The metric of $\AdS_5\times\text{S}^5$ in the Einstein frame reads
(see Appendix \ref{app:nc} for our notation and conventions)
\begin{align}\label{AdSSmetric} 
\de s^2&= Z^{-\frac{1}{2}}\eta_{\mu\nu}\de x^\mu\de x^\nu
+ Z^{\frac{1}{2}}\delta_{ij}\de y^i\de y^j\col\\
 Z(r)&=\frac{R^4}{r^4}\col\qquad 
r^2=y^i y^i\col\qquad 
R^4= 4 \pi g_\text{s}N\alpha'^2\col
\end{align}
where $\mu,\nu = 0,1,2,3$, $i,j=4, \dots, 9$.
The other fields of the background read simply
\begin{equation}\label{F5taudef} 
\begin{aligned}
F_{0123i}&=\e^{-\hat\phi}\partial_i Z^{-1}\col\\
\hat\tau&=\hat C_0+i\e^{-\hat\phi}=\text{const.}
\col
\end{aligned}
\end{equation} 
where we put a `hat' on $C_0$, $\tau$
to denote the unperturbed $\AdS_5\times\text{S}^5$ quantities.
The unperturbed dilaton $\hat\phi$ is related to the string coupling 
constant $g_\text{s}$ as $\e^{\hat\phi}=g_\text{s}$.

The $5$-form field strength $F_5$ follows from the $4$-form 
potential $C_4$ that reads
\begin{equation}\label{C4}
\hat C_{0123}=Z^{-1}
\end{equation}
by taking the exterior derivative and then imposing the condition 
of self-duality.

On the field theory side,  $\mathcal{N}=4$ supersymmetry is broken by
adding  mass terms for  the three adjoint chiral multiplets to the superpotential
\begin{equation} \label{DeltaW}
\Delta W= \frac{1}{g^2_{\text{YM}}} (m_1 \tr \Phi_1^2+m_2 \tr
\Phi_2^2+m_3 \tr \Phi_3^2)
\col
\end{equation}
where $g_\text{YM}^2=4\pi g_\text{s}$.
For generic masses, the theory has $\mathcal{N}=1$ supersymmetry, while
 for $m_1=0$, $m_2=m_3$ it has   $\mathcal{N}=2$ supersymmetry.

As shown in \cite{Polchinski:2000uf
}, on the gravity side the perturbation by the 
relevant mass operators \eqref{DeltaW}
corresponds to a non-trivial  $G_3$ flux, which is constructed from an 
imaginary anti-self dual tensor $T_3$, i.e.\ $T_3$ fulfills
\begin{equation}\label{T3iasdrel}
(\star_6+i)T_3=0\col
\end{equation}
where $\star_6$ is the six-dimensional Hodge star in flat space.
This condition ensures that $T_3$ forms a $\gr{\overline{10}}$ representation
of the $SO(6)$ isometry group of $\text{S}^5$, and hence transforms in the 
same way as the fermion mass matrix in the dual gauge theory.
The tensor field $G_3$ with the necessary asymptotic behaviour
to be dual to the mass 
perturbation is given by
\begin{equation}\label{G3expl}
G_3
=\e^{-\hat\phi}\frac{\zeta}{3}\de(ZS_2)\col
\end{equation}
where $\zeta$ is a numerical constant 
($\zeta=
-3\sqrt{2}$ 
in a proper normalization scheme \cite{Polchinski:2000uf}
).
The $2$-form potential $S_2$ is constructed from the components of
$T_3$ as follows,
\begin{equation}\label{S2def}
S_2=\frac{1}{2}T_{ijk}y^i\de y^j\wedge\de y^k\pnt
\end{equation}

To present the explicit form of $T_3$ it is
advantageous to work in a basis of three complex coordinates $z^p$ for 
the transverse space directions $y^i$ which is defined by 
\begin{equation}\label{cplxbasis}
z^p=\frac{1}{\sqrt{2}}(y^{p+3}+iy^{p+6})\col\qquad p=1,2,3 \, .
\end{equation}
The $z^p$ coordinates are dual to the three complex scalars $\phi^p$ 
of the chiral multiplets $\Phi^p$.

In this basis, a constant anti-selfdual antisymmetric $3$-tensor $T_3$
for a diagonal mass matrix with eigenvalues $m_p$, $p=1,2,3$ is given by
\begin{equation}
\begin{aligned}
T_3&=m_1\de z^1\wedge\de\bar z^2\wedge\de\bar z^3
+m_2\de\bar z^1\wedge\de z^2\wedge\de\bar z^3
+m_3\de\bar z^1\wedge\de\bar z^2\wedge\de z^3
\pnt
\end{aligned}
\end{equation}
In components the tensor $T_3$ reads
\begin{equation}\label{Tcplx}
T_{pqr}= T_{\bar{p}\bar{q}\bar{r}}=T_{\bar{p}qr} =0 \col \qquad
T_{p\bar{q}\bar{r}}= \epsilon_{pqr} m_p \pnt
\end{equation}
$S_2$ is proportional to the potential of $G_3$, which
up to quadratic order in the mass perturbation (where only the constant 
$\hat\tau$ enters the definition of $G_3$) reads, 
\begin{equation}\label{complex2form}
C_2-\hat\tau
B
=\e^{-\hat\phi}\frac{\zeta}{3}ZS_2\pnt
\end{equation}
The above given complex expression decomposes into real and imaginary part as
\begin{equation}\label{tildeC2B2inS2}
\tilde C_2=C_2-C_0B
=\e^{-\hat\phi}\frac{\zeta}{3}Z\re S_2\col\qquad
B
=-\frac{\zeta}{3}Z\im S_2\pnt
\end{equation}
As shown in \cite{Polchinski:2000uf}, the solution corresponding to the mass peturbation in 
the $\gr{\overline{10}}$ obeys
\begin{equation}\label{starminiG3indS2}
\hat Z^{-1}(\star_6-i)G_3
=-i\e^{-\hat\phi}\frac{2\zeta}{9}\de S_2
\pnt
\end{equation}

%
%
\subsection{Backreactions}
The unperturbed background is given by the 
$\AdS_5\times\text{S}^5$ metric \eqref{AdSSmetric} 
with the $5$-form field strength and constant
axion dilaton as in \eqref{F5taudef}.

A non-vanishing mass perturbation parameterized by $G_3$ 
starts at linear order in the masses $m_p$.
At linear order in $m_p$ away from the $\text{D}3$-brane source
the background is then given by the unperturbed result, $G_3$ itself and 
an induced $6$-form potential.
This potential has to be included in an analysis of $5$-brane probes, and its 
RR part $C_6$, was determined to be
\cite{Polchinski:2000uf}
\begin{equation}\label{C6}
C_6=\frac{2}{3}B\wedge\hat C_4\pnt
\end{equation}

Beyond the linear approximation, at quadratic order in $m_p$ the corrections
also effect the metric, $4$-form potential $C_4$, and the complex dilation 
axion $\tau$. Furthermore, we will show that also an $8$-form potential is 
induced,  to which the $\text{D}7$-brane probe couples. 
The deformations at quadratic order for the metric, $C_4$ and $\tau$ 
have been computed in
 \cite{Freedman:2000xb} with an appropriate gauge choice.
At this order, the deformed metric reads
\begin{equation}\label{metriccorr}
\de s^2=(Z^{-\frac{1}{2}}+h_0)\eta_{\mu\nu}\de x^\mu\de x^\nu
+\Big[(5 Z^{\frac{1}{2}}+p)I_{ij}
+(Z^{\frac{1}{2}}+q)\frac{y^iy^j}{r^2}
+w W_{ij}\Big]\de y^i\de y^j\col
\end{equation}
where the tensors $I_{ij}$ and $W_{ij}$ are given by
\begin{equation}\label{IWdef}
\begin{aligned}
I_{ij}&=\frac{1}{5}\Big(\delta_{ij}-\frac{y^iy^j}{r^2}\Big)\col\\
W_{ij}&=\frac{1}{|T_3|^2}\re(T_{ipk}\bar T_{jpl})\frac{y^ky^l}{r^2}-I_{ij}
\col\qquad
|T_3|^2=\frac{1}{3!}T_{ijk}\bar T_{ijk}\pnt
\end{aligned}
\end{equation}
It is important to remark that our definition of $|T_3|^2$ deviates from 
the one in \cite{Freedman:2000xb} by an extra factor $\frac{1}{3!}$, such that
\begin{equation}\label{T3abssquare}
|T_3|^2=m_1^2+m_2^2+m_3^2=M^2
\pnt
\end{equation}  
The function $h,w,p,q$ are given by\footnote{ 
We note two misprints in 
\cite{Freedman:2000xb}: Their eq.\ (125) to determine $w$ is ill 
written, though the final result matches; moreover their eq.\ (126) has an 
extra factor of 4 which contradicts their explicit results in
eqs. (71) and (143). With the latter two equations we 
coincide.}
\begin{equation}\label{wpqh0}
\begin{aligned}
w&
=-\frac{\zeta^2M^2R^2}{18}Z\col\qquad
p&
=-\frac{\zeta^2M^2R^2}{48}Z\col\qquad
q&
=\frac{\zeta^2M^2R^2}{1296}Z\col\qquad
h_0&
=\frac{7\zeta^2M^2R^2}{1296}\col
\end{aligned}
\end{equation}
and according to \cite{Freedman:2000xb} they satisfy
\begin{equation}\label{h0pqrel}
4h_0Z=q-p\pnt
\end{equation}
It is essential to note that the metric \eqref{metriccorr} 
has a curvature singularity at the origin, where the Ricci scalar is given by
\begin{equation}\label{Ricciscalar}
\mathcal{R}=M^2\frac{5}{2}\frac{R^2}{r^2}
\pnt
\end{equation}

For completeness we also state the expression for the dilaton $\phi$ 
 here \cite{Freedman:2000xb}. It is determined by
the equation of motion \eqref{EOMtaucomp} 
for the complex dilaton-axion $\tau$ defined as the combination in 
\eqref{F5taudef}.  
As shown in Appendix \ref{app:bases}, 
the dilaton solution obtained from  \eqref{EOMtaucomp} 
can be factorized into a purely radial and a 
purely angular part according to $\phi=\varphi Y_+$.
The explicit results taken from \eqref{Ypmdef} and \eqref{varphi}
then read\footnote{\cite{Freedman:2000xb} has the correct factor. 
\cite{LopesCardoso:2004ni} finds 18 times $\varphi(r)$ instead.}
\begin{equation}
\begin{aligned}
\varphi
&
=\frac{\zeta^2M^2R^2}{108}Z^{\frac{1}{2}}\col\qquad
Y_+=\frac{3}{M^2r^2}\big(m_2m_3(y_4^2-y_7^2)
+m_1m_3(y_5^2-y_8^2)+m_1m_2(y_6^2-y_9^2)\big)\pnt
\end{aligned}
\end{equation}

\subsection{Polarization of $\text{D}3$-branes}
\label{polarization}

For discussing the symmetries of this metric and for finding suitable
$\text{D}7$-brane embeddings, it is essential to discuss the infrared 
behaviour of the metric (\ref{metriccorr}). 
As has been found by \cite{Myers:1999ps}, 
a stack of $\text{D}p$-branes couples 
to higher $r$-form potentials ($r>p+1$) due to the non-commutativity of 
their matrix-valued positions. This coupling has an interpretation 
as a polarization of the $\text{D}p$-brane, with its worldvolume becoming
higher dimensional.
In the presence of potentials $B$ and $\tilde C_2$ which generate the 
non-vanishing $3$-form flux $G_3$,
the effective potential for the positions of the matrix-valued coordinates
$y^i$ is minimized if\footnote{This relation is valid only if higher powers
in $B$ and $\tilde C_2$ are suppressed, which according to the presence 
of the warp factor $Z$ in \eqref{tildeC2B2inS2} seems 
not to be the case close to $r=0$. However, one has to take into account that 
due to the strong backreaction at small $r$, $Z$ should be modified such that 
it does not become singular \cite{Polchinski:2000uf}.}
\begin{equation}
\comm{y^i}{y^j}
=i2\pi\alpha'c\zeta\im T_{ijk}\,y^k
\col
\end{equation}
$c=-\frac{2}{3}$.
The imaginary part of $T_{ijk}$, i.e.\ $B$ alone is therefore responsible 
for the polarization at this order. Using the expressions 
for $T_{ijk}$ in the real coordinates given in \eqref{Tinrealcoord}, 
one finds the concrete form of the polarizations.

We first discuss the $\mathcal{N}=2$ supersymmetric
case, where $m_1=0$, $m_2=m_3=m$. The only non-vanishing 
independent components are given by
\begin{equation}\label{tn2}
T_{456}=iT_{789}=iT_{567}=T_{489}=\frac{m}{\sqrt{2}}\pnt
\end{equation}
Inserting the non-vanishing imaginary parts into the equation for the 
embedding matrices $y^i$, gives rise to two $\mathfrak{su}(2)$ Lie 
algebras in the $y^5$, $y^6$, $y^7$ and $y^7$, $y^8$, $y^9$ directions. 
That means the 
$\text{D}3$-branes polarize into two $\text{S}^2$, having in common the 
$y^7$ direction.
The equations for the $2$-spheres read
\begin{equation}\label{2spheres}
(y^5)^2+(y^6)^2+(y^7)^2=r_0^2\col\qquad(y^7)^2+(y^8)^2+(y^9)^2=r_0^2\col
\end{equation}
from which it follows
\begin{equation}\label{invariant}
(y^5)^2+(y^6)^2+(y^8)^2+(y^9)^2=2(r_0^2-(y^7)^2)=\rho^2\pnt
\end{equation}
This equation defines a four-dimensional ball $\text{B}^4$ in 
the subplane spanned by $y^5$, $y^6$, $y^8$, $y^9$ with radial coordinate 
$\rho$.
\begin{figure}[H]
\begin{center}
\leavevmode
\epsfig{file=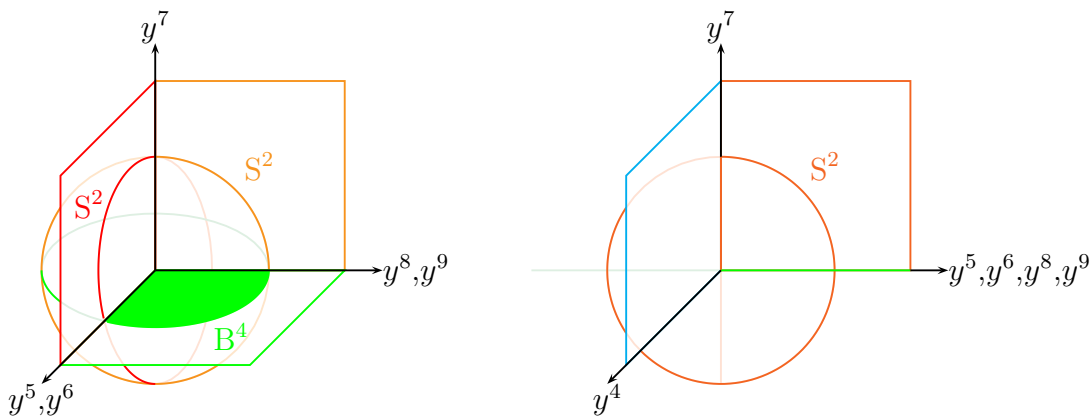, width=0.9\textwidth}
\vspace{3mm}
\caption{Directions of the polarization of the $\text{D}3$-branes in the 
$\mathcal{N}=2$ supersymmetric case $m_1=0$, $m_2=m_3=m$.}
\label{fig:D3polsymm}
\end{center}
\end{figure}

As shown in figure \ref{fig:D3polsymm}, 
the $\text{D}3$-branes are polarized into all their
transverse directions except of $y^4$ with the same radius 
$r_0=\pi\alpha'|c\zeta|m\sqrt{N^2-1}$, spanning a 
four-dimensional subspace.
In the subplanes spanned by $y^5$, $y^6$, $y^7$ (depicted in red) and
$y^8$, $y^9$, $y^7$ (depicted in orange), the coordinates are noncommutative, 
while in the subplane $y^5$, $y^6$, $y^8$, $y^9$ (depicted in green)
the two sets of coordinates 
commute. The volume into which the $\text{D}3$-branes polarize is therefore 
a four-dimensional ball in the subplane $y^5$, $y^6$, $y^8$, $y^9$, where 
different values of $y^7$ correspond to different $\text{S}^3$ orbits within 
the ball.
This configuration is symmetric under rotations of $y^5$, $y^6$, $y^8$, $y^9$
and hence should also lead to an embedding of a $\text{D}7$-brane which
is symmetric under these rotations, e.g.\ which does only depend on 
the radial coordinate $\rho$. 
The $\text{D}3$-branes are also smeared out 
along these directions of the $\text{D}7$-branes.
Since the $\text{D}3$-branes are only polarized into the direction $y^7$, 
this also means that the rotational invariance in the $y^4$, $y^7$ plane 
is lost. This corresponds to the breaking of the $U(1)$ symmetry. 

For comparison let us also consider the $\N=1^*$ case where all adjoint 
chiral masses are equal. Here, the tensor components of $T_3$ in 
\eqref{Tinrealcoord} become
\begin{equation}
T_{456}=iT_{789}=\frac{3m}{2\sqrt{2}}
\col\qquad
iT_{459}=T_{678}=-iT_{468}=-T_{579}=iT_{567}=T_{489}
=\frac{m}{2\sqrt{2}}
\pnt
\end{equation}
The $\text{D}3$-branes are also extended in the $y^4$-direction. 
There are $2$-spheres embedded in $y^4$, $y^5$, $y^9$. 
and in $y^4$, $y^6$, $y^8$, and as in the $\mathcal{N}=2$ case
in $y^5$, $y^6$, $y^7$ and $y^7$, $y^8$, $y^9$, 
While the prior two have radius smaller by a 
factor $\frac{1}{2}$ w.r.t.\ $r_0$ in the $\mathcal{N}=2$ case discussed 
above, the latter two have a radius bigger by a factor 
of $\frac{3}{2}$. Hence, the projection as discussed before in the 
$\mathcal{N}=2$ case cannot longer be a simple $B^4$. 
The equations for the spheres including $y^4$ read
\begin{equation}
(y^4)^2+(y^5)^2+(y^9)^2=\frac{r_0^2}{4}\col\qquad
(y^4)^2+(y^6)^2+(y^8)^2=\frac{r_0^2}{4}\pnt
\end{equation}
The polarization 
into the subspace $y^4$, $y^5$,  
$y^6$, $y^8$, $y^9$ then is similar to the 
one shown in the first picture in figure \eqref{fig:D3polsymm}, 
but with a radius which is 
smaller by a factor of $\frac{1}{2}$, and an exchange e.g.\ 
of $y^6$ and $y^9$. 
The situation is different for the two $2$-spheres having in common $y^7$. 
Their equations read
\begin{equation}
(y^5)^2+(y^6)^2+(y^7)^2=\frac{r_0^2}{4}\col\qquad
(y^7)^2+(y^8)^2+(y^9)^2=\frac{9r_0^2}{4}\pnt
\end{equation}
They are of different sizes. 
The above equations induce the relation
\begin{equation}
(y^5)^2+(y^6)^2+\frac{1}{9}(y^8)^2+\frac{1}{9}(y^9)^2
=\frac{1}{2}r_0^2-\frac{10}{9}(y^7)^2\pnt
\end{equation}
At $y^7=\frac{1}{2}r_0$ one has $y^5=y^6=0$, and the equation
becomes
\begin{equation}
(y^8)^2+(y^9)^2=2r_0^2\pnt
\end{equation}
At $y^7=0$, one finds a rotational ellipsoid.
\begin{figure}[H]
\begin{center}
\leavevmode
\epsfig{file=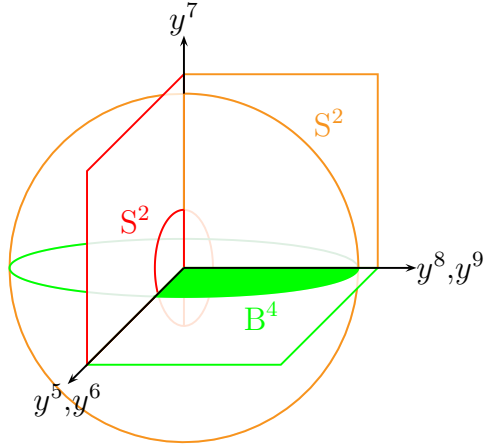, width=0.4\textwidth}
\vspace{3mm}
\caption{Directions of the polarization of the $\text{D}3$-branes in the equal
mass case $m_1=m_2=m_3=m$.}
\label{fig:D3polequalmass}
\end{center}
\end{figure}

Although topologically one still has a ball $\text{B}^4$, the difference 
in the length of the principal axes breaks the rotational symmetries in the 
$y^5$, $y^6$, $y^8$, $y^9$ plane. Still, the configuration is
symmetric under rotations in the 
$y^5$, $y^6$ and $y^8$, $y^9$ planes, but it is no longer symmetric under 
rotating these planes into each other. This breaking of the underlying 
$SU(2)\times SU(2)$ symmetry into 
$U(1)\times U(1)$ prevents one from finding an embedding 
of a $\text{D}7$-brane depending on the radial coordinate $\rho$ 
in this plane only.

\subsection{Symmetries and field theory action}

\begin{figure}[H]
\begin{center}
\leavevmode
\epsfig{file=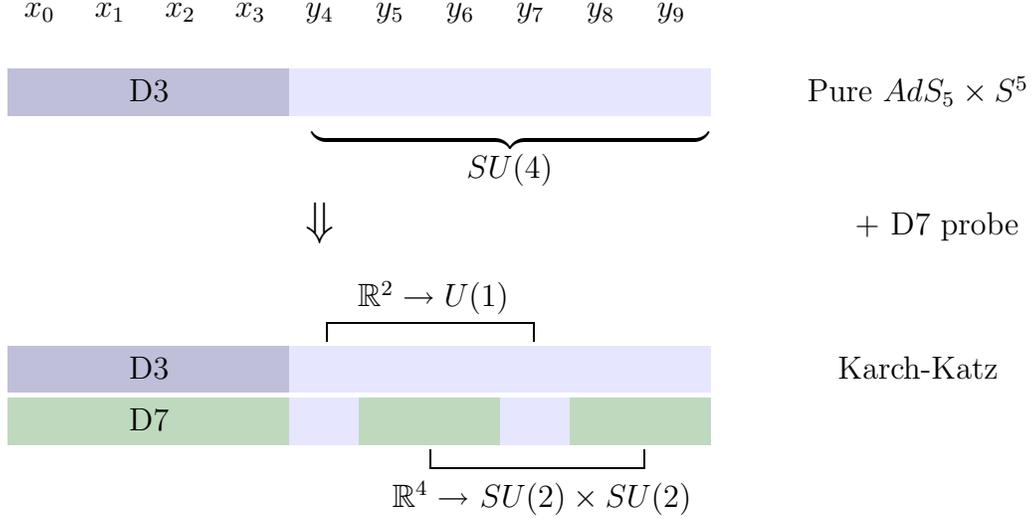, width=0.845\textwidth}
\vspace{3mm}
\caption{Symmetries for a $\text{D}7$-brane probe in $\AdS_5\times\text{S}^5$.}
\label{KKsym}
\end{center}
\end{figure}

\begin{figure}[H]
\begin{center}
\leavevmode
\hspace{2mm}
\epsfig{file=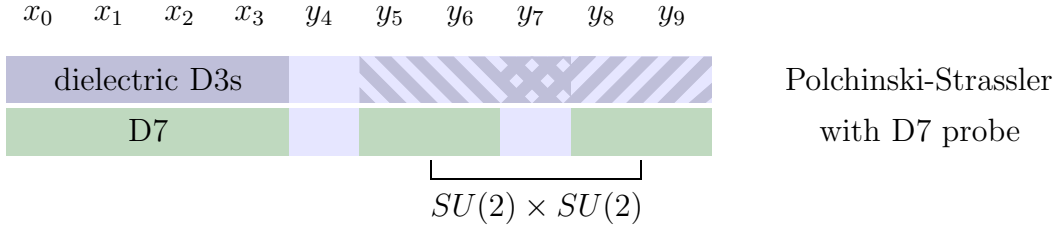, width=0.868\textwidth}
\vspace{3mm}
\caption{Symmetries for a $\text{D}7$-brane probe in the $\mathcal{N}=2$ 
Polchinski-Strassler background.}
\label{PSsym}
\end{center}
\end{figure}

We proceed by describing the $\text{D}7$-brane embedding and its symmetries. 
For comparison, we first recall the case of the undeformed 
$\AdS_5\times\text{S}^5$ background \cite{Karch:2002sh,krumy}, in which the
embedding of a $\text{D}7$-brane probe along 
$\AdS_5\times\text{S}^3$ with zero distance from the background 
generating $\text{D}3$-branes breaks the
original $SU(4)$ symmetry to $SU(2)\times SU(2)\times U(1)$, such
that there is a remaining $\N=2$ 
supersymmetry.\footnote{With a finite distance between the 
$\text{D}3$-branes and the $\text{D}7$-brane, also the $U(1)$ symmetry is 
broken.} This scenario is displayed in figure \ref{KKsym}. 
Let us now consider the symmetries of the Polchinski-Strassler
background with non-trivial $G_3$ flux at order $m^2$.
For simplicity we consider the case in which  $m_1=0$ and $m_2=m_3$ and, 
such that the background preserves $\N=2$
supersymmetry. As the discussion of  section \ref{polarization} shows
(see equations (\ref{tn2}) and (\ref{2spheres}) in particular), 
the background preserves
a global $SU(2) \times SU(2)$ symmetry in this case.
The background
preserves two 2-spheres which have one direction in common. As
(\ref{invariant}) shows, $\rho$ 
is an invariant under this $SU(2) \times SU(2) \simeq SO(4) $.  
It is thus convenient to embed the $\text{D}7$-brane probe into the directions
$y_5$, $y_6$, $y_8$, $y_9$. This embedding
preserves the symmetries of the background.
Note that the background does not have any further $U(1)$, which
corresponds to the fact that superconformal symmetry is broken by the
adjoint hyper mass terms. This
embedding is displayed in figure \ref{PSsym}.
We denote the real directions $5,6,8,9$ along the 
$\text{D}7$-brane with indices
$a,b$, and the directions $4,7$ perpendicular to it with $m,n$.
In the complex coordinates \eqref{cplxbasis}, $a,b=2,3$ and $m=1$.
These symmetries are consistent with
the symmetries of the $\N=2$ field theory in which the adjoint
hypermultiplet is massive. The corresponding classical $\N=2$ 
Lagrangian is 
\begin{equation}
\begin{aligned}
\mathcal{L} 
&=\im\bigg[\tau\int\de^2\theta\de^2\bar\theta
         \big(\tr (\bar \Phi_I e^V \Phi_I e^{-V}) 
          + Q^\dagger e^V Q + \tilde Q e^{V} \tilde Q^{\dagger}\big)\\
&\phantom{{}={}\im\bigg[}
+\tau\int\de^2\theta\big(\tr(\mathcal{W}^\alpha\mathcal{W}_\alpha)+W\big)
+\tau\int\de^2\bar\theta\big(\tr(\bar{\mathcal{W}}_{\dot\alpha}
\bar{\mathcal{W}}^{\dot\alpha})+\bar W\big)\bigg]
\col
\end{aligned}
\end{equation}
where the superpotential $W$ is
\begin{align}
W=\tr\big(\epsilon^{pqr}\Phi_p\Phi_q\Phi_r + m(\Phi_2^2+
\Phi_3^2)\big)+\tilde Q(m_\text{q}+\Phi_1)Q\pnt
\end{align}
The superfields $Q$ and $\tilde Q$
make up the $\mathcal{N}=2$ fundamental
hypermultiplet. Following the assignment of charges of \cite{krumy,Hong}, we
observe that this Lagrangian has an $SU(2) \times SU(2)_\text{R}$ symmetry,
where the first $SU(2)$ rotates the two complex scalars in each hypermultiplet
into each other. The mass terms explicitly break the $U(1)$ symmetry
of the $SU(2) \times U(1)$ superconformal group. This applies already
to the adjoint mass term before turning on the fundamental fields,
since the $U(1)$ charges of $\Phi_2$, $\Phi_3$ are zero, whereas a
superconformal superpotential requires a $U(1)$ charge of $2$. 
Thus the field theory symmetries agree with the supergravity symmetries.

\section{Forms}
\label{forms}

\subsection{Summary}

We have calculated the contributions to
 the background fields necessary for the $\text{D}7$-brane probe embedding
at order $\mathcal{O}(m^2)$. 
In particular the induced $C_8$ form, which has not been considered in the
literature, is  needed when adding $\text{D}7$-branes. Its computation
is given below in section \ref{C8subsection}.

To summarize, the background RR and NSNS forms read
\begin{gather}\label{bgforms}
\begin{gathered}
C_4=\e^{-\hat\phi}\Big(Z^{-1}+\frac{\zeta^2M^2R^2}{3^42^3}Z^{-\frac{1}{2}}\Big)
\de\vol(\mathds{R}^{1,3})+\frac{1}{2}B\wedge C_2\col\qquad
F_5=\de C_4+\star\de C_4\col
\end{gathered}
\\
\begin{aligned}
\tilde C_2&=\e^{-\hat\phi}\frac{\zeta}{3}Z\Re S_2\col\qquad
&B&=-\frac{\zeta}{3}Z\Im S_2\col\qquad
&G_3&=\de C_2-\tau\de B\col\\
C_6&=\frac{2}{3}B\wedge\hat C_4\col\qquad
&B_6&=\frac{2}{3}C_2\wedge\hat C_4\col\qquad 
&\star G_3&=\de C_6-\tau\de B_6\col\\
\end{aligned}
\\
\begin{gathered}
C_8=-\frac{1}{6}\big(
\e^{2\hat\phi}\tilde C_2\wedge\tilde C_2+B\wedge B\big)\wedge\hat C_4 \col
\end{gathered} 
\end{gather} 
where $\hat C_4$ denotes the unperturbed $4$-form potential given 
by the first term on the r.h.s.\ in the equation for $C_4$ above. 
The two types of corrections in $C_4$ are of order $\mathcal{O}(m^2)$. One 
of them has components along the spacetime directions spanned by $x^\mu$, the 
other comes from the redefinition of the $4$-form \eqref{C4redef} and
has no components in these directions. It turns out that only $\hat C_4$
is relevant for a $\text{D}7$-brane embedding up to order $\mathcal{O}(m^2)$. 
We also note that we can ignore the backreaction on $G_3$ itself, 
since it would be of order $\mathcal{O}(m^3)$.

\subsection{The $8$-form potential $C_8$}
\label{C8subsection}

We show that the backreaction of $G_3$ on the background at order $m^2$ 
induces a non-vanishing $8$-form potential $C_8$ with field strength 
$\tilde F_9$. 
This potential couples to the $\text{D}7$-brane charge and hence has to be 
considered in an embedding of $\text{D}7$-branes
in the Polchinski-Strassler background.

The physical field strengths are defined in \eqref{Ftildedef}. They are not
all independent but are related to their corresponding Hodge duals 
according to \eqref{Ftildedualities}. From these equations one finds, after
transforming to Einstein frame with \eqref{ESHdrel}, that the $8$-form 
potential $C_8$ obeys the equation
\begin{equation}\label{C8eom}
\de C_8=\star\de C_0+\de B\wedge C_6\pnt
\end{equation}
It therefore depends on the non-constant corrections to $C_0$ which start at
order $m^2$ as well as on 
non-vanishing potentials $B$ and $C_6$.
Taking the exterior derivative, thereby using that the $6$-form $C_6$ satisfies
\begin{equation}
\de C_6=-\star\tilde F_3+\de B\wedge C_4\col
\end{equation}
one finds
\begin{equation}
\de^2 C_8=\de\star\de C_0+H_3\wedge\star\tilde F_3\pnt
\end{equation}
This expression should vanish identically due to the nilpotency of the 
exterior derivative. 
Using the equation of motion for the axion found as the real part of 
\eqref{EOMtau} which up to quadratic order in the mass perturbation 
is given by
\begin{equation}
\de\star\de C_0=\frac{\e^{\hat\phi}}{2}\im(G_3\wedge\star G_3)
=\e^{\hat\phi}\im G_3\wedge\star\re G_3=-H_3\wedge\star\tilde F_3\col
\end{equation}
this vanishing is evident.
The reversed sign in the
relation $\tilde F_7=-\star\tilde F_3$ of \eqref{Ftildedualities}
is thereby crucial for the consistency.

Inserting the expression for $C_6$ \eqref{C6} into \eqref{C8eom} and 
then using \eqref{Hdstarrel}
one finds that $C_8$ decomposes as 
\begin{equation}\label{C8decomp}
C_8=\omega_4\wedge\de\vol(\mathds{R}^{1,3})\col
\end{equation}
where $\omega_4$ is a $4$-form which has to be determined from the 
equation for the remaining components transverse to the worldvolume of the
$\text{D}3$-branes
\begin{equation}\label{omega4eom}
\de\omega_4
=\star_6\de C_0
-\e^{-\hat\phi}\frac{2\zeta}{9}\de B\wedge\im S_2
=\star_6\de C_0
+\frac{2\zeta}{9}\im G_3\wedge\im S_2
\pnt
\end{equation} 
The first term in \eqref{omega4eom} is found from the equation of motion for 
the complex dilaton-axion $\tau$ which reads after using \eqref{Hdstarrel} and
neglecting terms beyond quadratic order in the masses
\begin{equation}
\de\star_6\de C_0
=-\frac{\zeta}{18}(G_3\wedge\de S_2+\bar
G_3\wedge\de\bar S_2)\pnt
\end{equation}
To find the r.h.s.\ we have also used the identity
\begin{equation}
\hat Z^{-1}(G_3\wedge\star_6 G_3-\bar
G_3\wedge\star_6\bar G_3)=G_3\wedge\hat Z^{-1}(\star_6-i)G_3-\bar
G_3\wedge\hat Z^{-1}(\star_6+i)\bar G_3
\end{equation}
together with \eqref{starminiG3indS2} and with \eqref{G3expl}.
Considering the Bianchi identity $\de G_3=0$ this equation can be integrated
easily and becomes
\begin{equation}
\star_6\de C_0=\frac{\zeta}{9}\re(G_3\wedge S_2)
\pnt
\end{equation}
Inserting this result, the
equation \eqref{omega4eom} for $\omega_4$ then assumes the form
\begin{equation}
\de\omega_4
=\frac{\zeta}{9}\re(G_3\wedge\bar S_2)
=\e^{-\hat\phi}\frac{\zeta^2}{27}\re(\de(\hat ZS_2)\wedge\bar S_2)\pnt
\end{equation}
After some elementary manipulations the Bianchi identity 
$\de^2\omega_4=0$ becomes
\begin{equation}\label{BIC8final}
\de(\de\hat Z\wedge S_2\wedge\bar S_2)=0\pnt
\end{equation}
Therefore $\de\hat Z\wedge S_2\wedge\bar S_2$ must follow from a $4$-form 
potential.
The equation of motion \eqref{omega4eom} can be transformed to 
\begin{equation}
\begin{aligned}\label{EOMomega4final}
\de\omega_4
&=\e^{-\hat\phi}\frac{\zeta^2}{54}\big(
\de\hat Z\wedge S_2\wedge\bar S_2
+\de(\hat ZS_2\wedge\bar S_2)\big)
\pnt
\end{aligned}
\end{equation}
The Bianchi identity \eqref{BIC8final} thereby ensures that a potential for
the first term and hence a $4$-form $\omega_4$ must exist.
As shown in Appendix \ref{app:C8}, the first term in \eqref{EOMomega4final}
can be rewritten as
\begin{equation}
\de\hat Z\wedge S_2\wedge\bar S_2=-2\de(\hat ZS_2\wedge\bar S_2)
\pnt
\end{equation} 
Using also the reexpression of $S_2$ in terms of $B$ and $\tilde C_2$ 
\eqref{tildeC2B2inS2}, the $4$-form potential therefore reads
\begin{equation}
\begin{aligned}
\omega_4
&=-\e^{-\hat\phi}\frac{\zeta^2}{54}\hat ZS_2\wedge\bar S_2
=-\frac{\e^{-\hat\phi}}{6}\hat Z^{-1}
(\e^{2\hat\phi}\tilde C_2\wedge\tilde C_2+B\wedge B)
\col
\end{aligned}
\end{equation}
and together with \eqref{C8decomp} it determines $C_8$. Using the expression 
for $\hat C_4$ \eqref{C4}, one then finds 
\begin{equation}\label{C8}
C_8=-\frac{1}{6}(\e^{2\hat\phi}\tilde C_2\wedge\tilde C_2+B\wedge B)
\wedge\hat C_4\pnt
\end{equation}

\section{$\text{D}7$-brane action}
\label{analytics}

With these ingredients we are now able to calculate the action for the
probe $\text{D}7$-brane in this background, which we present below. Moreover we
derive the equation of motion from this action and discuss possible
solutions for the brane embedding.

\subsection{The action}

The action for a $\text{D}7$-brane in the Einstein frame is given by
\begin{align} \label{action}
S&=S_\mathrm{DBI}+S_\mathrm{CS}\col\\
S_\mathrm{DBI}&=-T_7\e^{-2\hat\phi}
 \int\de^8\xi\e^{\phi}
\sqrt{\big|\det P\big[g-\e^{-\frac{\phi-\hat\phi}{2}}B\big]
+2\pi\alpha'\e^{-\frac{\phi-\hat\phi}{2}}F\big|}\col\label{DBIaction} \\
S_\mathrm{CS} & = -
  \mu_7 \int \sum\limits_{r=1}^{4}P\big[C_{2r}\wedge\e^{-B}\big]
\wedge\e^{2\pi\alpha'F}\label{CSaction} \col
\end{align}
where $T_7=\mu_7$. We use the conventions of 
\cite{Polchinski:2000uf}, 
in which the transformation between the Einstein and string frame 
metric contains just the non-constant part $\phi-\hat\phi$ of the dilaton 
$\phi$, where $\e^{\hat\phi}=g_\text{s}$ is the string coupling constant.
See appendix \ref{app:nc} for details.

In static gauge, the pullback of a generic $2$-tensor $E$ is defined by
\begin{equation}\label{PofEstatic}
\begin{aligned}
P[E]_{ab}&=E_{ab}+\partial_aX^mE_{mb}+\partial_bX^nE_{an}
+\partial_aX^m\partial_bX^nE_{mn}\pnt
\end{aligned}
\end{equation}

Note the minus sign in \eqref{CSaction}. In principle for the $AdS_5 \times
S^5$ background, there is a convention choice here corresponding 
to the sign choice in  the projector $ 
\varepsilon = \pm \Gamma
 \varepsilon
$ for the supersymmetries preserved by the brane. However in the $\N=2$
Polchinski-Strassler background, there remains only one choice consistent 
with the supersymmetries of the background. The correct choice corresponds to
the minus sign in (\ref{CSaction}). This is also in agreement with \cite{Grana,Grana2,Minasian}.\footnote{We are grateful to Andreas Karch
  for pointing this out to us. Compare also with the sign choice in 
\cite{SakaiSonnenschein}. In principle this can be checked using kappa
symmetry, an involved calculation which we leave for a forthcoming publication.
See also \cite{Kuperstein,Arean,Bandos}.}
 
Using the explicit expression for the induced forms $C_6$ and $C_8$  as in \eqref{C6} \eqref{C8},
%
the Chern-Simons part at order $\mathcal{O}(m^2)$ reduces to  ($P[F]=F$)
\begin{equation}\label{CSactionexpl}
\begin{aligned}
S_\CS
&=-\mu_7\int P\Big[\hat C_4\wedge
\Big(-\frac{1}{3}B\wedge(B+2\pi\alpha'F)
+2\pi^2\alpha'^2F\wedge F
-\frac{1}{6}\e^{2\hat\phi}\tilde 
C_2\wedge\tilde 
C_2\Big)\Big]\pnt
\end{aligned}
\end{equation}

We now expand to quadratic order in the mass perturbation around the
unperturbed metric \eqref{AdSSmetric}. For the total $\text{D}7$-brane 
action we find in appendix \ref{app:DBICSexpand}
\begin{equation}\label{Sexpand}
\begin{aligned}
S
&=-T_7\e^{-\hat\phi}\int\de^8\xi
\sqrt{\det P[\delta]_{ab}}\Big[1+\tilde\phi
+\frac{1}{2}Z^{\frac{1}{2}}\tilde g_{\mu\mu}
+\frac{1}{2}Z^{-\frac{1}{2}}P[\delta]^{ab}P[\tilde g]_{ab}\\
&\phantom{={}-T_7\e^{-\hat\phi}\int\de^8\xi\sqrt{\det P[\delta]_{ab}}\Big[}
+\frac{1}{2}Z^{-1}\Big(
\Big(1-\frac{2}{3}\star_4\Big)P[B]\cdot P[B]\\
&\phantom{={}-T_7\e^{-\hat\phi}\int\de^8\xi\sqrt{\det P[\delta]_{ab}}\Big[
+\frac{1}{2}Z^{-1}\Big(}
-4\pi\alpha'\Big(1+\frac{1}{3}
\star_4\Big) F\cdot P[B]\\
&\phantom{={}-T_7\e^{-\hat\phi}\int\de^8\xi\sqrt{\det P[\delta]_{ab}}\Big[
+\frac{1}{2}Z^{-1}\Big(}
+4\pi^2\alpha'^2(1+\star_4)F\cdot F
&\phantom{={}-T_7\e^{-\hat\phi}\int\de^8\xi\sqrt{\det P[\delta]_{ab}}\Big[
+\frac{1}{2}Z^{-1}\Big(}
-P[B]\cdot\star_4P[B]
-\frac{4}{3}P[B]\cdot\star_4F\\
&\phantom{={}-T_7\e^{-\hat\phi}\int\de^8\xi\sqrt{\det P[\delta]_{ab}}\Big[
+\frac{1}{2}Z^{-1}\Big(}
-\frac{1}{3}\e^{2\hat\phi}P[\tilde C_2]\cdot\star_4
P[\tilde C_2]\Big)\Big)\Big]
\col
\end{aligned}
\end{equation}
where throughout the paper with a `tilde' we denote the order 
$\mathcal{O}(m^2)$ 
corrections\footnote{By notational abuse, this does not apply to $\tilde C_2$
and the redefined field strengths $\tilde F_r$.}  
to the unperturbed quantities which carry a `hat'.
We should stress that here the four-dimensional inner product $\cdot$ 
as well as the Hodge star $\star_4$ are understood to be computed 
with the pullback metric according to \eqref{iponD7def} and
\eqref{HdonD7def}, respectively.

We are going to discuss the equation of motion derived from this
action in detail below. However first let us mention why 
we can neglect the backreaction of the $\text{D}7$-brane on the 
background. First, as in all probe approximations within AdS/CFT, 
we have a large number $N$ of background 
generating $\text{D}3$-branes compared to only a single $\text{D}7$-brane.
In the limit $N\to\infty$, $g_\text{s}N=\text{fixed}$,
the backreaction on the unperturbed $\AdS_5\times\text{S}^5$ part 
is negligible. It is of order $g_\text{s}N_\text{f}$, where
$N_\text{f}$ is a fixed number of $\text{D}7$-branes \cite{Karch:2002sh}. 
However for the Polchinski-Strassler background, we also have
to be sure that the backreaction on the 
perturbation parameterized by $G_3$ is negligible. 
Otherwise, we would have to consider its effect
 in the equations of motion from the type $\twob$ supergravity 
action $S_\twob$ before determining $G_3$ and the correction of the 
background.
In other words, the backreaction of the $\text{D}7$-brane must not be 
of the same order as the perturbation of the background by $G_3$.
We see that this is indeed the case:
Since $S_\twob\sim\e^{-2\hat\phi}$ whereas $S\sim\e^{-\hat\phi}$, the
$\text{D}7$-brane contributes to the background equations of motion at 
relative order $g_\text{s}$.

\subsection{Gauge field sources}
\label{gfsources}

The action \eqref{Sexpand} contains a linear coupling of the gauge field $F$ 
to the NSNS field $B$, generating a source term in the equation of motion 
for $F$. Since the corresponding $F$ is proportional to 
$\frac{1}{2\pi\alpha'}$, it cannot be neglected in the 
analysis.\footnote{We thank Rob Myers for bringing this fact to our attention.}
The equation of motion for $F$ reads
\begin{equation}\label{Feom}
\begin{aligned}
\de\Big(\hat C_4\wedge\Big(2\pi\alpha'(\star_4+1)F
-\Big(\star_4+\frac{1}{3}\Big)P[B]\Big)\Big)=0
\pnt
\end{aligned}
\end{equation}
Integrating the above equation, and decomposing it with \eqref{star4proj}
into its components purely along (anti-)holomorphic directions and along 
mixed directions, one finds
\begin{equation}
\begin{aligned}
2Z^{-1}\Big(2\pi\alpha' F_{(1,1)}^\text{P}-\frac{2}{3}P[B]_{(1,1)}\Big)
=\omega_{(1,1)}\col\qquad
-\frac{2}{3}Z^{-1}P[B]_{(2,0)}
=\omega_{(2,0)}
\col
\end{aligned}
\end{equation}
where we have omitted to indicate with an index P also the primitive part of 
$P[B]_{(1,1)}$ since it is primitive anyway. 
The $2$-form $\omega_2$ is closed. A separation of this condition into 
its individual components leads to the equations
\begin{equation}\label{omegaclosecond}
\partial\omega_{(2,0)}=0\col\qquad
\bar\partial\omega_{(2,0)}+\partial\omega_{(1,1)}=0\col
\end{equation}
where $\partial$ and $\bar\partial$ is the holomorphic and the antiholomorphic 
part of the exterior derivative operator $\de{}$. The first equation 
is trivially satisfied, since the subspace on which the above forms 
are defined is only four-dimensional. 
From the second equation one derives 
\begin{equation}\label{FinBdgl}
\begin{aligned}
\partial\Big(Z^{-1}\Big(2\pi\alpha' F_{(1,1)}^\text{P}
-\frac{2}{3}P[B]_{(1,1)}\Big)\Big)
=\frac{1}{3}\bar\partial\big(Z^{-1}P[B]_{(2,0)}\big)
\pnt
\end{aligned}
\end{equation}
The above equation is part of the full equation of motion, which reads
\begin{equation}
\begin{aligned}
{}&\partial\Big(4\pi\alpha' Z^{-1}F_{(1,1)}^\text{P}
+\frac{\zeta}{9}\Big(4P[\im S_2]_{(1,1)}
-i\bar\partial\varphi\Big)\Big)\\
&
+\bar\partial\Big(4\pi\alpha' Z^{-1}F_{(1,1)}^\text{P}
+\frac{\zeta}{9}\Big(4P[\im S_2]_{(1,1)}
+i\partial\bar\varphi\Big)\Big)\\
&
-\frac{\zeta}{9}
(\partial+\bar\partial)\Big(2P[\im S_2]_{(2,0)}+2P[\im S_2]_{(0,2)}
+i(\partial\varphi-\bar\partial\bar\varphi)\Big)
=0
\pnt
\end{aligned}
\end{equation}
We have extended the above equation by introducing a 
$1$-form $\varphi$ with only holomorphic components, which in total does
not give any contribution. The reason for this procedure will become
clear below.

Without a priori knowledge of the 
embedding it seems to be impossible to integrate the above equation 
without further input or assumptions. The reason for this is
the non-vanishing of the last term in the above equation. It is present
because the coefficients in the linear combination with $\star_4$ 
acting on $P[B]$ in the equation or motion \eqref{Feom} do not coincide with 
the ones 
in front of $F$. This term would be absent if also a projector $1+\star_4$
acted on $P[B]$.
One can easily integrate the equation of motion under the assumption that 
the last term above is vanishing, e.g.\
\begin{equation}\label{PimS2intcond}
\begin{aligned}
(\partial+\bar\partial)\Big(P[\im S_2]_{(2,0)}+P[\im S_2]_{(2,0)}
+\frac{i}{2}(\partial\varphi-\bar\partial\bar\varphi)\Big)=0
\pnt
\end{aligned}
\end{equation}
In this case a solution for $F$ is immediately found as
\begin{equation}\label{Fsol}
\begin{aligned}
4\pi\alpha' Z^{-1} F
=-\frac{\zeta}{9}(2(1+\star_4)P[\im S_2]
-i(\bar\partial\varphi-\partial\bar\varphi))
+\de A_0
\col
\end{aligned}
\end{equation}
where we have introduced an exact form $\de A_0$ which is a solution of the 
vacuum equations of motion for $F$.

In Appendix \ref{app:bases} we show that the pullback of the imaginary part 
of $S_2$ decomposes as
\begin{equation}\label{PimS2decomp2}
\begin{aligned}
P[\im S_2]_{(2,0)}&=
3\im S_{(2,0)}^\parallel-\frac{i}{2}\partial\theta\col\\
P[\im S_2]_{(1,1)}&=
3\im S_{(1,1)}^\parallel-\frac{i}{2}(\bar\partial\theta-\partial\bar\theta)
\col\\
P[\im S_2]_{(0,2)}&=
3\im S_{(0,2)}^\parallel+\frac{i}{2}\bar\partial\bar\theta
\col
\end{aligned}
\end{equation}
where with $\parallel$ we indicate the components along the directions
of the $\text{D}7$-brane explicitly given in \eqref{S2decomp}. 
Furthermore, $\theta$ is a $1$-form defined by
\begin{equation}\label{thetadef}
\begin{aligned}
\theta=\theta_{(1,0)}&=
(\bar T_{m\bar ab}\bar z^az^m+\bar T_{\bar mab}z^a\bar z^m
-T_{\bar m\bar ab}\bar z^a\bar z^m)\de z^b
\pnt
\end{aligned}
\end{equation}
The components of the condition \eqref{PimS2intcond} can be separated, and one 
uses the first of the above relations \eqref{PimS2decomp2} to obtain
\begin{equation}\label{intcond}
\begin{aligned}
\bar\partial\Big(P[\im S_2]_{(2,0)}
+\frac{i}{2}\partial\varphi\Big)
&=\bar\partial
\Big(3\im S_{(2,0)}^\parallel-\frac{i}{2}\partial(\theta-\varphi)\Big)
=0
\pnt
\end{aligned}
\end{equation}
The first term in the above equation explicitly reads with the definition 
of $S_2$ in \eqref{S2def} in the complex basis
\begin{equation}
\im S_{(2,0)}^\parallel
=\frac{i}{4}\bar T_{\bar mab}\de z^a\wedge\de z^b=0\pnt
\end{equation}
It vanishes since it is proportional to $m_1=0$.
The condition \eqref{intcond} is then easily satisfied for 
an appropriately chosen $\varphi$, e.g.\ given by
\begin{equation}
\varphi=\theta+\partial h\pnt
\end{equation}
We have introduced a function $h$ whose holomorphic derivative is 
part of the homogeneous solution.\footnote{A corresponding antiholomorphic 
derivative of a function has not been considered, 
since we want to keep $\varphi$ a $1$-form in only holomorphic directions.} 
Inserting the result for $\varphi$,
the solution for $F$ \eqref{Fsol} hence becomes
\begin{equation}
\begin{aligned}
4\pi\alpha' Z^{-1} F
=-\frac{\zeta}{9}(2(1+\star_4)P[\im S_2]
-i(\partial\bar\theta-\bar\partial\theta))
+\de A_0
\pnt
\end{aligned}
\end{equation}
One then uses the second relation in 
\eqref{PimS2decomp2} to eliminate the $\theta$-dependent terms. 
The result for $F$ then reads
\begin{equation}
\begin{aligned}
4\pi\alpha' Z^{-1} F
&=-\frac{\zeta}{3}(1+\star_4)(P[\im S_2]
-\im S_2^\parallel))+\de A_0
\pnt
\end{aligned}
\end{equation}
Using the relation \eqref{tildeC2B2inS2} to reexpress the imaginary parts
of $S_2$ in terms of $B$, one finally finds
\begin{equation}\label{Fimpl}
\begin{aligned}
2\pi\alpha' F
&=\frac{1}{2}(1+\star_4)(P[B]-B^\parallel)+\frac{Z}{2}\de A_0
\pnt
\end{aligned}
\end{equation}
It is worth to remark that by using the expression for the pullback 
in static gauge, the linear combination keeps only the derivative 
terms of the embedding coordinates that come from the pullback.
This is a particularity for the coefficients in the linear combination with
$\star_4$ in front of $P[B]$ in the equation of motion \eqref{Feom}.

\subsection{Expansion of the embedding}

At sufficient distance from the polarized brane source, the 
Polchinski-Strassler background is given by $\AdS_5\times\text{S}^5$ with 
corrections at quadratic order in the mass perturbation.
One can therefore expand the full embedding coordinates 
$X^m\equiv y^m$, $m=4,7$
into a known unperturbed part $\hat X^m$, which is the constant 
embedding in $\AdS_5\times\text{S}^5$ and into a perturbation $\tilde X^m$, 
i.e.\ 
\begin{equation}\label{embedexp}
X^m=\hat X^m+\tilde X^m\pnt
\end{equation}
This decomposition is inserted into the complete action. 
Then one expands in powers of 
$\tilde X^m$, thereby including all corrections to the background such that 
the resulting equations of motion contain all terms up to quadratic order in 
the mass perturbation. This yields differential equations for $\tilde X^m$.
In the case of the $\mathcal{N}=2$ embedding to be discussed below, it 
becomes an ordinary differential equation of second order that can be solved 
analytically. However,
the solution found in this way is accurate only in a regime where the bare 
embedding coordinates dominate the correction, i.e.\ where 
$\hat X^m\gtrsim\tilde X^m$.

Inserting \eqref{embedexp} and the corresponding decompositions for all 
background fields, 
as well as the expressions \eqref{C6} and \eqref{C8}
for the induced $C_6$ and $C_8$, 
one finds that for a constant unperturbed embedding the 
action \eqref{Sexpand}
becomes up to quadratic order in the perturbation
\begin{equation}\label{Sembed}
\begin{aligned}
S&=
-T_7\e^{-\hat\phi}\int\de^8\xi\Big[
1+\tilde\phi+\frac{1}{2}Z^{\frac{1}{2}}\tilde g_{\mu\mu}
+\frac{1}{2}Z^{-\frac{1}{2}}\tilde g_{aa}
+\frac{1}{2}(\partial_a\tilde X^m)^2
+Z^{-\frac{1}{2}}\partial_a\tilde X^m\tilde g_{ma}\\
&\phantom{=-T_7\e^{-\hat\phi}\int\de^8\xi\Big[}
+\frac{1}{2}Z^{-1}\Big(\Big(1-\frac{2}{3}\star_4\Big)
B\cdot (B+4\partial\tilde XB)\\
&\phantom{=-T_7\e^{-\hat\phi}\int\de^8\xi\Big[+\frac{1}{4}Z^{-1}\Big(}
-4\pi\alpha'\Big(1+\frac{1}{3}\star_4\Big) 
F\cdot\big(B+2\partial\tilde XB\big)
+4\pi^2\alpha'^2(1+\star_4)F\cdot F\\
&\phantom{=-T_7\e^{-\hat\phi}\int\de^8\xi\Big(+\frac{1}{4}Z^{-1}\Big(}
-\frac{1}{3}\e^{2\hat\phi}\star_4\tilde C\cdot
\big(\tilde C+4\partial\tilde X\tilde C\big)
\Big)\Big]\pnt
\end{aligned}
\end{equation}
Since the action at this order does not include terms that depend on 
the mass perturbation and include the perturbation $\tilde X^m$ at quadratic 
order, the Hodge star and the inner product as defined in \eqref{Hdform}, 
\eqref{ipdef} become 
the ordinary ones in flat space for the constant unperturbed embedding. 
It is also important to remark that in the above expansion terms that are 
quadratic in the mass perturbation but in addition 
linear in the perturbation of the 
embedding $\tilde X^m$ have been kept. This ensures that in the equations of 
motion for the fluctuations all terms that are quadratic in the perturbation
do appear. 

In the above equation we have used the abbreviations
\begin{equation}
(\partial\tilde XB)_{ab}=\partial_a\tilde X^mB_{mb}\col\qquad
(B\partial\tilde X)_{ab}=\partial_b\tilde X^mB_{am}\pnt
\end{equation}
They are also of use for an expansion of the expression 
\eqref{Fimpl} for the gauge field strength $F$, which becomes
\begin{equation}
\begin{aligned}
2\pi\alpha' F
=\frac{1}{2}(1+\star_4)(\partial\tilde XB+B\partial\tilde X)
\col
\end{aligned}
\end{equation}
and in which we have neglected $A_0$ that does not 
contribute to the source terms. The 
previously mentioned dependence of $F$ only on 
derivative terms of the embedding is now obvious.
Since $F$ is thus of linear order in 
the mass perturbation and in the corrections to the embedding, one 
can directly neglect all terms quadratic in $F$ as well as all terms
that contain $F$ and additional dependence of linear order in $\tilde X$.
The action thus becomes
\begin{equation}\label{SembedFexpl}
\begin{aligned}
S
&=-T_7\e^{-\hat\phi}\int\de^8\xi\Big[
1+\tilde\phi+\frac{1}{2}Z^{\frac{1}{2}}\tilde g_{\mu\mu}
+\frac{1}{2}Z^{-\frac{1}{2}}\tilde g_{aa}
+\frac{1}{2}(\partial_a\tilde X^m)^2
+Z^{-\frac{1}{2}}\partial_a\tilde X^m\tilde g_{ma}\\
&\phantom{=-T_7\e^{-\hat\phi}\int\de^8\xi\Big[}
+\frac{1}{2}Z^{-1}\Big(\Big(1-\frac{2}{3}\star_4\Big)
B\cdot B+\frac{4}{3}(1-4\star_4)B\cdot\partial\tilde XB\\
&\phantom{=-T_7\e^{-\hat\phi}\int\de^8\xi\Big(+\frac{1}{4}Z^{-1}\Big(}
-\frac{1}{3}\e^{2\hat\phi}\star_4\tilde C\cdot
\big(\tilde C+4\partial\tilde X\tilde C\big)
\Big)\Big]
\col
\end{aligned}
\end{equation}
where we have also made use of the fact that $\star_4^2=1$.

To present the equations of motion, it is advantageous to transform to 
polar coordinates. The two directions transverse to the $\text{D}7$-brane
become
\begin{equation}
\begin{aligned}
X^4&=u\cos\psi
=\hat u\cos\hat\psi-\hat u\tilde\psi\sin\hat\psi+\tilde u\cos\hat\psi\col\\
X^7&=u\sin\psi
=\hat u\sin\hat\psi+\hat u\tilde\psi\cos\hat\psi+\tilde u\sin\hat\psi\col
\end{aligned}
\end{equation}
where in the final expressions we have expanded up to linear order in the 
perturbations $\tilde u$ and $\tilde\psi$ of the radius and angular 
dependence, respectively.

As shown in appendix \ref{app:eomex}, the equations of motions
assume the form
\begin{equation}\label{diffeq}
\frac{1}{\rho^3}\partial_\rho(\rho^3\partial_\rho f(\rho))=g(\rho)\col\qquad
g(\rho)=\frac{\hat u}{\hat r^4}\Big(B_f+\frac{\hat u^2}{\hat r^2}C_f\Big)\col\qquad
\hat r^2=\hat r^2(\rho)=\rho^2+\hat u^2\col
\end{equation}
where $f=u$ or $f=\psi$ and $B_f$ and $C_f$ are constants that 
depend on the unperturbed embedding coordinates $\hat u$, $\hat\psi$ and which
are of quadratic order in the mass perturbation. 
We should remark that without the 
inhomogenity, i.e.\ $B_f=C_f=0$, the above equation is the one found for the 
embedding of $\text{D}7$-branes in pure $\AdS_5\times\text{S}^5$ at large 
$r$. 
Furthermore, the constant embedding $\hat u=0$ therefore 
remains a solution also in presence of the mass perturbation.
Note that  
$\hat u$ is identified with the quark mass by
  virtue of $\hat u = 2 \pi \alpha' m_\text{q}$. 
\label{scalesRa}  

We discuss the case $\hat u\neq 0$.
For $f=u$ the constants $B$ and $C$ read
\begin{equation}\label{BCu}
\begin{aligned}
B_u=\frac{\zeta^2m^2R^4}{54}(-4+\cos2\hat\psi)\col\qquad
C_u=\frac{\zeta^2m^2R^4}{81}(-10+3\cos2\hat\psi)\pnt
\end{aligned}
\end{equation}
For  $f=\psi$ one has to identify
\begin{equation}\label{BCpsi}
\begin{aligned}
B_\psi=-\frac{\zeta^2m^2R^4}{54\hat u}\sin2\hat\psi\col\qquad
C_\psi=-\frac{2\zeta^2m^2R^4}{27\hat u}\sin2\hat\psi
\pnt
\end{aligned}
\end{equation}
The full solution of the differential equation  
\eqref{diffeq} reads
\begin{equation}
f(\rho)
=\hat f+\frac{A_f}{\rho^2}
-\frac{B_f}{4}\frac{\hat u}{\rho^2}(1+\ln\hat r^2)
+\frac{C_f}{8}\frac{\hat u^3}{\rho^2\hat r^2}
=\hat f+\frac{1}{8\rho^2}\Big(C_f\frac{\hat u^3}{r^2}
+2(4A_f-B_f\hat u)-2B_f\hat u\ln\hat r^2\Big)
\col
\end{equation}
where $\hat f$ is the unperturbed value of $f$ which is the value of $f$ at the
boundary at $\rho\to\infty$. $A_f$ is an integration constant that 
has to be fixed by the condition that the solution does not become singular at 
$\rho=0$.
In the unperturbed $\AdS_5\times\text{S}^5$ case one 
would have to set $A_f=0$. Here, in contrast, 
we have to allow $A_f\neq0$ to find a non-singular embedding.

For an embedding that becomes singular at $\rho=0$ the decomposition 
in \eqref{embedexp} used to expand the action and equations of motion
is only justified for $\rho\gtrsim mR^2$,
for which $\hat X^m\gtrsim\tilde X^m$ holds. This means a more restrictive 
constraint than the condition under which the Polchinski-Strassler background 
is given by an expansion around the 
$\AdS_5\times\text{S}^5$ background, requiring only that 
$r\gtrsim mR^2$.
For the embedding regular at $\rho=0$, however, there is no restriction on 
$\rho$. We only have to take care that it does not enter the regime 
where the perturbative description of the background breaks down. 
We will see that the $\text{D}7$-branes
avoid to enter the region of small $r$.
For the expansion \eqref{embedexp} we only have to keep in mind
that it is a good approximation only when $\hat X^m\gtrsim mR^2$ such that 
the corrections $\tilde X^m$ are small.
\begin{figure}[H]
\begin{center}
\leavevmode
\epsfig{file=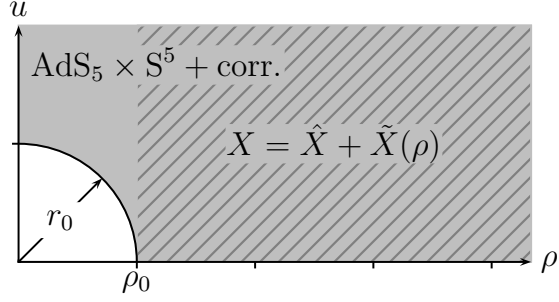, width=0.46\textwidth}
\vspace{3mm}
\caption{The two radial directions $\rho$ and $u$ in the six-dimensional space
perpendicular to $x^\mu$. In the grayscaled region, the perturbative treatment 
of the backreaction around $\AdS_5\times\text{S}^5$ is valid. 
In the hatched region, the expansion 
of the embedding $X^m$ into the constant embedding $\hat X$ plus a small 
$\rho$-dependent correction is valid. For regular embeddings 
the decomposition is valid also for $\rho<\rho_0$, provided $r\geq mR^2$. 
At radii smaller than $r_0\sim mR^2$, the backreaction becomes strong, and the 
background cannot be described by $\AdS_5\times\text{S}^5$ plus corrections.} 
\end{center}
\end{figure}

In the following we determine $A_f$ to cancel the divergence at $\rho\to0$
which leads to the regular solution.
We find in the limit $\rho^2\ll u^2$
\begin{equation}
f(\rho)
\sim\hat f-\frac{C_f}{8\hat u}-\frac{B_f}{4\hat u}
+\frac{1}{8\rho^2}\Big(8A_f-2B_f\hat u(1+\ln\hat u^2)+C_f\hat u\Big)
\pnt
\end{equation}
From this expansion it follows that there exist non-singular solutions 
in the special case
\begin{equation}
8A_f-2B_f\hat u(1+\ln\hat u^2)+C_f\hat u=0\pnt
\end{equation}
In this case the full solution, which is regular on $0\le\rho<\infty$, reads
\begin{equation}\label{frhoreg}
f(\rho)
=\hat f-C_f\frac{\hat u}{8\hat r^2}
-B_f\frac{\hat u}{4\rho^2}\ln\frac{\hat r^2}{\hat u^2}
\pnt
\end{equation}
It has the asymptotic behaviour
\begin{equation}
\begin{aligned}
\rho\to0:&\qquad f(\rho)\sim\hat f-\frac{1}{8\hat u}(2B_f+C_f)\col\\
\rho\to\infty:&\qquad
f(\rho)\sim\hat f-B_f\frac{\hat u}{4\rho^2}\ln\rho^2
-\frac{\hat u}{8\rho^2}\Big(C_f-2B_f\ln\hat u^2\Big)
\pnt
\end{aligned}
\end{equation}
Considering the values $B_\psi$ and $C_\psi$ which determine the 
angular dependence of the solution, no $\rho$ dependent perturbation is found 
for the unperturbed embedding with $\hat\psi=0$, where $u=X^4$, or 
$\hat\psi=\frac{\pi}{2}$, where $u=X^7$.
According to figure \ref{fig:D3polsymm}, $y^7$ is the only direction 
into which the $\text{D}3$-branes polarize and which is not parallel to 
the directions of the $\text{D}7$-brane.
The two choices $\hat\psi=0$ and $\hat\psi=\frac{\pi}{2}$ are singled out, 
since the polarization direction is respectively perpendicular or along 
the direction of the separation of the two types of branes. 
 
\subsection{Holographic renormalization}
\label{holoren}

The embedding solutions we have derived in the previous section
are not constant. We may therefore wonder if their boundary behaviour for
$\rho\rightarrow\infty$ might imply the presence of a VEV for the fermion
bilinear which would violate supersymmetry. In this section we
show that such a fermion condensate is absent. For this, suitable
finite counterterms have to be added to the action, such that it
vanishes when evaluated on a solution, as required by supersymmetry.
According to \cite{Skenderis,KarchOBannon}, this addition of counterterms
corresponds to a change of renormalization scheme.
If we were able to find the canonical coordinates for our background,
i.e.\ those in which the kinetic term is
canonically normalized in presence of the order 
$\mathcal{O}(m^2)$ corrections, an addition of finite counterterms
would not be necessary.\footnote{See also the discussion in 
\cite{EvansShockWaterson},
where the embedding functions become flat for a suitable coordinate choice.}

The holographic renormalization of the expanded $\text{D}7$-brane action
\eqref{Sembed} or correspondingly \eqref{SembedFexpl} is best performed 
in the coordinate system introduced in \cite{Skenderis}.
These are essentially Poincar\'e coordinates
in which the $\AdS_5$
metric \eqref{AdSSmetric} is given 
by
\begin{equation}\label{AdSSmetricpoincare}
\de s^2=\frac{1}{R^2\chi}\eta_{\mu\nu}\de x^\mu\de x^\nu
+\frac{R^2}{4\chi}\de\chi
\pnt
\end{equation}
The only replacement to be done is a redefinition of the holographic 
direction. Instead of $\rho$ one chooses 
\begin{equation}\label{hrcoorddef}
r=\frac{1}{\sqrt{\chi}}\col\qquad
\frac{\rho\de\rho}{r}=-\frac{\de\chi}{2\sqrt{\chi}^3}\col\qquad
\rho\de\rho=-\frac{\de\chi}{2\chi^2}\col\qquad
\partial_\rho=-2\rho\chi^2\partial_\chi
\pnt
\end{equation} 
In the special case $m_1=0$, $m_2=m_3=m$ 
the action in the new coordinates is derived in Appendix
\ref{app:osaction}. 
Keeping all terms up to order $\mathcal{O}(m^2)$ in the on-shell action
\eqref{piSchiaction}, the regularized action 
is found to be given by
\begin{equation}
\begin{aligned}
S_\text{reg}
&=-\frac{T_7}{2}\e^{-\hat\phi}\Omega_3\int\de\xi^4\,\Big[
-\frac{1}{2\hat\chi^2}+\frac{\hat u^2}{\hat\chi}
+\frac{\zeta^2m^2R^4}{108}\Big(
-\frac{5}{3\hat\chi}+\Big(c_0-\frac{5}{3}\Big)\hat u^2\ln\hat\chi
-c_0\hat u^4\hat\chi\Big)
\Big]_\epsilon^{\frac{1}{\hat u^2}}\\
&=-\frac{T_7}{2}\e^{-\hat\phi}\Omega_3\int\de\xi^4\,\Big[
\frac{\hat u^4}{2}+\frac{1}{2\epsilon^2}-\frac{\hat u^2}{\epsilon}\\
&\phantom{{}={}-\frac{T_7}{2}\e^{-\hat\phi}\Omega_3\int\de\xi^4\,\Big[}
-\frac{\zeta^2m^2R^4}{108}\Big(
\Big(c_0+\frac{5}{3}\Big)\hat u^2-\frac{5}{3\epsilon}
+\Big(c_0-\frac{5}{3}\Big)\hat u^2\ln\epsilon\hat u^2
-c_0\hat u^4\epsilon\Big)\Big]
\col
\end{aligned}
\end{equation}
where the constants found for both types of embeddings are defined in 
\eqref{cdef}.
Here we have used that the original integration interval $0\le\rho<\infty$
in the new variable $\hat\chi=\frac{1}{\hat r^2}$ translates 
to $\frac{1}{\hat u^2}\ge\hat\chi\ge 0$.

According to holographic renormalization, we have
to replace the boundary data $\hat u$ at the position
of the true boundary at $\hat\chi=0$ by the data on the regulator hypersurface
at $\hat\chi=\epsilon$. To this purpose we have 
to evaluate the solution
\eqref{uchireg} at $\hat\chi=\epsilon$ and invert it, making use of 
an expansion for small $\epsilon$. 
Denoting the value of the field at the location of the 
regulator hypersurface by $u_\epsilon$, 
one finds in this way
\begin{equation}\label{hatuinuepsilon}
\begin{aligned}
\hat u
&=u_\epsilon+\frac{\zeta^2m^2R^4}{216}\epsilon u_\epsilon
\Big(\frac{c_2}{2}-c_3+\frac{c_1}{1-\epsilon\hat u^2}
\ln\epsilon u_\epsilon^2\Big)
\pnt
\end{aligned}
\end{equation}
Concretely, one needs the following expressions
\begin{equation}
\begin{aligned}
\frac{\hat u^4}{2}
&=\frac{u_\epsilon^4}{2}
+\frac{\zeta^2m^2R^4}{108}\epsilon u_\epsilon^4
\Big(\frac{c_2}{2}-c_3+c_1\ln\epsilon u_\epsilon^2\Big)\\
\frac{\hat u^2}{\epsilon}
&=\frac{u_\epsilon^2}{\epsilon}
+\frac{\zeta^2m^2R^4}{108}u_\epsilon^2
\Big(\frac{c_2}{2}-c_3
+c_1(1+\epsilon u_\epsilon^2)\ln\epsilon u_\epsilon^2\Big)
\col
\end{aligned}
\end{equation}
which give non-divergent contributions that do not depend on $\epsilon$.
Inserting them into the action, 
up to order $\mathcal{O}(\epsilon)$ and $\mathcal{O}(m^2)$ 
the regularized action becomes
\begin{equation}
\begin{aligned}
S_\text{reg}
%
&=-\frac{T_7}{2}\e^{-\hat\phi}\Omega_3\int\de\xi^4\,\Big[
\frac{u_\epsilon^4}{2}
+\frac{1}{2\epsilon^2}
-\frac{u_\epsilon^2}{\epsilon}\\
&\phantom{{}={}-\frac{T_7}{2}\e^{-\hat\phi}\Omega_3\int\de\xi^4\,\Big[}
-\frac{\zeta^2m^2R^4}{108}\Big(
\Big(c_0+\frac{c_2}{2}-c_3+\frac{5}{3}\Big)u_\epsilon^2-\frac{5}{3\epsilon}
+\Big(c_0+c_1-\frac{5}{3}\Big)u_\epsilon^2\ln\epsilon u_\epsilon^2\\
&\phantom{{}={}-\frac{T_7}{2}\e^{-\hat\phi}\Omega_3\int\de\xi^4\,\Big[
+\frac{\zeta^2m^2R^4}{108}\Big(}
-\Big(c_0+\frac{c_2}{2}-c_3\Big)u_\epsilon^4\epsilon\Big)\Big]
\pnt
\end{aligned}
\end{equation}
Each set of constants $c_0,\dots,c_3$
defined in \eqref{cdef} for the two embeddings with constant 
angles fulfill $c_0+c_1-\frac{5}{3}=0$. Hence, the logarithmic 
term in the above expression is absent.
The counterterm action is defined as the negative of all terms in the 
above result which diverge in the limit $\epsilon\to0$.
It hence reads
\begin{equation}\label{Sct}
\begin{aligned}
S_\text{ct}
&=\frac{T_7}{2}\e^{-\hat\phi}\Omega_3\int\de\xi^4\,\Big[
\frac{1}{2\epsilon^2}
-\frac{u_\epsilon^2}{\epsilon}
+\frac{\zeta^2m^2R^4}{108}\frac{5}{3\epsilon}
\Big]
\pnt
\end{aligned}
\end{equation}
The subtracted action is given by the sum of the regularized action and the 
counterterm action. Up to linear order in $\epsilon$ it thus becomes
\begin{equation}
\begin{aligned}
S_\text{sub}
&=-\frac{T_7}{2}\e^{-\hat\phi}\Omega_3\int\de\xi^4\,\Big[
\frac{u_\epsilon^4}{2}
-\frac{\zeta^2m^2R^4}{108}\Big(
\Big(c_0+\frac{c_2}{2}-c_3+\frac{5}{3}\Big)u_\epsilon^2
-\Big(c_0+\frac{c_2}{2}-c_3\Big)u_\epsilon^4\epsilon\Big)\Big]
\pnt
\end{aligned}
\end{equation}
Interestingly, for the two embeddings with constant angles
the above combination of the constants defined in \eqref{cdef} is
universal and given by $c_0+\frac{c_2}{2}-c_3=-\frac{11}{3}$.
The value of the subtracted action hence is equal for both 
types of embeddings. 
It vanishes if we include a finite counterterm into the counterterm
action \eqref{Sct} which then reads
\begin{equation}
\begin{aligned}
S_\text{ct}
&=\frac{T_7}{2}\e^{-\hat\phi}\Omega_3\int\de\xi^4\,\Big[
\frac{1}{2\epsilon^2}
-\frac{u_\epsilon^2}{\epsilon}
+\frac{u_\epsilon^4}{2}
+\frac{\zeta^2m^2R^4}{108}\Big(
\frac{5}{3\epsilon}+2u_\epsilon^2\Big)\Big]
\pnt
\end{aligned}
\end{equation}
We then find
 $S_\text{sub}=S_\text{reg}+S_\text{ct}=0$ in agreement with supersymmetry, 
and hence the vanishing of the quark condensate.
Besides the finite counterterm already found in 
\cite{KarchOBannon}, we also have to add a finite counterterm proportional to 
$m^2u_\epsilon{}^2 $.
This finite term in $S_\text{ct}$ is the same for both types of embeddings 
considered. This seems to confirm that there should 
exist a canonical coordinate system in which it is not necessary to
add a finite 
counterterm in order to obtain $S_\text{sub}=0$.

\section{Full embedding and meson masses}
\label{numerics} 

We now move beyond the expansion \eqref{embedexp} for the embedding
an study the resulting embeddings numerically. Our results reflect the
anisotropy of the background. As discussed in section \ref{polarization},
this anisotropy is due to the fact that the shell of
polarized $\text{D}3$-branes extends into the $y_7$ but not into the $y_4$ 
direction.

For  moving beyond the expansion \eqref{embedexp} for the embedding, we
compute the $\text{D}7$-brane action \eqref{DBIaction} plus 
\eqref{CSactionexpl},  
\begin{equation}\label{fullaction}
\begin{aligned}
S&=-T_7\e^{-2\hat\phi}
\int\de^8\xi\e^{\phi}\sqrt{\big|\det\big(
P[g-\e^{-\frac{\phi-\hat\phi}{2}}B]
+2\pi\alpha'\e^{-\frac{\phi-\hat\phi}{2}}F\big)\big|} \\
&\phantom{{}={}}
-\mu_7\int P\Big[\hat C_4\wedge
\Big(-\frac{1}{3}B\wedge(B+2\pi\alpha'F)
+2\pi^2\alpha'^2F\wedge F
-\frac{1}{6}\e^{2\hat\phi}\tilde 
C_2\wedge\tilde 
C_2\Big)\Big]
\end{aligned}
\end{equation}
in the background \eqref{metriccorr}, \eqref{bgforms}, \eqref{tildeC2B2inS2},
\eqref{varphi}, \eqref{Ypmdef}, \eqref{Fimpl} evaluated 
with $m_1=0$, $m_2=m_3$, $m_1=0$
for a generic
embedding along both $y_4(\rho)$ and $y_7(\rho)$. We then solve the
resulting equations of motion numerically and discuss the two
embeddings $y_4=y^4(\rho)$, $y_7=0$ and
$y_7=y^7(\rho)$, $y_4=0$, respectively.

The action \eqref{Sexpand} in polar coordinates becomes,
with $r^2=\rho^2+y_4^2+y_7^2$, 
\begin{align}\label{y4y7action}
S&=-T_7\e^{-\hat\phi}
\int\de^4\xi\de\Omega_3\de\rho\Big[
\rho^3 \sqrt{1+ y'^2_4+y'^2_7}\\
&\phantom{{}={}}
+\frac{\rho^3 m^2 R^4}{72r^4\sqrt{1+ y'^2_4+y'^2_7}}\Big(
\rho^2(2+y'^2_4+y'^2_7)(10+9y'^2_7)\nonumber\\
&\phantom{{}={}
+\frac{\rho^3 m^2 R^4}{72r^4\sqrt{1+ y'^2_4+y'^2_7}}\Big(}
+4 \rho(5 y_4y'_4+5 y_7 y'_7+9 y_7y'_7(y'^2_4+y'^2_7))
\nonumber\\
&\phantom{{}={}
+\frac{\rho^3 m^2 R^4}{72r^4\sqrt{1+ y'^2_4+y'^2_7}}\Big(}
+2 y_4^2(14+19y'^2_4+23y'^2_7)
+2 y_7^2(14 +5y'^2_4+y'^2_7)
-16 y'_4y'_7 y_4 y_7 \nonumber\\
&\phantom{{}={}
+\frac{\rho^3 m^2 R^4}{72r^4\sqrt{1+ y'^2_4+y'^2_7}}\Big(}
 +6\sqrt{1+ y'^2_4+y'^2_7}
(-4 y_4^2-8 y_7^2+4 \rho(y_4y'_4+3 y_7 y'_7)+\rho^2y'^2_7 )
\Big)
\Big] \pnt \nonumber
\end{align}

\subsection{$y_4$ embedding} 

The equations of motion for $y_4$ and $y_7$ 
arising from the action \eqref{y4y7action} allow for solutions of the
form
$y_4=f(\rho)$, $y_7=0$ as well as $y_4=0$, $y_7=f(\rho)$.

\begin{figure}[H]
\begin{center}
\epsfig{file=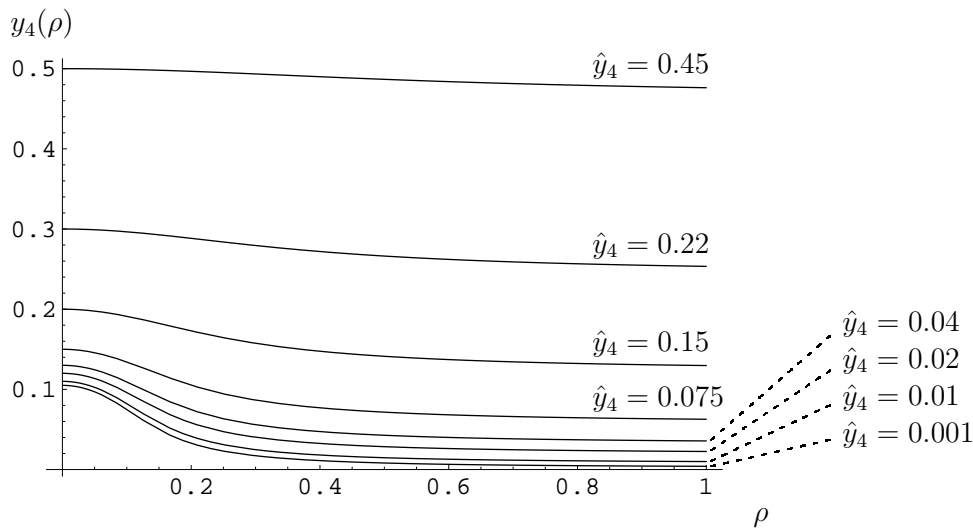, width=0.8\textwidth}
\vspace{3mm}
\caption{Embeddings along $y_4$. As discussed in section \ref{polarization},
the background does not form a
  D-brane shell in this direction. The $\text{D}7$-brane probe is repelled by
 the singularity at the origin. 
The AdS radius has been set to $R=1$ and the adjoint deformation to
$m=0.2$. Lengths are dimensionless and measured in units of $R$. The 
dimensionless boundary value $\hat y_4$ determines the dimensionful 
quark mass $m_\text{q}$ according to 
$m_\text{q}=\frac{R}{2\pi\alpha'}\hat y_4$.}
\label{y4embeddings}
\end{center}
\end{figure}

As discussed in subsection \ref{gfsources}, the pullback of $B$ to the
$\text{D}7$-brane worldvolume vanishes for
the embedding $y_4=y(\rho)$, $y_7=0$, $r^2=\rho^2+y^2$. 
In this case we have for
\eqref{fullaction} 
\begin{equation}\label{y4action}
\begin{aligned}
S&=-T_7\e^{-\hat\phi}\int\de^4\xi\de\Omega_3\de\rho\Big[
\rho^3 \sqrt{1+y'^2}\\
&\phantom{{}={}}
+\frac{\rho^3m^2R^4}{36r^4\sqrt{1+y'^2}}
\Big(10\rho^2+14y^2+5\rho^2y'^2+19y^2y'^2+10\rho yy'
-12 \sqrt{1+y'^2} (y^2-\rho y y')\Big) 
 \Big]
\pnt
\end{aligned}
\end{equation}
Solving the corresponding equation of motion for $y_4$ numerically, we
obtain
the embeddings shown in figure \ref{y4embeddings}. The quark mass
$m_\text{q}$ is identified with the boundary value $\hat y_4$ of
the embedding coordinate $y_4$ by virtue of 
$\hat y_4=y_4(\rho \rightarrow\infty)= 2 \pi \alpha' m_\text{q}$. 
For simplicity we 
choose $R=1$ in the following such that all coordinates are dimensionless,
given in units of $R$.
First of all we observe in figure \ref{y4embeddings} that although our
order $m^2$ approximation to the full Polchinski-Strassler background
breaks down at $r \sim mR^2 = 0.2$, the embeddings remain physical
within the metric given
beyond this value, since they display monotonic behaviour in 
$r=\sqrt{\rho^2 +y_4^2}$.  
Moreover we observe that the solutions are repelled by the pointlike
singularity\footnote{As discussed around \eqref{Ricciscalar}, 
in the perturbative treatment up to order $\mathcal{O}(m^2)$ of the corrections
to the background we find a singularity at $r=0$.}
at the origin. As discussed in section \ref{polarization}, 
the background shell of polarized $\text{D}3$-branes does not extend into the
$y^4$ direction.  
Note that $y^4=0$, corresponding to
vanishing quark mass, is also a solution for $\rho >0$. 
Although the solutions for generic quark mass are not constant, they
are nevertheless supersymmetric, as discussed using the methods of
holographic renormalization in subsection 
\ref{holoren}.\footnote{Compare also with
  the discussion of supersymmetric non-constant solutions in
  \cite{EvansShockWaterson}.}

\subsection{$y_7$ embedding}

\begin{figure}[H]
\begin{center}
\leavevmode
\epsfig{file=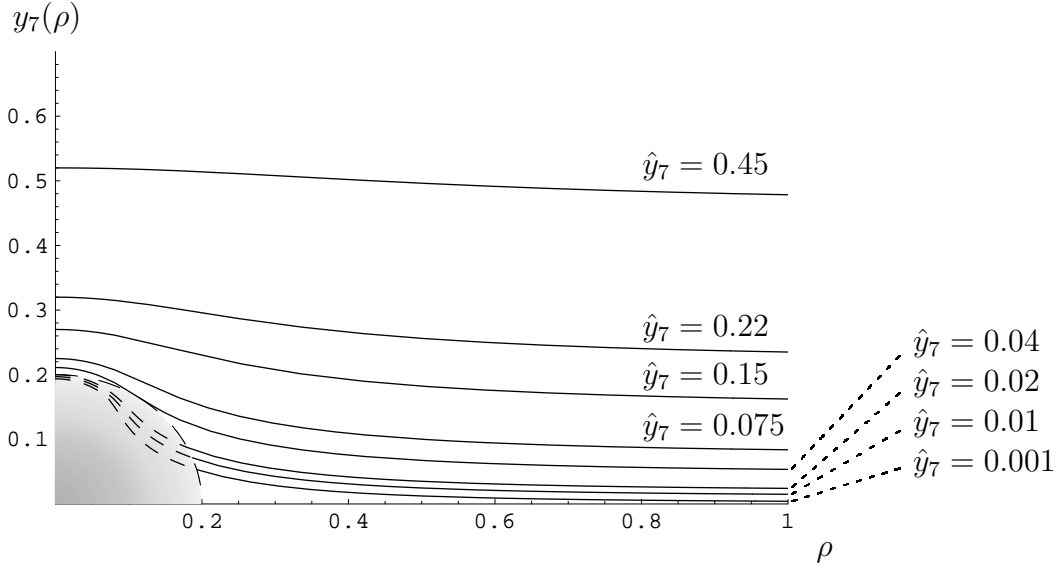, width=0.87\textwidth}
\vspace{3mm}
\caption{Embedding in the $y_7$ direction. 
The grey area corresponds to $r\leq m R^2$, into which
also the shell of dielectric $\text{D}3$-branes present in
the background is expected to expand ($R=1$ for the radius, 
$m=0.2$ for the adjoint masses, the 
dimensionless boundary value $\hat y_4$ determines the dimensionful 
quark mass $m_\text{q}$ according to 
$m_\text{q}=\frac{R}{2\pi\alpha'}\hat y_4$).}
\label{y7embedding}
\end{center}
\end{figure}

On the other hand,
if we choose the embedding $y_7=y(\rho)$, $y_4=0$, $r^2=\rho^2+y^2$ 
instead, it is $P[\tilde C_2]$ which vanishes, and we have for 
\eqref{fullaction} 
\begin{equation}
\begin{aligned}
S&=-T_7\e^{-\hat\phi}\int\de^4\xi\de\Omega_3\de\rho\Big[
\rho^3\sqrt{1+y'^2}\\
&\phantom{{}={}} 
+\frac{\rho^3m^2R^4}{72r^4\sqrt{1+y'^2}}
\Big(\rho^2 (2+y'^2) (10+9y'^2)+2 y^2(14+ y'^2)+4\rho y y' (5+9 y'^2)\\
&\phantom{{}={}+\frac{\rho^3m^2R^4}{72r^4\sqrt{1+y'^2}}\Big(} 
+6 \sqrt{1+y'^2} (-8y^2+12 \rho y y'+\rho^2 y'^2)\Big)\Big]
\pnt
\end{aligned}
\end{equation}
The corresponding $\text{D}7$-brane probe embeddings are shown in figure
\ref{y7embedding}. 
The quark mass $m_\text{q}$ is again identified with the 
boundary value of $y_7$, with the same coefficients as given below
figure \ref{y4embeddings}.  

Figure \ref{y7embedding} shows that 
the $\text{D}7$-brane probes remain outside the shell for large
values of the quark mass. 
For very small values of $m_\text{q}< 0.04$, 
the approximation of
the background metric to order $m^2$ in the adjoint masses breaks
down: The embeddings are no longer monotonic functions of
$r^2=\rho^2+y^2$ for $r < mR^2$. 

Following the discussion of section \ref{polarization}, we expect the
brane shell originating from the polarization of the background to
expand into the $y_7$ direction. From the original probe calculation
of \cite{Polchinski:2000uf},  we expect the radius of the shell to be 
$ r_0 \sim k m R^2$,  
of the same order as our expansion parameter $mR^2$.  $k$ is a
number of order one related to the flux of $F_2$ on the $\text{D}7$
probe through the $S_2$ wrapped by the $\text{D}7$. 
A definite statement
 about the repulsion of the $\text{D}7$ probe  by the shell in the
 background appears to be difficult since the shell is expected to be
 of the same size as our expansion parameter. Nevertheless, our result
 as displayed in figure
\ref{y7embedding} provides at least an  indication that for small
values of $m_\text{q}$, the $\text{D}7$
probes embedded in the $y_7$ direction merge with the background shell of
polarized $\text{D}3$-branes at $r_0 \sim mR^2= 0.2$. This is supported
further
by the comparison with the embeddings in the $y^4$ direction, in which
the shell does not form, as shown in figure
\ref{y4embeddings}. Consider for instance the embeddings with boundary
value $\hat y_4=0.04$ in both figures. We
see that $y_4(\rho)$ with boundary value $\hat y_4=0.04$ in figure 
\ref{y4embeddings} takes values smaller than $mR^2=0.2$, whereas
$y^7(\rho)$ with boundary value $\hat y_7=0.04$ 
as shown in figure \ref{y7embedding} is bounded from
below by $mR^2=0.2$ -- in fact, this solution does not enter the region
with $r<mR^2$.

\subsubsection{Meson mass}

Finally, let us discuss some aspects of meson masses in the $\N=2$
Polchinski-Strassler background, as obtained from small fluctuations
about the embedding.

Let us first consider what is expected from field theory for the
dependence of the meson mass on the quark mass.
The
contributions to the meson mass arise essentially from the VEV's of 
those contributions to
the Lagrangian which break the $U(1)$ symmetry \cite{Donoghue}. 
In our case these contributions are
\begin{gather} \label{Lbreaking}
{\cal L}_\text{breaking}=m^2\phi_\text{a}^2+m\bar\psi_\text{a}\psi_\text{a}+ 
m_\text{q} \bar \psi_\text{f}\psi_\text{f} + m_\text{q}^2 \phi_\text{f}^2
\col
\end{gather}
where `a' stands for adjoint and `f' for fundamental. Within QCD, $M^2
\propto \langle {\cal L}_\text{breaking} \rangle$ implies the famous 
Gell-Mann-Oakes-Renner relation \cite{GMOR}. However in our case, VEV's for
fermion bilinears are forbidden by supersymmetry (they are F terms of a
chiral multiplet and a non-vanishing VEV would imply that the vacuum
is not SUSY invariant). Therefore, only scalar VEV's may contribute to
the meson mass and we have
$M^2=b\,m^2 +c\,m_\text{q}^2$, with $b$, $c$ some constants.

\begin{figure}[H]
\begin{center}
\epsfig{file=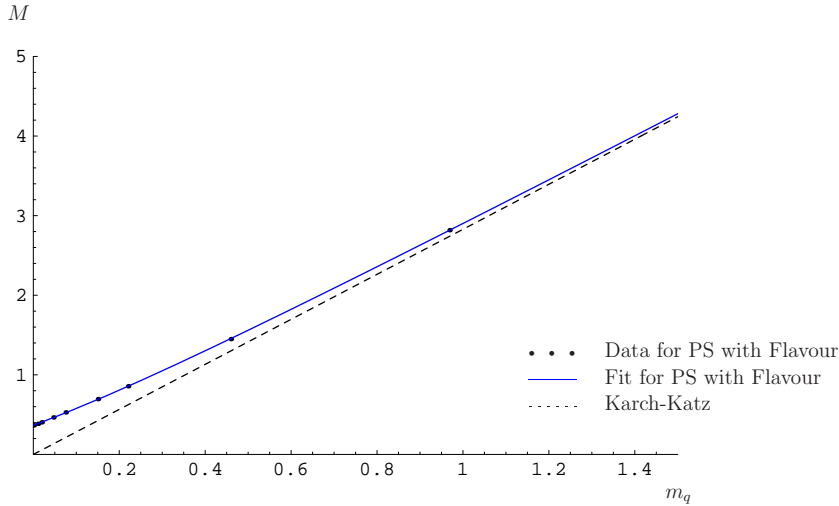, width=0.69\textwidth}
\vspace{2mm}
\caption{$y_7$ embedding: Meson mass in terms of quark mass $m_\text{q}$ 
(measured in units of $\frac{R}{2\pi\alpha'}$) 
for adjoint mass $m=0.2$.}
\label{Mm7}
\end{center}
\end{figure}
\begin{figure}[H]
\begin{center}
\leavevmode
\put(0,200){$M^2$}
\put(330,0){$m_\text{q}^2$}
\epsfig{file=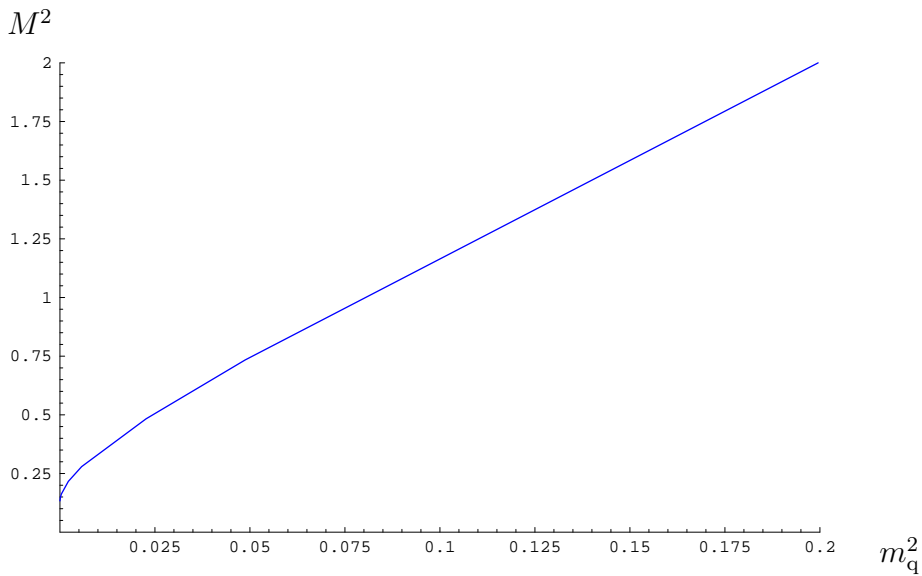, width=0.69\textwidth}
\vspace{2mm}
\caption{Square of the meson mass of figure \ref{Mm7} 
in terms of the square of the quark
  mass  $m_\text{q}$ 
(measured in units of $\frac{R}{2\pi\alpha'}$) for adjoint mass $m=0.2$.}
\label{Msquared}
\end{center}
\end{figure}

For the supergravity 
computation of the meson spectrum, we consider -- as an example --
radial fluctuations around
the solution $y_7=y_7(\rho)$, $y_4=0$  of the form
\begin{gather} \label{y4fluc}
\delta y^7(\rho,x)=\sin (k\cdot x) h(\rho)\col\qquad M^2=-k^2
\pnt
\end{gather}
We insert the ansatz \eqref{y4fluc} 
into the action \eqref{fullaction} and obtain the
equations of motion linearized in $h$. 
The values for $M$ for which
the solution is regular correspond to the meson masses.
The result of this computation for the lowest-lying meson mode 
is plotted in figure \ref{Mm7}. 
The spectrum shows a mass gap and is in agreement with
the behaviour $M = \sqrt{ c\,m_\text{q}^2 +b \, m^2}$ expected from field
theory, at least for $m_\text{q}\geq 0.04$ (the quark masses $m_\text{q}$ are 
given in units of $\frac{R}{2\pi\alpha'}$).
Note that due to our approximation of the gravity background to second
order in the adjoint masses, the meson mass calculation breaks down
for $m_\text{q}<0.04$, where the embeddings become unphysical, as may be seen
from figure \ref{y7embedding}.  For large values of $m_\text{q}$, 
the meson mass approaches the AdS result $M \propto m_\text{q}$. 
It is also instructive to plot the square of the meson mass versus the
square of the quark mass. This is done in figure \ref{Msquared}. 
\enlargethispage{\baselineskip}
For $m_\text{q}\geq 0.04$ this shows the expected linear behaviour, and for
$m_\text{q}< 0.04$ the expected breakdown of our approximation, which is
already seen for the embedding in figure  \ref{y7embedding}. 

A detailed analysis of the meson spectra for both radial and angular
fluctuations around both possible embeddings is beyond the scope of
this paper. Due to the
fact that the $U(1)$ symmetry in the two directions perpendicular to
the $\text{D}7$ probe is broken already by the background, not just by
the brane embedding itself, there may potentially 
be a mixing of $\delta y^4$ and $\delta y^7$ fluctuations. We leave a detailed
study of the meson spectrum for the future.

\section{Conclusions}
\label{conclusions}

By embedding a $\text{D}7$-brane probe into the $\N=2$
Polchinski-Strassler background, we have provided a model of
holography with flavour  -- involving $\text{D}7$-brane probes in a 
non-conformal background  -- which is well under
control both in the ultraviolet and in the infrared. In particular since the
background itself forms a $\text{D}7$-like structure in the infrared via 
the blow-up of
$\text{D}3$-branes, adding flavour via $\text{D}7$-brane probes 
appears to be natural. 
Our embeddings preserve the supersymmetry of the background. 
The meson mass displays a mass gap reflecting the presence of the
adjoint masses, in
agreement with field theory expectations. 

These appealing physical interpretations are encouraging in view of
generalizations of our results. It appears to be feasible to embed a 
$\text{D}7$-brane probe
also in the standard $\N=1$ Polchinski-Strassler background \cite{Sieg}.
Moreover from
the view of applications to strongly coupled non-supersymmetric gauge theories 
it would be very interesting to consider the $\N=0$ 
background where also
the gauginos acquire a mass \cite{Taylor}. Moreover, as mentioned in the 
introduction, for the $\N=2$ case there is the possibility of inducing
spontaneous supersymmetry breaking via non-commutative instanton solutions 
on the $\text{D}7$-brane probe. 

From a mathematical viewpoint, it would be interesting to investigate our
embeddings using $\kappa$ symmetry in order to confirm the choice of sign of
the Chern-Simons contribution to the action in \eqref{CSaction}. A further
avenue is to investigate the holonomy and spinor structure along the lines of
\cite{LopesCardoso:2004ni,DallAgata,Petrini}. 
Moreover, it would be interesting to 
study how the Donagi-Witten field theory \cite{Donagi}
in the infrared is modified
by the presence of the $\text{D}7$-brane probe. A further
  interesting avenue is to make contact with model building \cite{Uranga}.

We conclude that embedding $\text{D}7$-brane probes into the 
Polchinski-Strassler 
background is a promising approach for studying holography with flavour both
conceptionally and in view of applications. 


\section*{Acknowledgments}
We are grateful to Gabriel Lopes Cardoso,
Marco Caldarelli, Nick Evans, Marialuisa Frau, Johannes Gro\ss e,
Gabriele Honecker, 
Andreas Karch, Ingo Kirsch, Igor Klebanov, Dietmar Klemm, 
Alberto Lerda, David Mateos,  
Rob Myers,  Carlos Nu\~nez, Leo Pando Zayas, Angel Paredes,
Jeong-Hyuck Park, Alfonso Ramallo, Felix Rust, Stephan Stieberger, Stefan
  Theisen, Dimitrios Tsimpis and 
Alberto Zaffaroni for discussions and useful comments.\\    
 The work of J.\ E.\ and C.\ S.\ has been funded in part by DFG (Deutsche
Forschungsgemeinschaft) within the Emmy Noether programme, grant
ER301/1-4. The work of C.\ S.\ has also been funded in part by the European
Union within the Marie Curie Research-Training-Network, 
grant MRTN-CT-2004-005104.
The work of R.\ A.\ has been funded by DFG within the
`Schwerpunktprogramm Stringtheorie', grant ER301/2-1.

 
\appendix

\section{Notation and conventions}
\label{app:nc}

The Hodge duality operator $\star_d$ maps an $r$-form $\omega_r$ to 
a $(d-r)$-form $\star_d\omega_r$. The latter has the components
\begin{equation}\label{Hdform}
\star_d\omega_{a_1\dots a_{d-r}}=\frac{\sqrt{|\det g_{ab}|}}{r!}
\epsilon_{a_1\dots a_{d-r}}^{\hphantom{a_1\dots a_{d-r}}b_1\dots b_r}
\omega_{b_1\dots b_r}\col
\end{equation}
where we have defined
\begin{equation}
\epsilon_{12\dots d}=1\pnt
\end{equation}
The wedge product of an $r$-form $\omega_r$ and a $(d-r)$-form 
$\lambda_{d-r}$ in a $d$-dimensional space $\Sigma^d$ 
with metric $g_{ab}$ with $\tau$
negative eigenvalues and volume form
\begin{equation}
\de\vol(\Sigma^d)=\sqrt{|\det g_{ab}|}\de^d\xi
\end{equation}
reads
\begin{equation}\label{wedgeprodasip}
\begin{aligned}
\omega_r\wedge\lambda_{d-r}
&=\omega_r\cdot(\star_d\lambda)_r\de\vol(\Sigma^d)\col\qquad 
\star_d^2=(-1)^{r(d-r)+\tau}\pnt
\end{aligned}
\end{equation}
The inner product of two $r$-forms $\omega_r$ and
$\omega'_r$ is thereby defined as 
\begin{equation}\label{ipdef}
\omega_r\cdot\omega'_r=\frac{1}{r!}g^{a_1b_1}\dots g^{a_rb_r}\omega_{a_1\dots
    a_r}\omega'_{b_1\dots b_r}\pnt
\end{equation}
The components of the Hodge dual of an $r$-form are given in \eqref{Hdform}.
The wedge product therefore becomes
\begin{equation}\label{ddimwedgeprod}
\begin{aligned}
\omega_r\wedge\lambda_{d-r}
&=\frac{\det g_{ab}}{r!(d-r)!}
\epsilon^{a_1\dots a_rc_1\dots c_{d-r}}
\omega_{a_1\dots a_r}\lambda_{c_1\dots c_{d-r}}\de\xi^d\\
&=\frac{1}{r!(d-r)!}
\epsilon_{a_1\dots a_rc_1\dots c_{d-r}}
\omega_{a_1\dots a_r}\lambda_{c_1\dots c_{d-r}}\de\xi^d\col
\end{aligned}
\end{equation}
where in the last equality the independence of the wedge product of the 
(curved) metric has been used such that there the summation is understood 
as in flat space.

In type $\twob$ supergravity the physical field strengths are defined as
\begin{equation}\label{Ftildedef}
\tilde F_r=\de C_{r-1}-\de B\wedge C_{r-3}\col\qquad r=1,3,5\col
\end{equation}
where for $r=1$ the second term is zero. The missing forms 
of higher degree ($r=7,9$) are found by applying the $(D=10)$-dimensional Hodge
duality operator $\star=\star_{10}$ 
\begin{equation}\label{Ftildedualities}
\tilde F_9=\star_\S\tilde F_1\col\qquad\tilde F_7=-\star_\S\tilde
F_3\col\qquad\tilde F_5=\star_\S\tilde F_5\col 
\end{equation}
where the suffix S indicates the use of the string frame metric in 
the definition of the Hodge star operator \eqref{Hdform}.

$\tilde F_5$ is self dual. This is a particularity we have to take into 
account. 
Following the conventions of \cite{Freedman:2000xb}, the 
$5$-form field strength $\tilde F_5$ is derived from 
a redefined $4$-form potential $C_4$. We have to replace
\begin{equation}\label{C4redef}
C_4\to C_4+\frac{1}{2}B\wedge C_2\pnt
\end{equation}
Inserting this into \eqref{Ftildedef}, one finds
\begin{equation}\label{F5redef}
\tilde F_5=\star\tilde F_5=F_5-\frac{1}{2}C_2\wedge
H_3+\frac{1}{2}F_3\wedge B\pnt
\end{equation}

The Hodge star operator in \eqref{Ftildedualities} is evaluated 
with the string frame metric. 
Defining the relation
\begin{equation}\label{ESmetricrel}
g_{MN}^\E=\e^{-\frac{\phi-\hat\phi}{2}}g_{MN}^\S
\end{equation}
between the metric in the Einstein and the string frame,
the corresponding Hodge stars, acting on an $r$-form, are related via
\begin{equation}\label{ESHdrel}
\star_\E=\e^{\frac{\phi-\hat\phi}{2}(r-\frac{D}{2})}\star_\S\pnt
\end{equation}
We will skip the suffix E, denoting the Einstein frame, since this 
is the frame in which we work in the paper.

The only equation of motion which we will need
for our determination of $C_8$ is the one for the complex dilaton-axion 
$\tau$ 
\begin{equation}\label{EOMtau}
\begin{aligned}
\de\star\de\tau-\frac{1}{i\im\tau}\de\tau\wedge\star\de\tau-\frac{\e^{\hat\phi}}{2i}
G_3\wedge\star G_3=0\pnt
\end{aligned}
\end{equation} 
which with the covariant derivative $\mathcal{D}_M$ becomes in components 
\begin{equation}\label{EOMtaucomp}
\begin{aligned}
\mathcal{D}^2\tau
-\frac{1}{i\im\tau}\partial_M\tau\partial^M\tau
-\frac{\e^{\hat\phi}}{2i}G_3\cdot G_3&=0
\pnt
\end{aligned}
\end{equation}

Since also the corrections to the background respect the four-dimensional
Lorentz invariance, the metric always remains block diagonal w.r.t.\ 
the four directions longitudinal and the six directions transverse to the 
$\text{D}3$-brane. All further fields also do not contain mixed components, 
and hence the ten-dimensional Hodge star w.r.t.\ the unperturbed metric
\eqref{AdSSmetric} effectively decomposes as
\begin{equation}\label{Hdstarrel}
\star(\de x^{i_1}\wedge\dots\wedge\de
x^{i_r})=Z^{\frac{1-r}{2}}\star_6(\de x^{i_1}\wedge\dots\wedge\de
x^{i_r})\wedge\de\vol(\mathds{R}^{1,3})\col
\end{equation}
where $\star_6$ is the six-dimensional Hodge star w.r.t.\ flat Euclidean 
space.

In the expressions for the embedding of the $\text{D}7$-brane one has 
to use the inner product and Hodge star defined w.r.t.\ the pullback 
quantities. $P[\delta]_{ab}$  denotes the pullback of the Kronecker delta and 
$P[\delta]^{ab}$ its inverse. For two $r$-forms $\omega_r$, $\lambda_r$ 
they are defined as
\begin{equation}\label{iponD7def}
\omega_r\cdot\lambda_r
=\frac{1}{r!}P[\delta]^{a_1b_1}\dots P[\delta]^{a_rb_r}
\omega_{a_1\dots a_r}\lambda_{b_1\dots b_r}\col
\end{equation}
\begin{equation}\label{HdonD7def}
{\star_d}\omega_{a_1\dots a_{d-r}}
=\frac{\sqrt{\det P[\delta]_{ab}}}{r!}
P[\delta]^{c_1b_1}\dots P[\delta]^{c_rb_r}
\omega_{c_1\dots c_r}\epsilon_{b_1\dots b_ra_1\dots a_{d-r}}\pnt
\end{equation}
The Kronecker delta arises from the metric \eqref{AdSSmetric} in the six 
directions $y^i$ perpendicular to the $\text{D}3$-branes.

\section{The complex basis}
\label{app:bases}

It is convenient to introduce a complex basis with coordinates $z^p$ and
their complex conjugates $\bar z^p$, defined in \eqref{cplxbasis} and 
given by
\begin{equation}
z^p=\frac{1}{\sqrt{2}}(y^{p+3}+iy^{p+6})\col\qquad p=1,2,3\pnt
\end{equation}
In particular, for $\delta_{ij}$ one has in the complex basis
\begin{equation}\label{deltacplx}
\delta_{p\bar q}=\delta_{\bar pq}=\begin{cases}
1 &p=q \\
0 &p\neq q
\end{cases}
\col\qquad
\delta_{pq}=0\pnt
\end{equation}
A complex $2$-form can be written as
\begin{equation}
\begin{aligned}
\omega_2
&=\frac{1}{2}
\omega_{pq}\de z^p\wedge\de z^q
+\omega_{p\bar q}\de z^p\wedge\de\bar z^q
+\frac{1}{2}\omega_{\bar p\bar q}\de\bar z^p\wedge\de\bar z^q
\col
\end{aligned}
\end{equation}
where we have used $\omega_{p\bar q}=-\omega_{\bar qp}$.
The components of its complex conjugate fulfill
\begin{equation}
\bar\omega_{pq}=\overline{\omega_{\bar p\bar q}}\col\qquad
\bar\omega_{p\bar q}=\overline{\omega_{\bar pq}}\col\qquad
\bar\omega_{\bar p\bar q}=\overline{\omega_{pq}}\pnt
\end{equation}
Subtracting and adding to $\omega_2$ its complex conjugate, one finds
\begin{equation}
\begin{aligned}
\omega\pm\bar\omega_2
&=\frac{1}{2}
(\omega_{pq}\pm\overline{\omega_{\bar p\bar q}})\de z^p\wedge\de z^q
+(\omega_{p\bar q}\pm\overline{\omega_{\bar pq}})\de z^p\wedge\de\bar z^q
+\frac{1}{2}
(\omega_{\bar p\bar q}\pm\overline{\omega_{pq}})\de\bar z^p\wedge\de\bar z^q
\pnt
\end{aligned}
\end{equation}
The real and imaginary part of $\omega_2$ in the complex basis 
\eqref{cplxbasis} are then found to be
\begin{equation}\label{reimpartcplxbasis}
\begin{aligned}
\re\omega_{pq}
&=\frac{1}{2}(\omega_{pq}+\overline{\omega_{\bar p\bar q}})\col\\
\re\omega_{p\bar q}
&=\frac{1}{2}(\omega_{p\bar q}+\overline{\omega_{\bar pq}})\col\\ 
\re\omega_{\bar pq}
&=-\frac{1}{2}(\omega_{q\bar p}+\overline{\omega_{\bar qp}})\col\\ 
\re\omega_{\bar p\bar q}
&=\frac{1}{2}(\omega_{\bar p\bar q}+\overline{\omega_{pq}})\col
\end{aligned}
\qquad
\begin{aligned}
\im\omega_{pq}
&=-\frac{i}{2}(\omega_{pq}-\overline{\omega_{\bar p\bar q}})\col\\
\im\omega_{p\bar q}
&=-\frac{i}{2}(\omega_{p\bar q}-\overline{\omega_{\bar pq}})\col\\ 
\im\omega_{\bar pq}
&=\frac{i}{2}(\omega_{q\bar p}-\overline{\omega_{\bar qp}})\col\\ 
\im\omega_{\bar p\bar q}
&=-\frac{i}{2}(\omega_{\bar p\bar q}-\overline{\omega_{pq}})
\pnt
\end{aligned}
\end{equation}

In the complex basis \eqref{cplxbasis} the inner product of two 
$2$-forms reads
\begin{equation}
\omega_2\cdot\omega'_2
=\frac{1}{2}
(\omega_{pq}\omega'_{\bar p\bar q}+\omega_{\bar p\bar q}\omega'_{pq})
+\omega_{p\bar q}\omega'_{\bar pq}\col
\end{equation}
where summation over repeated indices is understood. 

With the convention $\epsilon_{456789}=1$ in the real basis, 
the non-vanishing components of the six-dimensional total
antisymmetric tensor density in the complex basis are given by
\begin{equation}
\epsilon_{123\bar 1\bar 2\bar 3}
=-i
\end{equation}
and permutations thereof.
A general component can then be represented as
\begin{equation}\label{epsilon6inepsilon3}
\epsilon_{pqr\bar s\bar t\bar u}
=-i\epsilon_{pqr}\epsilon_{\bar s\bar t\bar u}
=-i\big((\delta_{p\bar s}\delta_{q\bar t}-\delta_{p\bar t}\delta_{q\bar s})
\delta_{r\bar u}
+(\delta_{p\bar t}\delta_{q\bar u}-\delta_{p\bar u}\delta_{q\bar t})
\delta_{r\bar s}
+(\delta_{p\bar u}\delta_{q\bar s}-\delta_{p\bar s}\delta_{q\bar u})
\delta_{r\bar t}\big)\pnt
\end{equation}
Here a warning has to be made. The above representation in that form is 
\emph{valid only for the given order of unbared and bared components}, since 
the r.h.r. is not totally antisymmetric under permutations of bared and unbared
indices. In the generic case one has to adjust the global sign of the r.h.s.\ 
to take care of the order. For example, interchanging $r$ and $\bar s$ 
yields
\begin{equation}
\epsilon_{pq\bar sr\bar t\bar u}
=i\epsilon_{pqr}\epsilon_{\bar s\bar t\bar u}
=i\big((\delta_{p\bar s}\delta_{q\bar t}-\delta_{p\bar t}\delta_{q\bar s})
\delta_{ru}
+(\delta_{p\bar t}\delta_{q\bar u}-\delta_{p\bar u}\delta_{q\bar t})
\delta_{r\bar s}
+(\delta_{p\bar u}\delta_{q\bar s}-\delta_{p\bar s}\delta_{q\bar u})
\delta_{r\bar t}\big)\pnt
\end{equation}

An embedding of a $\text{D}7$-brane along $z^a$, $z^b$, $\bar z^a$, 
$\bar z^b$, $a,b=2,3$ and perpendicular 
to $z^m$, $\bar z^m$, $m=1$ induces a four-dimensional total antisymmetric 
tensor density on the parallel four directions.
One obtains from \eqref{epsilon6inepsilon3}
for the six-dimensional tensor density
\begin{equation}
\epsilon_{mab\bar m\bar c\bar d}
=-i\epsilon_{mab}\epsilon_{\bar m\bar c\bar d}
=-i(\delta_{a\bar c}\delta_{b\bar d}-\delta_{a\bar d}\delta_{b\bar c})
\pnt
\end{equation}
The four-dimensional $\epsilon$ tensor then reads 
\begin{equation}\label{epsilon4def}
\epsilon_{ab\bar c\bar d}=-\epsilon_{a\bar cb\bar d}
=-i\epsilon_{mab\bar m\bar c\bar d}
=-\epsilon_{mab}\epsilon_{\bar m\bar c\bar d}
=-\delta_{a\bar c}\delta_{b\bar d}+\delta_{a\bar d}\delta_{b\bar c}\col
\end{equation}
where the factor $i$ is chosen to ensure that in real coordinates the 
four-dimensional $\epsilon$ tensor is normalized to $1$.
With the above results, a four-dimensional Hodge star 
operator on the parallel four directions is defined.
Using the representation \eqref{epsilon4def}, one finds
that $\star_4$ acts on a $2$-form
 $\omega_2$ as follows
\begin{equation}\label{star42formcomp}
\begin{aligned}
\star_4\omega_{ab}
=-\omega_{ab}\col\qquad
\star_4\omega_{a\bar b}
=\omega_{a\bar b}-\delta_{a\bar b}\omega_{c\bar c}\col\qquad
\star_4\omega_{\bar ab}
=\omega_{\bar ab}-\delta_{a\bar b}\omega_{\bar cc}\col\qquad
\star_4\omega_{\bar a\bar b}
=-\omega_{\bar a\bar b}\col
\end{aligned}
\end{equation}
where a summation over $c$ is understood.
The above relations act differently on the components of $\omega_2$ which
are parallel to purely (anti)holomorphic directions and which point
in mixed directions. 
A generic $2$-form $\omega_2$ behaves as
\begin{equation}\label{orthstar4proj}
\frac{1}{2}(1-\star_4)\omega_2=\omega_{(2,0)}+\omega_{(0,2)}
-\frac{i}{2}\omega_{c\bar c}K\col\qquad
\frac{1}{2}(1+\star_4)\omega_2=\omega_{(1,1)}^\text{P}
\col
\end{equation}
where P denotes the primitive part of $\omega_2$, i.e.\ 
$\omega_2^\text{P}\cdot K=0$ 
and
\begin{equation}
K=i\delta_{a\bar b}\de z^a\wedge\de\bar z^b
\end{equation} 
is the K\"ahler form of the flat four-dimensional space. 
A general linear combination with $\star_4$ then acts as
\begin{equation}\label{star4proj}
(\alpha-\beta\star_4)\omega_2
=(\alpha+\beta)\Big(\omega_{(2,0)}+\omega_{(0,2)}
-\frac{i}{2}\omega_{c\bar c}K\Big)
+(\alpha-\beta)\omega_{(1,1)}^\text{P}
\pnt
\end{equation}

In the complex basis the exterior derivative operator $\de$ splits 
into its holomorphic and antiholomorphic derivative $\partial$ and
$\bar\partial$, respectively
\begin{equation}\label{dinpartial}
\de=\partial+\bar\partial
\pnt
\end{equation}
The nilpotency of $\de$ translates into the relations
\begin{equation}\label{dnilpotrel}
\partial^2=\bar\partial^2=0\col\qquad\partial\bar\partial=-\bar\partial\partial
\pnt
\end{equation}

The components of the tensor $T_3$ \eqref{Tcplx} in the real basis  
read
\begin{equation}\label{Tinrealcoord}
\begin{aligned}
T_{p+3\,q+3\,r+3}
&=\phantom{-} \frac{1}{2\sqrt{2}}\epsilon_{pqr}
(m_1+m_2+m_3)=\phantom{-} iT_{p+6\,q+6\,r+6}
\col\\
T_{p+3\,q+3\,r+6}
&=-\frac{i}{2\sqrt{2}}\epsilon_{pqr}
(m_1+m_2-m_3)=-iT_{p+6\,q+6\,r+3}
\col\\
T_{p+3\,q+6\,r+3}
&=-\frac{i}{2\sqrt{2}}\epsilon_{pqr}
(m_1-m_2+m_3)=-iT_{p+6\,q+3\,r+6}
\col\\
T_{p+6\,q+3\,r+3}
&=\phantom{-}\frac{i}{2\sqrt{2}}\epsilon_{pqr}
(m_1-m_2-m_3)=-iT_{p+3\,q+6\,r+6}
\pnt
\end{aligned}
\end{equation}
The $2$-form $S_2$ as defined in 
\eqref{S2def} reads
\begin{equation}\label{S2cplx}
\begin{aligned}
S_2&=\frac{1}{2}\epsilon_{pqr}
(m_pz^p\de\bar z^q\wedge\de\bar z^r
+m_q\bar z^p\de z^q\wedge\de\bar z^r
+m_r\bar z^p\de\bar z^q\wedge\de z^r)\\
&=\frac{1}{2}\epsilon_{pqr}
(m_pz^p\de\bar z^q
+2m_q\bar z^p\de z^q)\wedge\de\bar z^r
\pnt
\end{aligned}
\end{equation}
One finds from \eqref{S2cplx} that the components of $S_2$ are given by
\begin{equation}
S_{p\bar q}=\epsilon_{rpq}m_p\bar z^r\col\qquad 
S_{\bar pq}=\epsilon_{rpq}m_q\bar z^r\col\qquad 
S_{\bar p\bar q}=\epsilon_{rpq}m_rz^r
\pnt
\end{equation}
The components of the real and imaginary parts then read
\begin{equation}\label{S2reimcplx}
\begin{aligned}
\re S_{pq}&=\frac{1}{2}\epsilon_{rpq}m_r\bar z^r\col\\
\re S_{p\bar q}&=\frac{1}{2}\epsilon_{rpq}(m_p\bar z^r+m_qz^r)\col\\ 
\re S_{\bar pq}&=\frac{1}{2}\epsilon_{rpq}(m_q\bar z^r+m_pz^r)\col\\ 
\re S_{\bar p\bar q}&=\frac{1}{2}\epsilon_{rpq}m_rz^r\col
\end{aligned}
\qquad
\begin{aligned}
\im S_{pq}&=\frac{i}{2}\epsilon_{rpq}m_r\bar z^r\col\\
\im S_{p\bar q}&=-\frac{i}{2}\epsilon_{rpq}(m_p\bar z^r-m_qz^r)\col\\ 
\im S_{\bar pq}&=-\frac{i}{2}\epsilon_{rpq}(m_q\bar z^r-m_pz^r)\col\\ 
\im S_{\bar p\bar q}&=-\frac{i}{2}\epsilon_{rpq}m_rz^r\col
\end{aligned}
\end{equation}
where $p$, $q$
are not summed over. It is easily checked that the real and imaginary parts 
with mixed components are consistent with the antisymmetry of $S_2$.

Taking into account the split of the coordinates according to 
the presence of a $\text{D}7$-brane, the complex $2$-form 
$S_2$ as given in \eqref{S2cplx} decomposes in terms of the 
coordinates $z^a$, $z^b$, $\bar z^a$, $\bar z^b$  and $z^m$, $\bar z^m$ 
along and transverse to the $\text{D}7$-brane as  
\begin{equation}\label{S2decomp}
\begin{aligned}
S_2&=S_2^\parallel+S_2^\text{mixed}\col\\
S_2^\parallel
&=\frac{1}{2}\big(
T_{m\bar a\bar b}z^m\de\bar z^a\wedge\de\bar z^b
+2T_{\bar ma\bar b}\bar z^m\de z^a\wedge\de\bar z^b\big)\col\\
S_2^\text{mixed}
&=T_{b\bar a\bar m}z^b\de\bar z^a\wedge\de\bar z^m
+T_{\bar ba\bar m}\bar z^b\de z^a\wedge\de\bar z^m
+T_{\bar bm\bar a}\bar z^b\de z^m\wedge\de\bar z^a
\col
\end{aligned}
\end{equation}
where the components in purely transverse directions vanish.
We have restored the $3$-tensor $T_3$ according to \eqref{S2def}.
This is convenient for a later identification of the individual contributions
in terms of the $2$-form potentials.
The pullback into the four directions $z^a$, $z^b$, $\bar z^a$, $\bar z^b$ 
can be recast as follows
\begin{equation}
\begin{aligned}
P[S_2]
&=
\frac{3}{2}T_{m\bar a\bar b}z^m\de\bar z^a\wedge\de\bar z^b
+3T_{\bar ma\bar b}\bar z^m\de z^a\wedge\de\bar z^b\\
&\phantom{{}={}}
-(\partial+\bar\partial)(T_{m\bar a\bar b}z^m\bar z^a\de\bar z^b
+T_{\bar ma\bar b}\bar z^mz^a\de\bar z^b
+T_{\bar m\bar ab}\bar z^m\bar z^a\de z^b)
\col
\end{aligned}
\end{equation}
where $\partial$, $\bar\partial$ act along the four parallel directions.
We have rearranged some terms to complete exterior derivatives, which 
will turn out to be useful in the following.
The imaginary part of the above expression then decomposes
into its holomorphic and antiholomorphic components as follows
\begin{equation}\label{app:PimS2decomp2}
\begin{aligned}
P[\im S_2]_{(2,0)}&=
i\frac{3}{2}\bar S_{(2,0)}^\parallel-\frac{i}{2}\partial\theta
=
3\im S_{(2,0)}^\parallel-\frac{i}{2}\partial\theta\col\\
P[\im S_2]_{(1,1)}&=
-i\frac{3}{2}(S_{(1,1)}^\parallel-\bar S_{(1,1)}^\parallel)
-\frac{i}{2}(\bar\partial\theta-\partial\bar\theta)
=
3\im S_{(1,1)}^\parallel-\frac{i}{2}(\bar\partial\theta-\partial\bar\theta)
\col\\
P[\im S_2]_{(0,2)}&=
-i\frac{3}{2}S_{(0,2)}^\parallel
+\frac{i}{2}\bar\partial\bar\theta
=
3\im S_{(0,2)}^\parallel+\frac{i}{2}\bar\partial\bar\theta
\pnt
\end{aligned}
\end{equation}
We have thereby made use of the definition of the parallel components in 
\eqref{S2decomp},
as well as of the results $S_{(2,0)}=0$ and $\bar S_{(0,2)}=0$.
The $1$-form $\theta$ that appears above is defined as
\begin{equation}\label{app:thetadef}
\begin{aligned}
\theta=\theta_{(1,0)}&=
(\bar T_{m\bar ab}\bar z^az^m+\bar T_{\bar mab}z^a\bar z^m
-T_{\bar m\bar ab}\bar z^a\bar z^m)\de z^b
\pnt
\end{aligned}
\end{equation}

Inserting \eqref{S2reimcplx} into \eqref{tildeC2B2inS2}, one finds that 
the individual components of $\tilde C_2$ and $B$ in the complex basis 
are given by 
\begin{equation}\label{tildeC2B2cplx}
\begin{aligned}
\tilde C_{pq}&=\e^{-\hat\phi}\frac{\zeta}{6}Z\epsilon_{rpq}
m_r\bar z^r\col\\
\tilde C_{p\bar q}&=\e^{-\hat\phi}\frac{\zeta}{6}Z\epsilon_{rpq}
(m_p\bar z^r+m_qz^r)\col\\ 
\tilde C_{\bar pq}&=\e^{-\hat\phi}\frac{\zeta}{6}Z\epsilon_{rpq}
(m_q\bar z^r+m_pz^r)\col\\ 
\tilde C_{\bar p\bar q}&=\e^{-\hat\phi}\frac{\zeta}{6}Z\epsilon_{rpq}
m_rz^r\col
\end{aligned}
\qquad
\begin{aligned}
B_{pq}&=-i\frac{\zeta}{6}Z\epsilon_{rpq}
m_r\bar z^r\col\\
B_{p\bar q}&=i\frac{\zeta}{6}Z\epsilon_{rpq}
(m_p\bar z^r-m_qz^r)\col\\ 
B_{\bar pq}&=i\frac{\zeta}{6}Z\epsilon_{rpq}
(m_q\bar z^r-m_pz^r)\col\\ 
B_{\bar p\bar q}&=i\frac{\zeta}{6}Z\epsilon_{rpq}
m_rz^r\pnt
\end{aligned}
\end{equation}

In the special case $m_1=0$, $m_2=m_3=m$, 
the inner products in four dimensions hence become
\begin{equation}
\begin{aligned}
\tilde C_2\cdot\tilde C_2
&=\tilde C_{ab}\tilde C_{\bar a\bar b}
+\tilde C_{a\bar b}\tilde C_{\bar ab}
=\e^{-2\hat\phi}\frac{\zeta^2m^2}{18}Z^2
(2z^m\bar z^m+z^mz^m+\bar z^m\bar z^m)\col\\
\tilde C_2\cdot B
&=\tilde C_{ab}B_{\bar a\bar b}
+\tilde C_{\bar a\bar b}B_{ab}
+\tilde C_{a\bar b}B_{\bar ab}
+\tilde C_{b\bar a}B_{\bar ba}
=-i\e^{-\hat\phi}\frac{\zeta^2m^2}{18}Z^2(z^mz^m-\bar z^m\bar z^m)\col\\
B\cdot B
&=B_{ab}B_{\bar a\bar b}
+B_{a\bar b}B_{\bar ab}
=\frac{\zeta^2m^2}{18}Z^2
(2z^m\bar z^m-z^mz^m-\bar z^m\bar z^m)
\col
\end{aligned}
\end{equation}
and 
\begin{equation}
\begin{aligned}
\tilde C_2\cdot\star_4\tilde C_2
&=-\tilde C_{ab}\tilde C_{\bar a\bar b}
+\tilde C_{a\bar a}\tilde C_{b\bar b}
+\tilde C_{a\bar b}\tilde C_{\bar ab}
=\e^{-2\hat\phi}\frac{\zeta^2m^2}{18}Z^2
(2z^m\bar z^m+z^mz^m+\bar z^m\bar z^m)
\col\\
\tilde C_2\cdot\star_4B
&=-\frac{1}{2}(\tilde C_{ab}B_{\bar a\bar b}
+\tilde C_{\bar a\bar b}B_{ab})+\tilde C_{a\bar a}B_{b\bar b}
+\tilde C_{a\bar b}B_{\bar ab}
=i\e^{-\hat\phi}\frac{\zeta^2m^2}{18}Z^2(z^mz^m-\bar z^m\bar z^m)\col\\
B\cdot\star_4B
&=-B_{ab}B_{\bar a\bar b}
+B_{a\bar a}B_{b\bar b}
+B_{a\bar b}B_{\bar ab}
=\frac{\zeta^2m^2}{18}Z^2
(2z^m\bar z^m-z^mz^m-\bar z^m\bar z^m)
\pnt
\end{aligned}
\end{equation}

A combination that appears in the equations of motion for the 
embedding is then determined as
\begin{equation}\label{BBCCsum}
\begin{aligned}
{}&\frac{1}{2}Z^{-1}\Big(1-\frac{2}{3}\star_4\Big)B\cdot B
-\frac{1}{6}\e^{2\hat\phi}Z^{-1}\star_4\tilde C_2\cdot\tilde C_2
=-\frac{\zeta^2m^2}{54}Z(z^mz^m+\bar z^m\bar z^m)
\col
\end{aligned}
\end{equation}
where $\star_4$ is understood to act on the first form on its right.

Furthermore, one needs similar expressions where not all components are 
summed. They read
\begin{equation}\label{odformprodcplx}
\begin{aligned}
{}&\frac{Z^{-1}}{3}\big(
(1-4\star_4)(B_{ab}B_{m\bar b}+B_{a\bar b}B_{mb})
-\e^{2\hat\phi}\star_4(\tilde C_{ab}\tilde C_{m\bar b}
+\tilde C_{a\bar b}\tilde C_{mb})\big)
=\frac{\zeta^2m^2}{54}Z(2\bar z^a\bar z^m-\bar z^az^m)
\col\\
{}&\frac{Z^{-1}}{3}\big(
(1-4\star_4)(B_{\bar ab}B_{m\bar b}+B_{\bar a\bar b}B_{mb})
-\e^{2\hat\phi}\star_4(\tilde C_{\bar ab}\tilde C_{m\bar b}
+\tilde C_{\bar a\bar b}\tilde C_{mb})\big)
=\frac{\zeta^2m^2}{54}Z(2z^a\bar z^m-z^az^m)
\col
\end{aligned}
\end{equation}
where on the l.h.s.\ a sum over $b$ is understood, and $a$, $m$ take fixed 
values. One thereby first has to act with $\star_4$ on the right and then 
extract the required components 

The correction to the dilaton decomposes as
\begin{equation}\label{tildephi}
\tilde\phi=\varphi Y_+
\col
\end{equation}
where $\varphi=\varphi(r)$, and the spherical harmonic $Y_+$ arises as the real
part of the expression
\begin{equation}
\begin{aligned}
T_{ijk}V_{ijk}&=V_{ijk}V_{ijk}=\frac{3}{r^2}y^iy^lT_{ijk}T_{ljk}
=2M^2(Y_+-Y_-)\col
\end{aligned}
\end{equation}
where $M^2$ is defined as the sum of the squares of all masses as in 
\eqref{T3abssquare}.
The tensor $V_{ijk}$ is defined in \cite{Polchinski:2000uf}, and
for generic masses $Y_\pm$ are $SO(6)$ spherical harmonics with 
eigenvalue $-\frac{12}{R^2}$ which are explicitly given by
\begin{equation}\label{Ypmdef}
Y_\pm=\frac{3}{M^2r^2}\big(m_2m_3(z^1z^1\pm\bar z^1\bar z^1)
+m_1m_3(z^2z^2\pm\bar z^2\bar z^2)+m_1m_2(z^3z^3\pm\bar z^3\bar z^3)\big)\pnt
\end{equation}
The above tensor contraction appears on the r.h.s.\ in the equation of 
motion \eqref{EOMtaucomp} for the complex dilation-axion $\tau$. 
\begin{equation}
\begin{aligned}
G_3\cdot G_3&=\frac{\zeta^2}{3!}Z^{\frac{1}{2}}
\Big(T_{ijk}-\frac{4}{3}V_{ijk}\Big)
\Big(T_{ijk}-\frac{4}{3}V_{ijk}\Big)
=-\frac{4\zeta^2}{27}Z^{\frac{1}{2}}T_{ijk}V_{ijk}
\pnt
\end{aligned}
\end{equation}
With these results, it is easy to determine the radial dependent part 
in \eqref{tildephi} as
\begin{equation}\label{varphi}
\varphi
=\frac{\zeta^2M^2R^2}{108}\hat Z^{\frac{1}{2}}\pnt
\end{equation}

The tensors \eqref{IWdef} for the corrected metric \eqref{metriccorr} read in 
complex coordinates
\begin{equation}\label{Icplxbasis}
I_{pq}=-\frac{\bar z^p\bar z^q}{10z\bar z}\col\qquad
I_{p\bar q}=\frac{1}{5}\Big(\delta_{p\bar q}-\frac{\bar z^pz^q}{2z\bar z}\Big)
\col\qquad
\end{equation}
and
\begin{equation}
\begin{aligned}\label{Wcplxbasis}
W_{pq}
&=\frac{1}{4M^2z\bar z}(2\delta_{p\bar q}m_pm_r\bar z^r\bar z^r
-(m_p^2+m_q^2)\bar z^p\bar z^q)+\frac{\bar z^p\bar z^q}{10z\bar z}\col\\
W_{p\bar q}
&=\frac{1}{20}\Big(\delta_{p\bar q}-3\frac{\bar z^p z^q}{z\bar z}\Big)
+\frac{1}{4M^2z\bar z}((m_p^2+m_q^2)\bar z^pz^q-2m_pm_qz^p\bar z^q)
\pnt
\end{aligned}
\end{equation}
The remaining components are obtained by complex conjugation from the 
above expressions.
Taking the traces of the corrections in \eqref{metriccorr}
w.r.t.\ to the four-dimensional subspace, i.e.\ summing over $a=2,3$, 
thereby using that
\begin{equation}\label{rhourdef}
\rho^2=2z^a\bar z^a\col\qquad u^2=2z^m\bar z^m\col\qquad r^2=\rho^2+u^2\col
\end{equation}
one finds
\begin{equation}\label{trtildegtexpl}
\begin{aligned}
\tilde g_{aa}
&=\frac{1}{10}(6p+10q-w)
+\frac{u^2}{r^2}\frac{1}{10}(2p-10q+3w)
\pnt
\end{aligned}
\end{equation}
Furthermore, the required off-diagonal elements read
\begin{equation}\label{odtildegcplx}
\begin{aligned}
\tilde g_{am}
&=\frac{1}{10z\bar z}(5q-p+w)\bar z^a\bar z^m
-\frac{w}{4M^2z\bar z}(m_a^2+m_m^2)\bar z^a\bar z^m
\col\\
\tilde g_{\bar am}
&=\frac{1}{20z\bar z}(10q-2p-3w)z^a\bar z^m
+\frac{w}{4M^2z\bar z}((m_a^2+m_m^2)z^a\bar z^m-2m_am_m\bar z^az^m)
\col\\
\end{aligned}
\end{equation}
where the missing combinations are obtained by complex conjugation.
Furthermore, $m_a$ and $m_m$ indicate the masses corresponding
to the direction $a=2,3$ and $m=1$ in the complex basis, respectively. 
No summation over $a$ and $m$ is understood on the r.h.s.\ 
In particular, we need the specialization to $m_m=0$ and 
$m_a=m$ independent of $a$.

\section{Form relations to compute $C_8$}
\label{app:C8}

We will work with generic masses in the following. This keeps the expressions 
compact and leads to a result which is valid beyond the special case analyzed
in this paper. 
The determination of $C_8$ requires the explicit result for the wedge product 
of $S_2$ with its complex conjugate $\bar S_2$.
Using the explicit expression \eqref{S2cplx} for $S_2$ and its complex 
conjugate, as well as the representation of the product of two $\epsilon$ 
tensors in terms of Kronecker $\delta$s similar to \eqref{epsilon6inepsilon3}, 
the result can be recast into the form
\begin{equation}\label{S2wedgebarS2prelim}
\begin{aligned}
S_2\wedge\bar S_2
&=-\frac{1}{2}(m_pm_p-2m_qm_q)z^p\bar z^p\de z^q\wedge\de\bar z^q
\wedge\de z^r\wedge\de\bar z^r\\
&\phantom{={}}
-\frac{1}{2}(m_pm_p-m_qm_q)
(z^pz^q\de\bar z^q\wedge\de\bar z^p-\bar z^p\bar z^q\de z^q\wedge\de z^p)
\wedge\de z^r\wedge\de\bar z^r\\
&\phantom{={}}
+m_qm_q\bar z^pz^r\de z^q\wedge\de\bar z^q\wedge\de\bar z^r\wedge\de z^p
\pnt
\end{aligned}
\end{equation}
Introducing the diagonal mass matrix and its square
\begin{equation}\label{M2def}
M_{pq}=m_p\delta_{pq}\col\qquad
M^2_{pq}=M_{pr}M_{rq}=m_p^2\delta_{pq}\col
\end{equation}
one can rewrite
\eqref{S2wedgebarS2prelim} as
\begin{equation}\label{S2wedgebarS2}
\begin{aligned}
S_2\wedge\bar S_2
&=\Big[-\frac{1}{2}zM^2\bar z\de z\wedge\de\bar z
+z\bar z\de z\overset{M^2}{\wedge}\de\bar z
-z\de\bar z\wedge zM^2\de\bar z+\bar z\de z\wedge\bar zM^2\de z\Big]
\wedge\de z\wedge\de\bar z\\
&\phantom{={}}
+\de z\overset{M^2}{\wedge}\de\bar z\wedge z\de\bar z\wedge\bar z\de z
\pnt
\end{aligned}
\end{equation}
The following abbreviations have thereby been used
\begin{equation}
\begin{gathered}
z\bar z=z^p\bar z^p\col\qquad 
zM^2\bar z=m_p^2z^p\bar z^p\col\qquad
\de z\wedge\de\bar z=\de z^p\wedge\de\bar z^p\col\qquad
\de z\overset{M^2}{\wedge}\de\bar z=m_p^2\de z^p\wedge\de\bar z^p\col\\
z\de\bar z\wedge\bar z\de z=z^p\de\bar z^p\wedge\bar z^q\de z^q\col\qquad
z\de\bar z\wedge zM^2\de\bar z=m_q^2z^p\de\bar z^p\wedge z^q\de\bar z^q\col
\qquad
\end{gathered}
\end{equation}
where a summation over $p,q=1,2,3$ is understood on the r.h.s., and a similar
abbreviation holds for the complex conjugate of the last expression.
With the relations
\begin{equation}
\begin{aligned}
z\de\bar z\wedge zM^2\de\bar z
&=\de(z\bar z)\wedge zM^2\de\bar z
-\bar z\de z\wedge\de(zM^2\bar z)+\bar z\de z\wedge\bar zM^2\de z\col\\
\bar z\de z\wedge\de(zM^2\bar z)
&=-\de(zM^2\bar z\bar z\de z)-zM^2\bar z\de z\wedge\de\bar z
\end{aligned}
\end{equation}
the result can be rewritten as
\begin{equation}\label{S2wedgebarS2mod}
\begin{aligned}
S_2\wedge\bar S_2
&=\Big[-\frac{3}{2}zM^2\bar z\de z\wedge\de\bar z
+z\bar z\de z\overset{M^2}{\wedge}\de\bar z
-\de(z\bar z)\wedge zM^2\de \bar z-\de(zM^2\bar z\bar z\de z)\Big]
\wedge\de z\wedge\de\bar z\\
&\phantom{={}}
+\de z\overset{M^2}{\wedge}\de\bar z\wedge\de(z\bar z)\wedge\bar z\de z
\pnt
\end{aligned}
\end{equation}
Inserting the definition of $Z$ and the expression for $r^2$ in complex 
coordinates into the Bianchi identity \eqref{BIC8final}, 
one must be able to write
$\de(z\bar z)^{-2}\wedge S_2\wedge\bar S_2$ at least locally as an exact 
form.
The above expression can be seen as the special case for a more generic 
$5$-form with parameter $\beta$, which becomes
\begin{equation}
\begin{aligned}
\de (z\bar z)^\beta\wedge S_2\wedge\bar S_2
&=\Big[-\frac{3}{2}zM^2\bar z\de (z\bar z)^\beta\wedge\de z\wedge\de\bar z
+\frac{\beta}{\beta+1}\de (z\bar z)^{\beta+1}
\wedge\de z\overset{M^2}{\wedge}\de\bar z\\
&\phantom{{}={}\Big[}
+\beta\de((z\bar z)^{\beta-1}zM^2\bar zz\de\bar z\wedge\bar z\de z)\Big]
\wedge\de z\wedge\de\bar z
\pnt
\end{aligned}
\end{equation}
The first term can be rewritten 
such that it is the exterior derivative of a $4$-form potential
\begin{equation}\label{term1pot}
\begin{aligned}
{}&zM^2\bar z\de (z\bar z)^\beta
\wedge\de z\wedge\de\bar z\wedge\de z\wedge\de\bar z\\
&\qquad=\de\Big((z\bar z)^\beta\sum_{p\neq q\neq r}\Big(zM^2\bar z
-\frac{1}{\beta+1}z\bar z m_p^2\Big)\de z^q\wedge\de\bar z^q
\wedge\de z^r\wedge\de\bar z^r\Big)
\pnt
\end{aligned}
\end{equation}
As a check one can take the limit of equal masses $m=m_1=m_2=m_3$
to find an obvious identity.

One then finds immediately that the
form $\de(z\bar z)^\beta\wedge S_2\wedge\bar S_2$ follows from a $4$-form 
potential $\lambda_4$, i.e.\
\begin{equation}\label{lambda4def}
\de\lambda_4=\de(z\bar z)^\beta\wedge S_2\wedge\bar S_2\col
\end{equation}
which is given by
\begin{equation}
\begin{aligned}
\lambda_4
&=(z\bar z)^\beta\Big[-\frac{3}{2}\sum_{p\neq q\neq r}\Big(zM^2\bar z
-\frac{1}{\beta+1}z\bar z m_p^2\Big)\de z^q\wedge\de\bar z^q
+\frac{\beta}{\beta+1}z\bar z\de z\overset{M^2}{\wedge}\de\bar z\\
&\phantom{{}={}(z\bar z)^\beta\Big[}
-\de(zM^2\bar z\bar z\de z)\Big]
\wedge\de z^r\wedge\de\bar z^r
\pnt
\end{aligned}
\end{equation}
Replacing parts of the expression by $(z\bar z)^\beta S_2\wedge\bar S_2$ and
then integrating by parts and neglecting terms that can be written as an 
exterior derivative acting on a $3$-form, one finds
that an equivalent $4$-form potential $\lambda_4$, obeying \eqref{lambda4def},
is given by 
\begin{equation}\label{lambda4}
\begin{aligned}
\lambda_4
&=(z\bar z)^\beta S_2\wedge\bar S_2
+\frac{3}{2(\beta+1)}(z\bar z)^{\beta+1}
\sum_{p\neq q\neq r}(m_p^2-m_q^2-m_r^2)\de z^q\wedge\de\bar z^q
\wedge\de z^r\wedge\de\bar z^r
\pnt
\end{aligned}
\end{equation}
One could stop at this point and use this $\lambda_4$ in the special case 
$\beta=-2$ to compute $\omega_4$ from \eqref{EOMomega4final}. However 
it turns out that it is possible to find an even simpler 
$\lambda_4$ which is entirely expressed in terms of $S_2$ and $\bar S_2$. 
This is demonstrated in the following.

Taking the exterior derivative of \eqref{lambda4}
\begin{equation}
\begin{aligned}
\de\lambda_4
&=\de(z\bar z)^\beta\wedge S_2\wedge\bar S_2
+(z\bar z)^\beta\de(S_2\wedge\bar S_2)\\
&\phantom{{}={}}
+\frac{3}{2}(z\bar z)^{\beta}\de(z\bar z)\wedge
\sum_{p\neq q\neq r}(m_p^2-m_q^2-m_r^2)\de z^q\wedge\de\bar z^q
\wedge\de z^r\wedge\de\bar z^r
\col
\end{aligned}
\end{equation}
the last two terms have to cancel against each other to be in accord with 
\eqref{lambda4def}. This is guaranteed by the relation
\begin{equation}
\begin{aligned}
S_2\wedge\bar S_2
+\frac{3}{2}z\bar z
\sum_{p\neq q\neq r}(m_p^2-m_q^2-m_r^2)\de z^q\wedge\de\bar z^q
\wedge\de z^r\wedge\de\bar z^r
=\de\lambda_3
\col
\end{aligned}
\end{equation}
where $\lambda_3$ is a three form.
This result is a special case of a more general identity which holds for 
the product $(z\bar z)^\beta S_2\wedge\bar S_2$. Using the Leibnitz rule for 
the exterior derivative, one finds 
from \eqref{S2wedgebarS2mod} that the product becomes
\begin{equation}\label{zbarzbetaS2wedgeS2barnex}
\begin{aligned}
(z\bar z)^\beta S_2\wedge\bar S_2
&=(z\bar z)^\beta\Big[-\frac{3}{2}zM^2\bar z\de z\wedge\de\bar z
+\frac{\beta+3}{\beta+1}z\bar z\de z\overset{M^2}{\wedge}\de\bar z
\Big]
\wedge\de z\wedge\de\bar z\\
&\phantom{={}}
+zM^2\bar z\de(z\bar z)^\beta\wedge\bar z\de z\wedge\de z\wedge\de\bar z
+\de\sigma_3\\
\sigma_3
&=\frac{1}{\beta+1}(z\bar z)^\beta\Big(z\bar z
\de z\overset{M^2}{\wedge}\de\bar z\wedge\bar z\de z
-(z\bar zzM^2\de\bar z
+zM^2\bar z\bar z\de z)\wedge\de z\wedge\de\bar z\Big)
\pnt
\end{aligned}
\end{equation}
Using this expression, one can build the following linear combination 
with a constant $b$
\begin{equation}\label{S2wedgebarS2formrel}
\begin{aligned}
{}&(z\bar z)^\beta S_2\wedge\bar S_2
+b(z\bar z)^{\beta+1}
\sum_{p\neq q\neq r}(m_p^2-m_q^2-m_r^2)\de z^q\wedge\de\bar z^q
\wedge\de z^r\wedge\de\bar z^r\\
&\qquad=
(z\bar z)^\beta\Big[
-\frac{1}{2}\sum_{p\neq q\neq r}\Big(
3(m_p^2z^p\bar z^p+2m_q^2z^q\bar z^q)
-2bm_p^2z\bar z
\Big)
\de z^q\wedge\de\bar z^q\\
&\qquad\phantom{{}={}(z\bar z)^\beta\Big[}
+\Big(\frac{\beta+3}{\beta+1}-2b\Big)
z\bar z\de z\overset{M^2}{\wedge}\de\bar z\Big]\wedge\de z^r\wedge\de\bar z^r\\
&\qquad\phantom{{}={}}
+zM^2\bar z\de(z\bar z)^\beta\wedge\bar z\de z\wedge\de z\wedge\de\bar z
+\de\sigma_3
\pnt
\end{aligned}
\end{equation}
By using similar manipulations as the ones applied to obtain \eqref{term1pot}, 
the term in the last line can be rewritten as
\begin{equation}
\begin{aligned}
{}&zM^2\bar z\de(z\bar z)^\beta\wedge\bar z\de z\wedge\de z\wedge\de\bar z\\
&\qquad=
(z\bar z)^\beta\sum_{p\neq q\neq r}\Big(
\frac{\beta}{\beta+1}m_p^2z\bar z
+3(m_q^2-m_p^2)z^q\bar z^q\Big)
\wedge\de z^q\wedge\de\bar z^q\wedge\de z^r\wedge\de\bar z^r
+\de\omega_3
\col
\end{aligned}
\end{equation}
where
\begin{equation}
\begin{aligned}
&\omega_3
=\Big[(z\bar z)^\beta zM^2\bar z\bar z\de z
-\frac{1}{\beta+1}\sum_{p\neq q\neq r}m_p^2(z\bar z)^{\beta+1}
\bar z^q\de z^q\Big]\wedge\de z^r\wedge\de\bar z^r
\pnt
\end{aligned}
\end{equation}
Inserting this identity into the linear combination, one obtains
\begin{equation}
\begin{aligned}
{}&(z\bar z)^\beta S_2\wedge\bar S_2
+b(z\bar z)^{\beta+1}
\sum_{p\neq q\neq r}(m_p^2-m_q^2-m_r^2)\de z^q\wedge\de\bar z^q
\wedge\de z^r\wedge\de\bar z^r\\
&\qquad=\Big(\frac{\beta+3}{\beta+1}-2b\Big)
(z\bar z)^{\beta+1}\Big[\frac{1}{2}
\sum_{p\neq q\neq r}m_p^2z^q\bar z^q\de z^q\wedge\de\bar z^q
+\de z\overset{M^2}{\wedge}\de\bar z\Big]
\wedge\de z^r\wedge\de\bar z^r
+\de\lambda_3
\col
\end{aligned}
\end{equation}
where $\lambda_3=\sigma_3+\omega_3$.
It is obvious that for $b=\frac{\beta+3}{2(\beta+1)}$
the combination is an exact $4$-form, i.e.\
\begin{equation}
\begin{aligned}
{}&(z\bar z)^\beta S_2\wedge\bar S_2
+\frac{\beta+3}{2(\beta+1)}(z\bar z)^{\beta+1}
\sum_{p\neq q\neq r}(m_p^2-m_q^2-m_r^2)\de z^q\wedge\de\bar z^q
\wedge\de z^r\wedge\de\bar z^r
=\de\lambda_3
\pnt
\end{aligned}
\end{equation}
Since the potential $\lambda_4$ given by \eqref{lambda4} is only defined up to 
adding exact $4$-forms, the above result can be used to simplify the 
expression, expressing $\lambda_4$ in terms of $S_2$ and $\bar S_2$ only. 
One finally obtains
\begin{equation}\label{lambda4simpl}
\begin{aligned}
\lambda_4
=\frac{\beta}{\beta+3}(z\bar z)^\beta S_2\wedge\bar S_2\pnt
\end{aligned}
\end{equation}
In the special case $\beta=-2$ this result is the required $4$-form potential 
for the first term in \eqref{EOMomega4final}.

\section{Expansion of the Dirac-Born-Infeld and Chern Simons action}
\label{app:DBICSexpand}

With $B_{\mu\nu}=0$, $F_{\mu\nu}=0$, and a block-diagonal metric,  
the Dirac-Born-Infeld action \eqref{DBIaction} can be rewritten as
\begin{equation}\label{DBIaction2}
S_\DBI=-T_7\int\de^8\xi\e^{-\phi}\sqrt{\big|\det\e^{\frac{\phi-\hat\phi}{2}}
  g_{\mu\nu}\big|\det\big(P[E]_{ab}+2\pi\alpha' F_{ab}\big)}\col
\end{equation}
where $P$ denotes the pullback w.r.t.\ the full metric onto the four 
worldvolume directions $a,b,\dots$ of the $\text{D}7$-brane.
Furthermore, $E_{MN}$ is defined as
\begin{equation}
E_{MN}=\e^{\frac{\phi-\hat\phi}{2}}g_{MN}-B_{MN}\pnt
\end{equation}
Using the expansion of the determinant which up to second order is given by
\begin{equation}\label{sqrtdetexpansion}
\begin{aligned}
\sqrt{\det(M+\tilde M)}&=\sqrt{\det M}
\Big(1+\frac{1}{2}\tr M^{-1}\tilde M
+\frac{1}{8}(\tr M^{-1}\tilde M)^2
-\frac{1}{4}\tr M^{-1}\tilde MM^{-1}\tilde M
\Big)\col
\end{aligned}
\end{equation}
and the expression for the metric in \eqref{AdSSmetric},
the first determinant factor of \eqref{DBIaction2} expands up to 
quadratic order in the mass perturbation as
\begin{equation}\label{parallelexpansion}
\sqrt{\big|\det\e^{\frac{\phi-\hat\phi}{2}}g_{\mu\nu}\big|}
=\e^{\phi-\hat\phi}Z^{-1}
\Big(1+\frac{1}{2}Z^{\frac{1}{2}}\tilde g_{\mu\mu}\Big)
\col
\end{equation}
where summation over doubled indices is understood 
w.r.t.\ the flat Minkowski metric.

The combination $E_{MN}$ can be decomposed as
\begin{equation}
\begin{aligned}
E_{MN}
&=\hat g_{MN}+\frac{\tilde\phi}{2}\hat g_{MN}+\tilde g_{MN}-B_{MN}
\pnt
\end{aligned}
\end{equation}
`Hats' denote the unperturbed quantities, e.g.\ $\hat g_{MN}$ 
is the metric \eqref{AdSSmetric}, while a `tilde' denotes the correction 
starting at quadratic order in the perturbation.
Using \eqref{PofEstatic} for the pullback in static gauge, inserting it 
into the expansion \eqref{sqrtdetexpansion}, and keeping terms up to 
quadratic order in the perturbation, one finds
\begin{equation}
\begin{aligned}
{}&\sqrt{\det P[E]_{ab}}
=\sqrt{\det P[\hat g]_{ab}}\Big(1+\tilde\phi
+\frac{1}{2}P[\hat g]^{ab}P[\tilde g]_{ab}
-\frac{1}{4}P[\hat g]^{ab}P[\hat g]^{cd}P[B]_{bc}P[B]_{da}\Big)
\pnt
\end{aligned}
\end{equation}
Here $P[\hat g]^{ab}$ denotes the inverse of the pullback metric 
$P[\hat g]_{ab}$. 
Combining the above result with \eqref{parallelexpansion}, 
and restoring the dependence on $F_{ab}$ by replacing 
$B_{ab}\to B_{ab}-2\pi\alpha' F_{ab}$,   
the Dirac-Born-Infeld action \eqref{DBIaction2} reads
\begin{equation}\label{SDBIexpand}
\begin{aligned}
S_\DBI&=-T_7\e^{-\hat\phi}\int\de^8\xi
\sqrt{\det P[\delta]_{ab}}
\Big(1+\tilde\phi
+\frac{1}{2}Z^{\frac{1}{2}}\tilde g_{\mu\mu}
+\frac{1}{2}Z^{-\frac{1}{2}}P[\delta]^{ab}P[\tilde g]_{ab}\\
&\phantom{=-T_7\e^{-\hat\phi}\int\de^8\xi\sqrt{\det P[\delta]_{ab}}\Big(}
-\frac{1}{4}Z^{-1}P[\delta]^{ab}P[\delta]^{cd}
(P[B]-2\pi\alpha' F)_{bc}(P[B]-2\pi\alpha' F)_{da}\Big)
\col
\end{aligned}
\end{equation}
where we have cancelled factors of $Z$ by making use of the fact that for the 
six coordinates labelled by $i,j$ the unperturbed metric \eqref{AdSSmetric} 
fulfills $\hat g_{ij}=Z^{\frac{1}{2}}\delta_{ij}$. 

Similar to the Dirac-Born-Infeld action, also the
Chern-Simons action 
obtains corrections by the mass 
perturbation.  
With the induced forms $C_6$ and $C_8$ given in \eqref{C6} and \eqref{C8} 
respectively,
one finds up to order $m^2$ for the Chern-Simons action ($P[F]=F$)
\begin{equation}
\begin{aligned}
S_\CS
&=-\mu_7\int P\Big[\Big(\frac{1}{2}(\hat C_4+\tilde C_4)
\wedge(-B+2\pi\alpha'F)+C_6\Big)\wedge(-B+2\pi\alpha'F)+C_8\Big]\\
&=-\mu_7\int P\Big[\hat C_4\wedge
\Big(-\frac{1}{3}B\wedge(B+2\pi\alpha'F)
+2\pi^2\alpha'^2F\wedge F
-\frac{1}{6}\e^{2\hat\phi}\tilde 
C_2\wedge\tilde 
C_2\Big)\Big]\pnt
\end{aligned}
\end{equation}
Using the explicit expression for $\hat C_4$ \eqref{C4}
and the component expression for the wedge products \eqref{ddimwedgeprod}, 
one can reexpress the Chern-Simons part as
\begin{equation}\label{SCSexpand}
\begin{aligned}
S_\CS
&=\mu_7\e^{-\hat\phi}\int\de^8\xi\frac{1}{4}Z^{-1}\epsilon_{abcd}
\Big(\frac{1}{3}P[B]_{ab}P[B]_{cd}
+\frac{2}{3}\pi\alpha' P[B]_{ab}F_{cd}
-2\pi^2\alpha'^2F_{ab}F_{cd}\\
&\phantom{{}={}\mu_7\e^{-\hat\phi}
\int\de^8\xi \frac{1}{4}Z^{-1}\epsilon^{abcd}\Big(}
+\frac{1}{6}\e^{2\hat\phi}P[\tilde C_2]_{ab}P[\tilde C_2]_{cd}
\Big)
\pnt
\end{aligned}
\end{equation}
Using that $T_7=\mu_7$, the complete expanded action is the sum of 
\eqref{SDBIexpand} and \eqref{SCSexpand}. 
It reads
\begin{equation}
\begin{aligned}
S
&=-T_7\e^{-\hat\phi}\int\de^8\xi\Big[
\sqrt{\det P[\delta]_{ab}}\Big(1+\tilde\phi
+\frac{1}{2}Z^{\frac{1}{2}}\tilde g_{\mu\mu}
+\frac{1}{2}Z^{-\frac{1}{2}}P[\delta]^{ab}P[\tilde g]_{ab}\Big)\\
&\phantom{={}-T_7\e^{-\hat\phi}\int\de^8\xi\Big(}
+\frac{1}{4}Z^{-1}\Big(
\sqrt{\det P[\delta]_{ab}}P[\delta]^{ac}P[\delta]^{bd}
(P[B]-2\pi\alpha' F)_{ab}(P[B]-2\pi\alpha' F)_{cd}\\
&\phantom{={}-T_7\e^{-\hat\phi}\int\de^8\xi\Big(+\frac{1}{4}Z^{-1}\Big(}
-\epsilon_{abcd}\Big(
\frac{1}{3}P[B]_{ab}P[B]_{cd}
+\frac{2}{3}\pi\alpha' P[B]_{ab}F_{cd}
-2\pi^2\alpha'^2F_{ab}F_{cd}\\
&\phantom{={}-T_7\e^{-\hat\phi}\int\de^8\xi\Big(+\frac{1}{4}Z^{-1}\Big(
-\epsilon_{abcd}\Big(}
+\frac{1}{6}\e^{2\hat\phi}P[\tilde C_2]_{ab}P[\tilde C_2]_{cd}\Big)\Big)
\Big]
\pnt
\end{aligned}
\end{equation}
The result \eqref{Sexpand} is then found after using the
relations \eqref{wedgeprodasip}, \eqref{ipdef} and \eqref{ddimwedgeprod}.

\section{Perturbative expansion of the embedding}
\label{app:eomex}

An expansion of the embedding \eqref{embedexp}
into the unperturbed constant 
$\AdS_5\times\text{S}^5$ embedding $\hat X^m$ and a correction 
$\tilde X^m$, that turns out to be of order $\mathcal{O}(m^2)$, 
simplifies the problem further. Inserting this decomposition into 
\eqref{Sexpand}, the first simplification is that the pullbacks of the 
Kronecker $\delta$ become the Kronecker $\delta$ on the worldvolume of 
the $\text{D}7$-brane. Since the equations of motion are found by taking 
derivatives w.r.t.\ $\tilde X^m$ and $\partial_a\tilde X^m$, one has to 
keep those terms which contribute up to order $\mathcal{O}(m^2)$ 
to the equations, even if they are of higher order in the action.
The action found
in this way is given by \eqref{SembedFexpl}.
The equations of motion derived from it are given by
\begin{equation}\label{tildeXeom}
\begin{aligned}
{}&\partial_a\Big(
\partial_a\tilde X^m
+Z^{-\frac{1}{2}}\tilde g_{ma}
+\frac{1}{3}\big((1-4\star_4)B_{ab}B_{mb}
+\e^{2\hat\phi}(\star_4\tilde C)_{ab}\tilde C_{mb}
\big)\Big)\\
&=
\preparderiv{\tilde X^m}\Big(
\tilde\phi+\frac{1}{2}Z^{\frac{1}{2}}\tilde g_{\mu\mu}
+\frac{1}{2}Z^{-\frac{1}{2}}\tilde g_{aa}
+\frac{1}{2}Z^{-1}
\Big(\Big(1-\frac{2}{3}\star_4\Big)B\cdot B
-\frac{1}{3}\e^{2\hat\phi}\star_4\tilde C_2\cdot
\tilde C_2\Big)\Big)\Big|_{\tilde X^m=0}
\col
\end{aligned}
\end{equation}
where a sum over $a,b=5,6,8,9$ is understood and $m=4,7$ are the two 
directions transverse to the $\text{D}7$-brane. 
Transforming the summation on the l.h.s.\ to complex coordinates,
the above result reads
\begin{equation}\label{eomcplx}
\begin{aligned}
{}&2\partial_a\partial_{\bar a}\tilde{\bar z}^m
+\partial_a\big(Z^{-\frac{1}{2}}\tilde g_{m\bar a}\big)
+\partial_{\bar a}\big(Z^{-\frac{1}{2}}\tilde g_{ma}\big)\\
&
+\partial_a\Big(\frac{1}{3}Z^{-1}\big(
(1-4\star_4)(B_{\bar ab}B_{m\bar b}+B_{\bar a\bar b}B_{mb}) 
-\e^{2\hat\phi}((\star_4\tilde C)_{\bar ab}\tilde C_{m\bar b}
+(\star_4\tilde C)_{\bar a\bar b}\tilde C_{mb}\big)\Big)\\
&
+\partial_{\bar a}\Big(\frac{1}{3}Z^{-1}\big(
(1-4\star_4)(B_{ab}B_{m\bar b}+B_{a\bar b}B_{mb})
-\e^{2\hat\phi}((\star_4\tilde C)_{ab}\tilde C_{m\bar b}
+(\star_4\tilde C)_{a\bar b}\tilde C_{mb})\big)\Big)\\
&=
\preparderiv{\tilde z^m}\Big(
\tilde\phi+\frac{1}{2}Z^{\frac{1}{2}}\tilde g_{\mu\mu}
+\frac{1}{2}Z^{-\frac{1}{2}}\tilde g_{aa}
+\frac{1}{2}Z^{-1}\Big(
\Big(1-\frac{2}{3}\star_4\Big)B\cdot B
-\frac{1}{3}\e^{2\hat\phi}\star_4\tilde C_2\cdot
\tilde C_2\Big)\Big)\Big|_{\tilde z^m=\tilde{\bar z}^m=0}
\pnt
\end{aligned}
\end{equation}
The individual expressions that enter the above equation are given by the  
derivatives of the results computed in Appendix \ref{app:bases}.
From \eqref{BBCCsum} one finds with the definition of $\rho$, $u$ and $r$ 
in \eqref{rhourdef}
\begin{equation}\label{mderformprodcplx}
\begin{aligned}
{}&\partial_m\Big(\frac{1}{2}Z^{-1}
\Big(1-\frac{2}{3}\star_4\Big)B\cdot B
-\frac{1}{6}\e^{2\hat\phi}Z^{-1}\star_4\tilde C_2\cdot\tilde C_2\Big)
=-\frac{\zeta^2m^2}{27}Z
\Big(\Big(1-\frac{u^2}{r^2}\Big)z^m-2\frac{(\bar z^m)^3}{r^2}\Big)
\pnt
\end{aligned}
\end{equation}
The derivative of \eqref{odformprodcplx} reads
 \begin{equation}\label{mdivformprodcplx}
\begin{aligned}
{}&\frac{1}{3}
\partial_a\big(Z^{-1}\big(
(1+4\star_4)(B_{\bar ab} B_{m\bar b}+ B_{\bar a\bar b} B_{mb})
+\e^{2\hat\phi}\star_4(\tilde C_{\bar ab}\tilde C_{m\bar b}
+\tilde C_{\bar a\bar b}\tilde C_{mb})\big)\big)
+(a\leftrightarrow\bar a)
\\
&=\frac{2\zeta^2m^2}{27}Z(2\bar z^m-z^m)\frac{u^2}{r^2}
\pnt
\end{aligned}
\end{equation}
The gradient of the dilaton as given in \eqref{tildephi}
becomes in the case $m_1=0$, $m_2=m_3=m$
\begin{equation}\label{mdertildephicplx}
\begin{aligned}
\partial_m\tilde\phi
&=\frac{\zeta^2m^2}{18}Z\Big(
-2\frac{(\bar z^m)^3}{r^2}+\Big(1-\frac{u^2}{r^2}\Big)z^m\Big)
\col\\
\end{aligned}
\end{equation}
The derivative of the subtraces of the corrections to the metric
in \eqref{trtildegtexpl} become
\begin{equation}\label{mdertrtildegcplx}
\begin{aligned}
\frac{1}{2}\partial_m
\big(Z^{-\frac{1}{2}}(Z\tilde g_{\mu\mu}+\tilde g_{aa})\big)
&=\frac{1}{5R^2}\Big(3p-15q+2w-(2p-10q+3w)\frac{u^2}{r^2}\Big)
\bar z^m
\pnt
\end{aligned}
\end{equation}
Finally, the derivatives of the off-diagonal elements 
of the corrections to the metric in \eqref{odtildegcplx}
are found to be given by
\begin{equation}\label{mdivtildegcplx}
\begin{aligned}
{}&\partial_a(Z^{-\frac{1}{2}}\tilde g_{\bar am})
+\partial_{\bar a}(Z^{-\frac{1}{2}}\tilde g_{am})
=\frac{1}{5R^2}(20q-4p-\omega)\frac{u^2}{r^2}\bar z^m
\pnt
\end{aligned}
\end{equation}
Inserting the above equations into \eqref{eomcplx}, one obtains
\begin{equation}
\begin{aligned}
2\partial_a\partial_{\bar a}\tilde{\bar z}^m
&=\frac{\zeta^2m^2}{54}Z\Big(
-8\frac{u^2}{r^2}\bar z^m+\Big(1+3\frac{u^2}{r^2}\Big)z^m
-2\frac{(\bar z^m)^3}{r^2}\Big)\\
&\phantom{{}={}}
+\frac{1}{5R^2}\Big(3p-15q+2\omega+(2p-10q-2\omega)\frac{u^2}{r^2}\Big)
\bar z^m
\pnt
\end{aligned}
\end{equation}
The quantities on the r.h.s.\ that carry a `hat' have to be 
evaluated evaluated using the unperturbed 
embedding coordinates $\hat z^m$ and $\hat{\bar z}^m$ as required by
\eqref{eomcplx}.

As a final step, one inserts the explicit values for $p$, $q$ and $\omega$ 
given in \eqref{wpqh0}, to find the combinations
\begin{equation}
\begin{aligned}
3p-15q+2\omega
&=-\frac{10\zeta^2m^2R^2}{27}Z
\col\qquad
2p-10q-2\omega
&=\frac{10\zeta^2m^2R^2}{81}Z
\pnt
\end{aligned}
\end{equation}
The final result hence reads
\begin{equation}\label{tildebarzeom}
\begin{aligned}
2\partial_a\partial_{\bar a}\tilde{\bar z}^m
&=\frac{\zeta^2m^2}{54}\hat Z\Big(
-2\Big(2+\frac{10}{3}\frac{\hat u^2}{\hat r^2}\Big)\hat{\bar z}^m
+\Big(1+3\frac{\hat u^2}{\hat r^2}\Big)\hat z^m
-2\frac{(\hat{\bar z}^m)^3}{\hat r^2}
\Big)
\col
\end{aligned}
\end{equation}

The r.h.s.\ of \eqref{tildebarzeom} depends on $\rho=2z^a\bar z^a$ 
via $\hat r^2=\rho^2+\hat u^2$
only. It is therefore reasonable to assume that also the embedding coordinates
depend on $\rho$ only. The Laplace operator on the l.h.s.\ acts on 
a function $f(\rho)$ as
\begin{equation}\label{Laplaceopsimpl}
2\partial_a\partial_{\bar a}f
=f''+f'\frac{3}{\rho}
=\frac{1}{\rho^3}\partial_\rho\big(\rho^3\partial_\rho f\big)
\pnt
\end{equation}
Parameterizing the embedding coordinates in the complex basis as
\begin{equation}\label{cylembedcoord}
\begin{aligned}
\sqrt{2}z^m&=u\e^{i\psi}=(\hat u+\tilde u)\e^{i(\hat\psi+\tilde\psi)}
=(\hat u+\tilde u+i\hat u\tilde\psi)\e^{i\hat\psi}
\col\\
\sqrt{2}\bar z^m&=u\e^{-i\psi}=(\hat u+\tilde u)\e^{-i(\hat\psi+\tilde\psi)}
=(\hat u+\tilde u-i\hat u\tilde\psi)\e^{-i\hat\psi}
\col
\end{aligned}
\end{equation}
one finds the linear combinations
\begin{equation}\label{barztildezlincomb}
\begin{aligned}
\hat z^m\partial_a\partial_{\bar a}\tilde{\bar z}^m
+\hat{\bar z}^m\partial_a\partial_{\bar a}\tilde z^m
&=\hat u\partial_a\partial_{\bar a}\tilde u
\col\\
\hat z^m\partial_a\partial_{\bar a}\tilde{\bar z}^m
-\hat{\bar z}^m\partial_a\partial_{\bar a}\tilde z^m
&=-i\hat u^2\partial_a\partial_{\bar a}\tilde\psi
\pnt
\end{aligned}
\end{equation}
We therefore first multiply \eqref{tildebarzeom} by $\hat z^m$ and 
use $\hat u^2=2\hat z^m\hat{\bar z}^m$, 
$\frac{\hat z^m}{\hat{\bar z}^m}=\e^{2i\hat\psi}$. This yields
\begin{equation}
\begin{aligned}
2\hat z^m\partial_a\partial_{\bar a}\tilde{\bar z}^m
&=\frac{\zeta^2m^2}{108}\hat Z\hat u^2
\Big(-2\Big(2+\frac{10}{3}\frac{\hat u^2}{\hat r^2}\Big)
+\Big(1+2\frac{\hat u^2}{\hat r^2}\Big)
\cos2\hat\psi
+\Big(1+4\frac{\hat u^2}{\hat r^2}\Big)i\sin2\hat\psi
\Big)
\pnt
\end{aligned}
\end{equation}
Then we use the relations \eqref{barztildezlincomb} and \eqref{Laplaceopsimpl} 
to obtain
\begin{equation}
\begin{aligned}
\frac{1}{\rho^3}\partial_\rho\big(\rho^3\partial_\rho\tilde u\big)
&=\frac{\zeta^2m^2}{54}\hat Z\hat u
\Big(-2\Big(2+\frac{10}{3}\frac{\hat u^2}{\hat r^2}\Big)
+\Big(1+2\frac{\hat u^2}{\hat r^2}\Big)
\cos2\hat\psi\Big)
\col\\
\frac{1}{\rho^3}\partial_\rho\big(\rho^3\partial_\rho\tilde\psi\big)
&=-\frac{\zeta^2m^2}{54}\hat Z
\Big(1+4\frac{\hat u^2}{\hat r^2}\Big)\sin2\hat\psi
\pnt
\end{aligned}
\end{equation}
Collecting the terms with the same dependence on $\rho$, one immediately 
finds \eqref{diffeq} with the values \eqref{BCu} and \eqref{BCpsi}.

\section{Evaluation of the on-shell action}
\label{app:osaction}
The explicit expression of the action up to order $\mathcal{O}(m^2)$ follows
from \eqref{Sexpand} or \eqref{SembedFexpl}. After transforming to 
polar coordinates with radius $\rho$ it reads
\begin{equation}
\begin{aligned}
S
&=-T_7\e^{-\hat\phi}\Omega_3\int\de\xi^4\de\rho\rho^3\Big(
1+\frac{1}{2}(\partial_\rho\tilde u)^2
+\frac{\hat u^2}{2}(\partial_\rho\tilde \psi)^2\\
&\phantom{{}={}-T_7\e^{-\hat\phi}\Omega_3\int\de\xi^4\de\rho\rho^3\Big(}
+\frac{\zeta^2m^2}{108}\hat Z\Big(
-\Big(\frac{1}{3}+\cos2\hat\psi\Big)\hat u^2
+\frac{5}{3}\hat r^2\\
&\phantom{{}={}-T_7\e^{-\hat\phi}\Omega_3\int\de\xi^4\de\rho\rho^3\Big(
+\frac{\zeta^2m^2}{108}\hat Z\Big(}
-2\Big(3-\frac{8}{3}\frac{\hat u^2}{\hat r^2}\Big)\hat u\tilde u
+\frac{11}{3}\hat u\rho\partial_\rho\tilde u\\
&\phantom{{}={}-T_7\e^{-\hat\phi}\Omega_3\int\de\xi^4\de\rho\rho^3\Big(
+\frac{\zeta^2m^2}{108}\hat Z\Big(}
-2\hat u
\Big(\Big(1-2\frac{\hat u^2}{\hat r^2}\Big)
(1-\cos2\hat\psi)\tilde u+\hat u\sin2\hat\psi\,\tilde\psi)\Big)\\
&\phantom{{}={}-T_7\e^{-\hat\phi}\Omega_3\int\de\xi^4\de\rho\rho^3\Big(
+\frac{\zeta^2m^2}{108}\hat Z\Big(}
+2\hat u\rho
((1-\cos2\hat\psi)\partial_\rho\tilde u
+\hat u\sin2\hat\psi\partial_\rho\tilde\psi)
\Big)
\pnt
\end{aligned}
\end{equation}
We derive the action expressed in the new coordinate 
$\hat\chi=\frac{1}{\hat r^2}$ and restrict 
ourselves to the embeddings with constant $\psi$. For these embeddings
the inhomogeneity in the equation of motion for $\psi$ has to vanish
vanishes. According to \eqref{BCpsi} this is the case for the choices 
$\hat\psi=0$ or $\hat\psi=\frac{\pi}{2}$.
The action then simplifies to
\begin{equation}
\begin{aligned}
S
&=-T_7\e^{-\hat\phi}\Omega_3\int\de\xi^4\de\rho\rho^3\Big(
1+\frac{1}{2}(\partial_\rho\tilde u)^2\\
&\phantom{{}={}-T_7\e^{-\hat\phi}\Omega_3\int\de\xi^4\de\rho\rho^3\Big(}
+\frac{\zeta^2m^2}{108}\hat Z\Big(
c_0\hat u^2+\frac{5}{3}\hat r^2
-2\Big(c_1-c_2\frac{\hat u^2}{\hat r^2}\Big)\hat u\tilde u
+c_3\hat u\rho\partial_\rho\tilde u
\Big)\col
\end{aligned}
\end{equation}
where the coefficients are explicitly given by
\begin{equation}\label{cdef}
\begin{aligned}
\hat\psi&=0:
\qquad &c_0&=-\frac{4}{3}\col\qquad &c_1&=3\col\qquad &c_2&=\frac{8}{3}\col
\qquad &c_3&=\frac{11}{3}\col\\ 
\hat\psi&=\frac{\pi}{2}:
\qquad &c_0&=\frac{2}{3}\col\qquad &c_1&=5\col\qquad &c_2&=\frac{20}{3}\col
\qquad &c_3&=\frac{23}{3}
\pnt
\end{aligned}
\end{equation}
We then
evaluate it on the solution of the equation of motion.
Including the measure from the integration, 
in the coordinate $\hat\chi$  the kinetic term 
becomes with \eqref{hrcoorddef} and $\rho^2=\frac{1}{\hat\chi}-\hat u^2$
\begin{equation}
\de\rho\rho^3(\partial_\rho\tilde u)^2
=-\de\hat\chi\rho^42\hat\chi^2(\partial_{\hat\chi}\tilde u)^2
=-2\de\hat\chi(1-\hat\chi u^2)^2(\partial_{\hat\chi}\tilde u)^2
\pnt
\end{equation} 
Introducing $\hat\chi$ as independent coordinate, the action becomes
\begin{equation}
\begin{aligned}
S
&=\frac{T_7}{2}\e^{-\hat\phi}\Omega_3\int\de\xi^4
\de\hat\chi\,\Big(
\frac{1}{\hat\chi^3}-\frac{\hat u^2}{\hat\chi^2}
+2(1-\hat\chi\hat u^2)^2(\partial_{\hat\chi}\tilde u)^2\\
&\phantom{{}={}\frac{T_7}{2}\e^{-\hat\phi}\Omega_3\int\de\xi^4
\de\chi\,\Big(}
+\frac{\zeta^2m^2R^4}{108}
\frac{1-\hat\chi\hat u^2}{\hat\chi}\Big(
c_0\hat u^2+\frac{5}{3\hat\chi}
-2\hat u(c_1-c_2\hat\chi\hat u^2)\tilde u\\
&\phantom{{}={}\frac{T_7}{2}\e^{-\hat\phi}\Omega_3\int\de\xi^4
\de\chi\,\Big(+\frac{\zeta^2m^2R^4}{108}
\frac{1-\hat\chi\hat u^2}{\hat\chi}\Big(}
-2c_3\hat u(1-\hat\chi\hat u^2)\hat\chi
\partial_{\hat\chi}\tilde u\Big)\Big)
\pnt
\end{aligned}
\end{equation}
Before evaluating it on the solution of the equations of motion, 
it is advantageous to partially integrate some terms. 
The found result reads
\begin{equation}
\begin{aligned}
S
&=\frac{T_7}{2}\e^{-\hat\phi}\Omega_3\int\de\xi^4\bigg[
\int\de\hat\chi\,\Big(
\frac{1}{\hat\chi^3}-\frac{\hat u^2}{\hat\chi^2}
-2\tilde u\partial_{\hat\chi}
((1-\hat\chi\hat u^2)^2\partial_{\hat\chi}\tilde u)\\
&\phantom{{}={}\frac{T_7}{2}\e^{-\hat\phi}\Omega_3\int\de\xi^4\bigg[
\int\de\chi\,\Big(}
+\frac{\zeta^2m^2R^4}{108}
\frac{1-\hat\chi\hat u^2}{\hat\chi}\Big(
c_0\hat u^2+\frac{5}{3\hat\chi}
-2\hat u(c_1-(c_2-2c_3)\hat\chi\hat u^2)\tilde u\Big)\Big)\\
&\phantom{{}={}\frac{T_7}{2}\e^{-\hat\phi}\Omega_3\int\de\xi^4\bigg[}
+2\tilde u(1-\hat\chi\hat u^2)^2
\Big(\partial_{\hat\chi}\tilde u
-\frac{\zeta^2m^2R^4}{108}c_3\hat u
\Big)\bigg]
\pnt
\end{aligned}
\end{equation}
The equation of motion for $\tilde u$ in the coordinate $\hat\chi$ reads
\begin{equation}\label{uinchieom}
\begin{aligned}
4\partial_{\hat\chi}
((1-\hat\chi\hat u^2)^2\partial_{\hat\chi}\tilde u)
&=-\frac{\zeta^2m^2R^4}{54}\hat u
\frac{1-\hat\chi\hat u^2}{\hat\chi}\Big(
c_1-(c_2-2c_3)\hat\chi\hat u^2\Big)\\
&=-\frac{\zeta^2m^2R^4}{54}
\hat u\Big(\frac{c_1}{\hat\chi}
-(c_1+c_2-2c_3)\hat u^2+(c_2-2c_3)\hat\chi\hat u^4\Big)
\pnt
\end{aligned}
\end{equation}
The explicit (regular) solution of the above equation 
that correspond to \eqref{frhoreg} with $f=u$, but now given in $\hat\chi$ 
read
\begin{equation}\label{uchireg}
u=\hat u+\tilde u
=\hat u-\frac{\zeta^2m^2R^4}{216}\hat u
\Big(\Big(\frac{c_2}{2}-c_3\Big)\hat\chi+c_1\frac{\hat\chi}{1-\hat\chi\hat u^2}
\ln\hat\chi\hat u^2\Big)
\pnt
\end{equation}
From the above result we read off the constants $B_u$ and $C_u$ given in 
\eqref{BCu} with $\hat\psi=0,\frac{\pi}{2}$ in terms of 
$c_1$, $c_2$ and $c_3$. They are given by
\begin{equation}
B_u=-\frac{\zeta^2m^2R^4}{54}c_1\col\qquad
C_u=\frac{\zeta^2m^2R^4}{54}(c_2-2c_3)
\pnt
\end{equation}
Furthermore, the first derivative of the solution reads
\begin{equation}
\begin{aligned}
(1-\hat\chi\hat u^2)^2\partial_{\hat\chi}\tilde u
&=-\frac{\zeta^2m^2R^4}{216}
\hat u\Big(c_1\ln\hat\chi\hat u^2
+\Big(\frac{c_2}{2}-c_3\Big)(1-\hat\chi\hat u^2)^2
+c_1(1-\hat\chi\hat u^2)\Big)
\pnt
\end{aligned}
\end{equation}
To obtain it directly from the above differential equations, one has to 
add appropriate integration constants.
Using the equations of motion \eqref{uinchieom}, the action becomes
\begin{equation}\label{piSchiaction}
\begin{aligned}
S
&=\frac{T_7}{2}\e^{-\hat\phi}\Omega_3\int\de\xi^4\bigg[
\int\de\hat\chi\,\Big(
\frac{1}{\hat\chi^3}-\frac{\hat u^2}{\hat\chi^2}\\
&\phantom{{}={}\frac{T_7}{2}\e^{-\hat\phi}\Omega_3\int\de\xi^4\bigg[
\int\de\chi\,\Big(}
+\frac{\zeta^2m^2R^4}{108}
\frac{1-\hat\chi\hat u^2}{\hat\chi}\Big(
c_0\hat u^2+\frac{5}{3\hat\chi}
-\hat u(c_1-(c_2-2c_3)\hat\chi\hat u^2)\tilde u\Big)\Big)\\
&\phantom{{}={}\frac{T_7}{2}\e^{-\hat\phi}\Omega_3\int\de\xi^4\bigg[}
+2\tilde u(1-\hat\chi\hat u^2)^2
\Big(\partial_{\hat\chi}\tilde u
-\frac{\zeta^2m^2R^4}{108}c_3\hat u
\Big)\bigg]
\pnt
\end{aligned}
\end{equation}\label{mesonappendix}



\begin{thebibliography}{99}

\bibitem{Maldacenaoriginal}
 J.~M.~Maldacena,
  Adv.\ Theor.\ Math.\ Phys.\  {\bf 2} (1998) 231
  [Int.\ J.\ Theor.\ Phys.\  {\bf 38} (1999) 1113]
  [arXiv:hep-th/9711200].

\bibitem{Bertolini} 
 M.~Bertolini, P.~Di Vecchia, M.~Frau, A.~Lerda and R.~Marotta,
  Nucl.\ Phys.\ B {\bf 621} (2002) 157
  [arXiv:hep-th/0107057].

\bibitem{Grana}   M.~Gra\~na and J.~Polchinski,
  Phys.\ Rev.\ D {\bf 65} (2002) 126005
  [arXiv:hep-th/0106014].


\bibitem{Karch:2002sh}
A.~Karch and E.~Katz, 
{JHEP} {\bf 06} (2002) 043 
 [hep-th/0205236].



\bibitem{krumy}
M.~Kruczenski, D.~Mateos, R.C.~Myers and D.J.~Winters, 
JHEP {\bf 07} (2003) 049 
[hep-th/0304032]. 

\bibitem{meson}
I.~Kirsch,
  [arXiv:hep-th/0607205];\\
 F.~Bigazzi and A.~L.~Cotrone,
  [arXiv:hep-th/0606059];\\
  J.~Erdmenger, N.~Evans and J.~Gro\ss e,
  [arXiv:hep-th/0605241];\\
  A.~V.~Ramallo,
  Mod.\ Phys.\ Lett.\ A {\bf 21} (2006) 1481
  [arXiv:hep-th/0605261];\\
R.~C.~Myers and R.~M.~Thomson,
 JHEP {\bf 0609}, 066 (2006)  [arXiv:hep-th/0605017].

\bibitem{BEEGK}   J.~Babington, J.~Erdmenger, N.~J.~Evans, Z.~Guralnik
 and I.~Kirsch,
  Phys.\ Rev.\ D {\bf 69} (2004) 066007
  [arXiv:hep-th/0306018].

\bibitem{krumy2} M.~Kruczenski, D.~Mateos, R.~C.~Myers and D.~J.~Winters,
  ``Towards a holographic dual of large-N(c) QCD,''
  JHEP {\bf 0405} (2004) 041
  [arXiv:hep-th/0311270].

\bibitem{Sugimoto}   T.~Sakai and S.~Sugimoto,
  Prog.\ Theor.\ Phys.\  {\bf 113} (2005) 843
  [arXiv:hep-th/0412141]; \\
T.~Sakai and S.~Sugimoto,
  Prog.\ Theor.\ Phys.\  {\bf 114} (2006) 1083
  [arXiv:hep-th/0507073].


\bibitem{chiral}
D.~Gepner and S.~Pal,
  [arXiv:hep-th/0608229]; \\
  E.~Antonyan, J.~A.~Harvey and D.~Kutasov,
  [arXiv:hep-th/0608177];\\
Y.~H.~Gao, W.~S.~Xu and D.~Z.~Zeng,
  JHEP {\bf 0608} (2006) 018
  [arXiv:hep-th/0605138];
A.~Parnachev and D.~A.~Sahakyan,
  [arXiv:hep-th/0604173];\\
E.~Dudas and C.~Papineau,
  [arXiv:hep-th/0608054]; \\
 N.~J.~Evans and J.~P.~Shock,
  Phys.\ Rev.\ D {\bf 70}, 046002 (2004)
  [arXiv:hep-th/0403279].

\bibitem{AEEG}
   R.~Apreda, J.~Erdmenger, N.~Evans and Z.~Guralnik,
  Phys.\ Rev.\ D {\bf 71} (2005) 126002
  [arXiv:hep-th/0504151];\\
R.~Apreda, J.~Erdmenger and N.~Evans,
  JHEP {\bf 0605} (2006) 011
  [arXiv:hep-th/0509219];\\
 Z.~Guralnik, S.~Kovacs and B.~Kulik,
  JHEP {\bf 0503} (2005) 063
  [arXiv:hep-th/0405127].

\bibitem{thermal}
N.~Horigome and Y.~Tanii,
  [arXiv:hep-th/0608198];\\
P.~C.~Argyres, M.~Edalati and J.~F.~Vazquez-Poritz,
  [arXiv:hep-th/0608118];\\
 C.~P.~Herzog,
  [arXiv:hep-th/0605191];\\
  C.~P.~Herzog, A.~Karch, P.~Kovtun, C.~Kozcaz and L.~G.~Yaffe,
  JHEP {\bf 0607}, 013 (2006)
  [arXiv:hep-th/0605158];\\
A.~Karch and A.~O'Bannon,
  [arXiv:hep-th/0605120];\\
 T.~Albash, V.~Filev, C.~V.~Johnson and A.~Kundu,
  [arXiv:hep-th/0605088];\\
  D.~Mateos, R.~C.~Myers and R.~M.~Thomson,
  Phys.\ Rev.\ Lett.\  {\bf 97}, 091601 (2006) [arXiv:hep-th/0605046];\\
 K.~Ghoroku, A.~Nakamura and M.~Yahiro,
  Phys.\ Lett.\ B {\bf 638} (2006) 382
  [arXiv:hep-ph/0605026];\\
 K.~Y.~Kim, S.~J.~Sin and I.~Zahed,
  [arXiv:hep-th/0608046];\\
D.~Mateos, R.~C.~Myers and R.~M.~Thomson,
  [arXiv:hep-th/0610184];\\
  A.~Parnachev and D.~A.~Sahakyan,
  [arXiv:hep-th/0610247].

\bibitem{AdSQCD}
J.~Erlich, E.~Katz, D.~T.~Son and M.~A.~Stephanov,
  Phys.\ Rev.\ Lett.\  {\bf 95}, 261602 (2005)
  [arXiv:hep-ph/0501128]; \\
 L.~Da Rold and A.~Pomarol,
  JHEP {\bf 0601}, 157 (2006)
  [arXiv:hep-ph/0510268];\\
  S.~J.~Brodsky and G.~F.~de Teramond,
  Phys.\ Rev.\ Lett.\  {\bf 96} (2006) 201601
  [arXiv:hep-ph/0602252];\\
 N.~Evans and A.~Tedder,
  [arXiv:hep-ph/0609112].


\bibitem{Hong}
  S.~Hong, S.~Yoon and M.~J.~Strassler,
  JHEP {\bf 0404} (2004) 046
  [arXiv:hep-th/0312071].



\bibitem{KS}
  I.~R.~Klebanov and M.~J.~Strassler,
  JHEP {\bf 0008} (2000) 052
  [arXiv:hep-th/0007191].


\bibitem{MN}
  J.~M.~Maldacena and C.~Nu\~nez,
  Phys.\ Rev.\ Lett.\  {\bf 86} (2001) 588
  [arXiv:hep-th/0008001].




\bibitem{SakaiSonnenschein} T.~Sakai and J.~Sonnenschein,
  JHEP {\bf 0309} (2003) 047
  [arXiv:hep-th/0305049].


\bibitem{KSMN}
P.~Ouyang,
  Nucl.\ Phys.\ B {\bf 699} (2004) 207
  [arXiv:hep-th/0311084];\\
C.~Nu\~nez, A.~Paredes and A.~V.~Ramallo,
  JHEP {\bf 0312} (2003) 024
  [arXiv:hep-th/0311201];\\
  X.~J.~Wang and S.~Hu,
  JHEP {\bf 0309} (2003) 017
  [arXiv:hep-th/0307218].


\bibitem{NunezParedes}   R.~Casero, C.~Nunez and A.~Paredes,
  Phys.\ Rev.\ D {\bf 73} (2006) 086005
  [arXiv:hep-th/0602027];\\
  A.~Paredes,
  [arXiv:hep-th/0610270].

\bibitem{beyond}
 B.~A.~Burrington, J.~T.~Liu, L.~A.~Pando Zayas and D.~Vaman,
  JHEP {\bf 0502} (2005) 022
  [arXiv:hep-th/0406207];\\
I.~Kirsch and D.~Vaman,
  Phys.\ Rev.\ D {\bf 72} (2005) 026007
  [arXiv:hep-th/0505164];\\
 J.~Erdmenger and I.~Kirsch,
  JHEP {\bf 0412} (2004) 025
  [arXiv:hep-th/0408113].


\bibitem{Polchinski:2000uf}
  J.~Polchinski and M.~J.~Strassler,
  [arXiv:hep-th/0003136].


\bibitem{GKP}
 S.~B.~Giddings, S.~Kachru and J.~Polchinski,
  Phys.\ Rev.\ D {\bf 66} (2002) 106006
  [arXiv:hep-th/0105097].

\bibitem{Kachru}
  S.~Kachru, M.~B.~Schulz and S.~Trivedi,
  JHEP {\bf 0310} (2003) 007
  [arXiv:hep-th/0201028].


\bibitem{Freedman:2000xb}
  D.~Z.~Freedman and J.~A.~Minahan,
  JHEP {\bf 0101} (2001) 036
  [arXiv:hep-th/0007250].

\bibitem{LopesCardoso:2004ni}
  G.~Lopes Cardoso, G.~Curio, G.~Dall'Agata and D.~L\"ust,
  JHEP {\bf 0409}, 059 (2004)
  [arXiv:hep-th/0406118].


\bibitem{Warner2000}
  K.~Pilch and N.~P.~Warner,
  Nucl.\ Phys.\ B {\bf 594} (2001) 209
  [arXiv:hep-th/0004063];\\
 K.~Pilch and N.~P.~Warner,
  Adv.\ Theor.\ Math.\ Phys.\  {\bf 4} (2002) 627
  [arXiv:hep-th/0006066].


\bibitem{Pilch:2003jg}
  K.~Pilch and N.~P.~Warner,
  Nucl.\ Phys.\ B {\bf 675} (2003) 99
  [arXiv:hep-th/0306098].


\bibitem{Brandhuber}

  A.~Brandhuber and K.~Sfetsos,
  Phys.\ Lett.\ B {\bf 488} (2000) 373
  [arXiv:hep-th/0004148]. 


\bibitem{Donagi}
  R.~Donagi and E.~Witten,
  Nucl.\ Phys.\ B {\bf 460} (1996) 299
  [arXiv:hep-th/9510101].



\bibitem{Minasian} 
  M.~Mari\~no, R.~Minasian, G.~W.~Moore and A.~Strominger,
  JHEP {\bf 0001} (2000) 005
  [arXiv:hep-th/9911206].



\bibitem{Nekrasov} 
N.~Nekrasov and A.~S.~Schwarz,
  Commun.\ Math.\ Phys.\  {\bf 198} (1998) 689
  [arXiv:hep-th/9802068].

\bibitem{SeibergNC}
  N.~Seiberg and E.~Witten,
  JHEP {\bf 9909} (1999) 032
  [arXiv:hep-th/9908142].


\bibitem{Lerda} 
  M.~Billo, M.~Frau, F.~Fucito and A.~Lerda,
  [arXiv:hep-th/0606013].


\bibitem{Intriligator}
  K.~Intriligator, N.~Seiberg and D.~Shih,
  JHEP {\bf 0604} (2006) 021
  [arXiv:hep-th/0602239].


\bibitem{Bertolini2}
 R.~Argurio, M.~Bertolini, C.~Closset and S.~Cremonesi,
  JHEP {\bf 0609} (2006) 030
  [arXiv:hep-th/0606175];\\
 R.~Argurio, M.~Bertolini, S.~Franco and S.~Kachru,
  [arXiv:hep-th/0610212].



\bibitem{Johannes} 

J.~Erdmenger, J.~Gro\ss e and Z.~Guralnik,
  JHEP {\bf 0506} (2005) 052
  [arXiv:hep-th/0502224].


\bibitem{Myers:1999ps}
  R.~C.~Myers,
  JHEP {\bf 9912} (1999) 022
  [arXiv:hep-th/9910053].


\bibitem{Grana2}  M.~Gra\~na and J.~Polchinski,
  Phys.\ Rev.\ D {\bf 63} (2001) 026001
  [arXiv:hep-th/0009211].


\bibitem{Kuperstein}
S.~Kuperstein,
  JHEP {\bf 0503} (2005) 014
  [arXiv:hep-th/0411097].


\bibitem{Arean}
 D.~Arean, D.~E.~Crooks and A.~V.~Ramallo,
  JHEP {\bf 0411} (2004) 035
  [arXiv:hep-th/0408210].



\bibitem{Bandos}
 I.~Bandos and D.~Sorokin,
  [arXiv:hep-th/0607163].


%



\bibitem{Skenderis} 
 S.~de Haro, S.~N.~Solodukhin and K.~Skenderis,
  Commun.\ Math.\ Phys.\  {\bf 217}, 595 (2001)
  [arXiv:hep-th/0002230].


\bibitem{KarchOBannon} 
  A.~Karch, A.~O'Bannon and K.~Skenderis,
  JHEP {\bf 0604} (2006) 015
  [arXiv:hep-th/0512125].


\bibitem{Donoghue} J.F.~Donoghue, E.~Golowich and B.R.~Holstein,
  Dynamics of the Standard Model. Cambridge University Press, 1992.


\bibitem{GMOR}   M.~Gell-Mann, R.~J.~Oakes and B.~Renner,
  Phys.\ Rev.\  {\bf 175} (1968) 2195.



\bibitem{EvansShockWaterson}
 N.~Evans, J.~Shock and T.~Waterson,
  JHEP {\bf 0503} (2005) 005
  [arXiv:hep-th/0502091].

\bibitem{Sieg}
C.~Sieg, work in progress.

\bibitem{Taylor}
  M.~Taylor,
  [arXiv:hep-th/0103162].

\bibitem{DallAgata}
  G.~Dall'Agata,
  Nucl.\ Phys.\ B {\bf 695} (2004) 243
  [arXiv:hep-th/0403220].


\bibitem{Petrini}
  R.~Minasian, M.~Petrini and A.~Zaffaroni,
  [arXiv:hep-th/0606257].

\bibitem{Uranga} 
  J.~F.~G.~Cascales, F.~Saad and A.~M.~Uranga,
  JHEP {\bf 0511} (2005) 047
  [arXiv:hep-th/0503079].

\end{thebibliography}
\end{document}